\documentclass[12pt,oneside,english]{amsart}
\usepackage[T1]{fontenc}
\usepackage[utf8]{inputenc}
\usepackage[letterpaper]{geometry}
\usepackage{booktabs}
\usepackage{amstext}
\usepackage{amsthm}
\usepackage{amssymb}
\usepackage{setspace}
\usepackage{mathtools}
\usepackage[authoryear]{natbib}
\usepackage[shortlabels]{enumitem}
\usepackage[breaklinks,hidelinks]{hyperref}
\usepackage[ruled]{algorithm2e}
\makeatletter

\numberwithin{equation}{section}
\newtheoremstyle{exampstyle}
  {\topsep} % Space above
  {\topsep} % Space below
  {\itshape} % Body font
  {} % Indent amount
  {\bfseries} % Theorem head font
  {.} % Punctuation after theorem head
  {.5em} % Space after theorem head
  {} % Theorem head spec (can be left empty, meaning `normal')

\newtheorem{definition}{Definition} %[section]
\theoremstyle{exampstyle}
\newtheorem{lemma}{Lemma}

\theoremstyle{exampstyle}
\theoremstyle{exampstyle}
\newtheorem{cor}{Corollary}
\theoremstyle{exampstyle}

\newtheorem{assumption}{Assumption}
\newtheorem{thm}{Theorem}
\newcommand{\mockalph}[1]{}
\newcommand{\blind}{0}

\def\Cov{\mathrm{Cov}}
\def\Var{\mathrm{V}}
\include{def}
\def\dispmuskip{\thinmuskip= 3mu plus 0mu minus 2mu \medmuskip=  4mu plus 2mu minus 2mu \thickmuskip=5mu plus 5mu minus 2mu}
\def\textmuskip{\thinmuskip= 0mu                    \medmuskip=  1mu plus 1mu minus 1mu \thickmuskip=2mu plus 3mu minus 1mu}
\def\beq{\dispmuskip\begin{equation}}    \def\eeq{\end{equation}\textmuskip}
\def\beqn{\dispmuskip\begin{displaymath}}\def\eeqn{\end{displaymath}\textmuskip}
\def\bea{\dispmuskip\begin{eqnarray}}    \def\eea{\end{eqnarray}\textmuskip}
\def\bean{\dispmuskip\begin{eqnarray*}}  \def\eean{\end{eqnarray*}\textmuskip}
\newcommand{\wh}{\widehat}
\newcommand{\wt}{\widetilde}
\newcommand{\ov}{\overline}
\newcommand{\E}{\mathrm{E}}

\def\IF{\text{\rm IF}}
\usepackage{graphicx}
\usepackage{subcaption}
\usepackage{babel}
\usepackage{amsmath}
\usepackage{bbm}

\usepackage{algpseudocode}

\newcommand*{\patchAmsMathEnvironmentForLineno}[1]{%
      \expandafter\let\csname old#1\expandafter\endcsname\csname #1\endcsname
      \expandafter\let\csname oldend#1\expandafter\endcsname\csname end#1\endcsname
      \renewenvironment{#1}%
         {\linenomath\csname old#1\endcsname}%
         {\csname oldend#1\endcsname\endlinenomath}}%
    \newcommand*{\patchBothAmsMathEnvironmentsForLineno}[1]{%
      \patchAmsMathEnvironmentForLineno{#1}%
      \patchAmsMathEnvironmentForLineno{#1*}}%
    \AtBeginDocument{%
    \patchBothAmsMathEnvironmentsForLineno{equation}%
    \patchBothAmsMathEnvironmentsForLineno{align}%
    \patchBothAmsMathEnvironmentsForLineno{flalign}%
    \patchBothAmsMathEnvironmentsForLineno{alignat}%
    \patchBothAmsMathEnvironmentsForLineno{gather}%
    \patchBothAmsMathEnvironmentsForLineno{multline}%
    }
\usepackage[mathlines,displaymath]{lineno}

\mathchardef\mhyphen="2D

\makeatother

\usepackage{babel}

\bigskip

\usepackage{xr}

\begin{document}
%\linenumbers

\title[Exact Subsampling MCMC]{The Block-Poisson Estimator for Optimally Tuned \\Exact Subsampling MCMC}
\if0\blind
{
\author{Matias Quiroz$ ^{1,2}$, Minh-Ngoc Tran$ ^3$, Mattias Villani$ ^{4,5}$, \\ Robert Kohn$ ^6$ and Khue-Dung Dang$ ^1$}
%\author[1,2]{Matias Quiroz}
%\author[3]{Minh-Ngoc Tran}
%\author[1]{Robert Kohn}
%\author[4]{Mattias Villani}
%\author[1]{Khue-Dung Dang}
%\affil[1]{School of Economics, UNSW Business School,
%University of New South Wales}
%\affil[2]{Research Division, Sveriges Riksbank}
%\affil[3]{Discipline of Business
%Analytics, University of Sydney}
%\affil[4]{Division
%of Statistics and Machine Learning, Department of Computer and Information
%Science, Link\"{o}ping University}

\thanks{$^1$\textit{School of Mathematical and Physical Sciences, University of Technology Sydney.} $^2$\textit{Research Division, Sveriges Riksbank.} $^3$\textit{Discipline of Business
Analytics, University of Sydney.}  $^4$\textit{Department of Statistics, Stockholm University.} $^5$\textit{Department of Computer and Information
Science, Link\"{o}ping University.} $^6$\textit{School of Economics, UNSW Business School,
University of New South Wales.}}
}\fi

\begin{abstract}
Speeding up Markov Chain Monte Carlo (MCMC) for datasets with many
observations by data subsampling has recently received considerable
attention. A pseudo-marginal MCMC method is proposed that estimates the likelihood by data subsampling using a block-Poisson estimator. The estimator is a product of Poisson estimators, allowing us to update a single block of subsample indicators in each MCMC iteration so that a desired correlation is achieved between the logs of successive likelihood estimates. This is important since pseudo-marginal MCMC with positively correlated likelihood estimates can use substantially smaller subsamples without adversely affecting the sampling efficiency. The block-Poisson estimator is unbiased but not necessarily positive, so the algorithm runs the MCMC on the absolute value of the likelihood estimator and uses an importance sampling correction to obtain consistent estimates of the posterior mean of any function of the parameters.
Our article derives guidelines to select the optimal tuning parameters for our method and shows that it compares very favourably to regular MCMC without subsampling, and to two other recently proposed exact subsampling approaches in the literature.
\\
\noindent \textsc{Keywords}: Bayesian inference, Control
variates, Data subsampling, Exact inference, Poisson Estimator,
Pseudo-marginal MCMC.  \newpage{}
\end{abstract}

\maketitle

\section{Introduction\label{sec:Introduction}}

Standard Markov Chain Monte Carlo (MCMC) algorithms for Bayesian inference require a large number of likelihood function evaluations and are therefore prohibitively expensive for datasets with many observations. This is a real obstacle for industrial applications where huge amounts are recorded using modern sensor technology, and in the natural and social sciences where datasets are increasingly larger; digitalization is also turning text into data and also the humanities are now concerned with big data. A recent literature therefore attempts to speed up MCMC algorithms by using random subsets of the data; see \citet{korattikara2014austerity,bardenet2014towards,bardenet2015markov,maclaurin2014firefly,quiroz2016speeding, quiroz2018review} and \citet{bierkens2016zig}.
Section \ref{sec:Previous-research} briefly reviews these approaches
and highlights possible pitfalls.

\citet{bardenet2015markov} provide an excellent review of subsampling approaches and propose a positive unbiased estimator of the likelihood in a pseudo-marginal framework \citep{andrieu2009pseudo} to accelerate the Metropolis-Hastings (MH) algorithm \citep{metropolis1953equation,hastings1970monte}.
Their unbiased likelihood estimator is constructed from a sequence of unbiased
log-likelihood estimates from small batches of observations used in a Rhee-Glynn type debiasing estimator \citep{rhee2013unbiased}. To ensure positiveness, \citet{bardenet2015markov} use a lower bound for the log-likelihood estimates in all batches \citep{jacob2015nonnegative}, but they find that the large variability of their estimator causes the chain to mix poorly.

We propose an alternative unbiased estimator of the likelihood having several attractive properties and demonstrate that it can be successfully used for subsampling MCMC. This block-Poisson estimator is a product of Poisson estimators (Wagner \citeyear{wagner1988unbiased};
\citealp{papaspiliopoulos2009methodological}), where each estimator is based on a block of observations. Our estimator has the following four key features. First, the product form makes it possible to only update the subsamples in some of the blocks in each iteration to target a given correlation between the logs of the estimated likelihood at the current and proposed MH draws. \cite{deligiannidis2015correlated} show that such dependent pseudo-marginal schemes are very efficient because they can use substantially noisier likelihood, and hence smaller subsamples, without adversely affecting the sampling efficiency of the chain. Our blocking strategy makes it possible to directly control the correlation, and to update the subsample without proposing moves for all the subsample indicators in the dataset. Second, the block-Poisson estimator has a lower variance than the Rhee-Glynn estimator in \citet{bardenet2015markov}. Third, the block-Poisson estimator uses variance reducing control variates proposed in \citet{bardenet2015markov} and \citet{quiroz2016speeding}. Fourth, the block-Poisson estimator uses a soft lower bound for its constituent batch estimators, rather than a strict lower bound as in \citet{bardenet2015markov}. Section \ref{subsec:On-the-lower} explains why a soft lower bound is computationally more efficient than a strict bound. However, the soft bound makes it possible to obtain negative likelihood estimates, which cannot be used within the usual pseudo-marginal MCMC framework in \citet{andrieu2009pseudo}. \citet{lyne2015russian} propose an ingenious solution to the problem of negative likelihood estimates by running the pseudo-marginal sampler on the absolute value of the likelihood estimate followed by an importance sampling correction step to estimate the posterior mean of any function of the model parameters. We refer to this pseudo-marginal Metropolis-Hastings with importance sampling sign correction as \emph{signed PMMH}.

It is well known that pseudo-marginal algorithms need to carefully tune the number of particles (subsamples) and several recent papers develop practical guidelines for this choice for strictly positive likelihood estimators; see \citet{pitt2012some}, \citet{doucet2012efficient} and \citet{sherlock2013efficiency}. A major contribution of our article is the derivation of easily implemented guidelines for the optimal number of subsamples (or more generally particles) for the signed PMMH based on the block-Poisson estimator. The optimal number of subsamples is chosen to minimize the asymptotic variance of the importance sampling estimator for a given computational budget and therefore balances: i) the inefficiency in the MCMC on the absolute measure, ii) the computational cost of the likelihood estimator and iii) the probability of negative estimates in signed PMMH. We show that the asymptotic variance of the importance sampling estimator can be obtained in closed form when the likelihood is estimated by the block-Poisson estimator. The guidelines are derived under idealized conditions, but are demonstrated to be quite accurate in empirical experiments. The proposed framework to derive these optimality results applies more generally than subsampling; for example it can be applied to doubly intractable problems using the Exponential auxiliary variable construction in \citet{lyne2015russian}.

Our article demonstrates that subsampling MCMC using the block-Poisson estimator is
an efficient sampler that mixes well and is more efficient than a MH sampler on the full sample. It also shows empirically that our exact subsampling MCMC is much more efficient than Firefly Monte Carlo \citep{maclaurin2014firefly}, a highly cited exact subsampling algorithm. We also show empirically that our approach scales better with respect to the dimension of the parameter space than the zig-zag sampler in \citet{bierkens2016zig}. Our approach is also fully automatic without requiring manual derivations of model specific upper bounds; see Section \ref{subsec:zigzag_experiment}.

The rest of the article is organized as follows. Section \ref{sec:Previous-research} briefly reviews the main
subsampling approaches proposed in the recent literature. Section
\ref{sec:Unbiased-Likelihood-estimator} introduces the block-Poisson estimator and derives its key properties.
Section \ref{sec:Methodology} outlines the proposed sampling algorithm and provides guidelines on the selection of tuning parameters to obtain an optimal implementation. Section \ref{sec:Experiments} demonstrates the methodology and Section \ref{sec:Conclusions} concludes.
 The paper has an online supplement containing proofs, additional material and extensions. We refer to equations, sections, lemmas in the main paper as (1), Section~1, Lemma~1 etc., and to equations, sections and lemmas, etc in the supplement as (S1), Section~S1 and Lemma~S1, etc.

\section{Previous research\label{sec:Previous-research}}

Previous research in scalable MCMC by data subsampling is either approximate
\citep[e.g.][]{korattikara2014austerity,bardenet2014towards,bardenet2015markov,quiroz2016speeding}
or exact \citep[e.g.][]{maclaurin2014firefly,bierkens2016zig}.
\cite{banterle2014accelerating}, \cite{payne2015bayesian} and \cite{quiroz2015delayed} propose using delayed acceptance MCMC  where the acceptance probability in the first stage is computed from on a subsample. However, the second stage is computed on the full sample and therefore the methods are not fully subsampling approaches.

The algorithms in \citet{korattikara2014austerity} and \citet{bardenet2014towards,bardenet2015markov}
replace the computationally costly MH ratio with a hypothesis
test based on a fraction of the data; \citet{bardenet2015markov} and \cite{quiroz2016speeding} evaluate these methods.

\citet{quiroz2016speeding} estimate the MH ratio based on a random
subsample of the data and use a dependent pseudo-marginal approach to sample from the
posterior. They derive an efficient unbiased log-likelihood estimator
with control variates and apply an approximate bias-correction to get an {\it approximately} unbiased likelihood estimator. Their target MCMC distribution is a perturbation of the posterior which they show is within $O(n^{-1}m^{-2})$ of the true posterior, where $n$ is the sample size and $m$ the subsample size. However, their bias-correction implies that the approximation error is a function of the variance of the log-likelihood estimator, and targeting a large variance may therefore degrade the posterior approximation.
The approximation error is small in their examples even when the variance
of the log-likelihood estimator is large, but there is no guarantee that this will hold in other applications. In contrast, our proposed
method is exact in the sense that it provides simulation consistent estimates of posterior expectation of any function of the parameters, regardless of the estimator's variability. See Section \ref{app:ExactVsApprox} of the supplement for a comparison with the approximate approach in \citet{quiroz2016speeding}.

The Firefly Monte Carlo algorithm in \citet{maclaurin2014firefly} introduces an auxiliary variable for each observation which determines if it should be included in the evaluation of the posterior in a given MCMC iteration. A lower bound for each likelihood term caters for the excluded observations. The authors suggest using Gibbs sampling with the original parameters and the auxiliary variables in two different blocks. The method has been documented to be very inefficient; see e.g. \citet{bardenet2015markov,li2020improving,quiroz2016speeding}.

The zig-zag sampler \citep{bierkens2016zig} and the bouncy particle sampler \citep{bouchard2018bouncy} are two recent exact non-reversible subsampling MCMC algorithms based on piecewise deterministic Markov processes. While the samples from the zig-zag and bouncy particle algorithms converge in distribution to the posterior, our method only gives simulation consistent estimates of the posterior mean of functions of the parameters. Section \ref{subsec:zigzag_experiment} compares our approach with the zig-zag sampler.

\section{The block-Poisson estimator\label{sec:Unbiased-Likelihood-estimator}}

\subsection{The estimator and its properties\label{subsec:Unbiased-estimator}}
Suppose that the observations $y\coloneqq(y_1\dots, y_n)$ are conditionally independent given a parameter vector $\theta \in \Theta \subset \mathbb{R}^p$. The likelihood can then be written as
\begin{equation}
L(\theta)\coloneqq p(y|\theta)=\exp(\ell(\theta)),\quad \ell(\theta)=\sum_{k=\text{1}}^{n}\ell_{k}(\theta),\quad\text{where }\ell_{k}(\theta) \coloneqq \log p(y_{k}|\theta,x_{k})\label{eq:cross_sec_likelihoods}
\end{equation}
is the log-likelihood contribution of the $k$th observation with covariates $x_k$. Our method can also be used for longitudinal data where subjects are subsampled instead of individual observations. It also applies to stationary time series data by subsampling in the frequency domain; see \cite{salomone2020spectral}.

The block-Poisson estimator of the likelihood in Definition \ref{BlockPoisDef} below relies on an unbiased estimator of the log-likelihood, $\ell(\theta)$. \citet{quiroz2016speeding} propose estimating $\ell(\theta)$ by a \emph{difference estimator} based on computationally efficient control variates. The control variate $q_{k}(\theta)$ for the $k$th observation is
such that the difference $d_{k}(\theta) = \ell_{k}(\theta) - q_{k}(\theta)$ is small. The difference estimator uses the equation
\begin{equation}
    \ell(\theta) = q(\theta) + d(\theta), \text{ where } q(\theta)=\sum_{k=1}^n q_{k}(\theta) \text{ and } d(\theta)=\sum_{k=1}^n d_{k}(\theta)
\end{equation}
to estimate $\ell(\theta)$ by unbiasedly estimating $d(\theta)$. Since the control variates capture the size of each $\ell_k(\theta)$, it is efficient to use a simple sample mean estimator from a subsample of size $m$ drawn with replacement to estimate $d$ as
\begin{equation}
    \widehat{d}_{m}=\frac{n}{m}\sum_{i=1}^{m}d_{u_{i}},
\end{equation}
where the $u_i$ are iid from the distribution $\Pr(u_{i}=k)=1/n$ for $k=1,\dots,n$, and we suppress the dependence on $\theta$ to simplify notation. The variance of the difference estimator is
\begin{equation*}
%\label{eq:variance_log_likelihood_estimator}
\sigma_{\widehat{d}}^2 \coloneqq \mathrm{V}[\widehat{d}_m]=\frac{\gamma}{m}, \text{ where } \gamma \coloneqq n^2 \sigma^2_{d_{u_i}}\text{ with } \sigma^2_{d_{u_i}} \coloneqq \mathrm{V}[d_{u_i}].
\end{equation*}

There are several proposals for unbiasedly estimating the likelihood in the literature; see e.g. the Rhee-Glynn estimator \citep{rhee2013unbiased}, Russian roulette estimators \citep{lyne2015russian} and the Poisson estimator \citep{wagner1988unbiased}. We now propose
 a modified Poisson estimator which is constructed to be particularly useful for the block pseudo-marginal algorithm developed in Section \ref{sec:Methodology}.
\begin{definition}\label{BlockPoisDef}
The block-Poisson likelihood estimator is defined as
\begin{equation}
\widehat{L}_{B}(\theta) \coloneqq \exp(q(\theta)) \prod_{l=1}^{\lambda} \xi_l, \, \text{ where } \hspace{0.2cm}\xi_l = \exp\left(\frac{a+\lambda}{\lambda}\right)\prod_{h=1}^{\mathcal{X}_l}\left(\frac{\widehat{d}_{m}^{\,\,(h,l)}-a}{\lambda}\right), \label{eq:UnbiasedLikelihoodEstimator}
\end{equation}
where $\lambda$ is a positive integer, $a$ is a real number, $\widehat{d}_{m}^{\,\,(h,l)}$ is some unbiased estimator of $d=\ell-q$ from a small batch of $m$ observations, and  $\mathcal{X}_1,\dots,\mathcal{X}_\lambda$ are independent $\mathrm{Pois}(1)$ variables. The rightmost product is defined as $1$ when $\mathcal{X}_l=0$.
\end{definition}
Important properties of the estimator are listed in Lemma \ref{lem:PoissonEstimator} below. The block-Poisson estimator in \eqref{eq:UnbiasedLikelihoodEstimator} is a product over $\lambda$ Poisson estimators; this makes it possible to update the subsampled observations only in a subset of the products, thereby generating a correlation between estimates in successive MCMC iterations; see Section \ref{sec:Methodology}. The product form also makes it possible to use fixed, non-random, number of $\lambda$ factors in the estimator, which simplifies the properties of the estimator. \cite{deligiannidis2015correlated} show that inducing dependence between likelihood estimates over the MCMC iterations makes it possible to use a much more variable likelihood estimator; this means that much smaller subsamples can be used so that each MCMC iteration is much faster. The reason for using the product of Poisson estimators is to induce correlation, not to reduce the variance of the estimator; see Section \ref{subsec:signed-block-algorithm}. In fact, Part (v) of Lemma \ref{lem:PoissonEstimator} below shows that the variance of the block-Poisson estimator is the same as the standard Poisson estimator \citep{papaspiliopoulos2009methodological}
\begin{eqnarray}
\widehat{L}_P(\theta) & \coloneqq & \exp(q(\theta)) \exp(a+\lambda)\prod_{h=1}^{G}\left(\frac{\widehat{d}_{m}^{\,\,(h)}-a}{\lambda}\right),\quad G\sim\mathrm{Pois}(\lambda),\label{eq:UnbiasedLikelihoodEstimatorPoisson}
\end{eqnarray}
where the product is $1$ if $G=0$. Lemma \ref{lem:PoissonEstimator} states some useful properties of the $\widehat{L}_{B}(\theta)$ estimator.
\begin{lemma}\label{lem:PoissonEstimator}
Assume $\sigma^2_{\widehat{d}} < \infty$. Then, for any $\theta \in \Theta,$
\begin{enumerate}[topsep=0pt, label={\emph{(\roman*)}}]
\item $\mathrm{E}[\widehat{L}_{B}]=L$.
\item If $\widehat{d}_m^{\,\,(h, l)} \geq a$ almost surely for all $h$ and $l$, then $\widehat{L}_{B}$ is almost surely non-negative.
\item $\mathrm{V}[|\widehat{L}_B|] < \infty $
\item For a fixed $\lambda$, $\mathrm{V}[\widehat{L}_B]$ is minimized at $a = d-\lambda$.
\item $\mathrm{V}[\widehat{L}_B(\theta)] = \mathrm{V}[\widehat{L}_P(\theta)]$.
\end{enumerate}
\end{lemma}
\vspace{0.1cm}
Although  $\wh L_B$ is unbiased by Part (i) of Lemma~\ref{lem:PoissonEstimator}, Part (ii) shows
 that the estimator is only positive with probability 1 if $a$ is a lower bound of all $\widehat{d}_{m}^{\,\,(h, l)}$. Section \ref{subsec:On-the-lower} shows that it is prohibitively expensive to choose $a$ in this way.
      We therefore adopt the approach in \cite{lyne2015russian} and carry out the pseudo-marginal MCMC on $|\widehat{L}_B|$.
       Part (iii) ensures that $|\widehat{L}_B|$ has a finite variance. Part (iv) motivates the simplification of setting $a = d-\lambda$ so that it is only necessary to optimize performance with respect to $m$ and $\lambda$. Under this simplifying assumption,
\begin{equation}
\widehat{L}_{B}(\theta) = \exp(q(\theta)) \prod_{l=1}^{\lambda} \xi_l, \, \text{ where } \hspace{0.2cm} \xi_l = \exp\left(\frac{d}{\lambda}\right)\prod_{h=1}^{\mathcal{X}_l}\left(\frac{\widehat{d}_{m}^{\,\,(h,l)}-d}{\lambda} + 1\right). \label{eq:SimplifiedLikelihoodEstimator}
\end{equation}

Section \ref{sec:Methodology} derives the optimal tuning parameters $m$ and $\lambda$ by studying a performance measure involving the variance of the log of the estimator. Since the MCMC is based on $|\widehat{L}_B|$, it is necessary to evaluate the variance of $\log |\widehat{L}_B|$. Lemma \ref{lem:VarAbs} gives an expression for this variance assuming that the batch means $\widehat{d}_{m}^{\,\,(h,l)}$ are Gaussian; when $m$ is moderately large this is motivated by the Central Limit Theorem (CLT). For small $m$, Lemma \ref{lem:VarAbsL_finite_mixture} gives the corresponding result when the $\widehat{d}_{m}^{\,\,(h,l)}$ follow a finite mixture of normals distribution.
\begin{lemma}
\label{lem:VarAbs}If $\widehat{d}_{m}^{\,\,(h,l)}\overset{iid}{\sim}\mathcal{N}(d,\frac{\gamma}{m})$
for all $h$ and $l$, then the variance of $\log|\widehat{L}_B|$
when $a=d-\lambda$ is
\begin{align*}
\sigma^2_{\log |\widehat{L}_B|} & =\lambda(\nu^{2}+\eta^{2}),
\end{align*}
where
\[
\eta \coloneqq \log\left(\sqrt{\frac{\gamma}{m\lambda^{2}}}\right)+\frac{1}{2}\left(\log2+\mathrm{E}_{J}\left[\psi^{(0)}(1/2+J)\right]\right)
\]
 and
\[
\nu^{2} \coloneqq \frac{1}{4}\left(\mathrm{E}_{J}\left[\psi^{(1)}(1/2+J)\right]+\mathrm{V}_{J}\left[\psi^{(0)}(1/2+J)\right]\right),
\]
with $J\sim\mathrm{Pois}\left(\frac{m\lambda^{2}}{2\gamma}\right)$
and $\psi^{(q)}$ the polygamma function of order $q$. Furthermore,
$\sigma^2_{\log |\widehat{L}_B|}<\infty$ for all
$m>0$, $\lambda>0$ and $\gamma>0$.
\end{lemma} The infinite sums in the expectation and variance with respect to the Poisson random variable $J$ in Lemma \ref{lem:VarAbs} are accurately approximated by truncation since $\Pr(J = j)$ decreases very quickly to zero as $j$ increases and the polygamma functions are either bounded or grow much slower than the rate of decrease of $\Pr(J = j)$, see the proof of Lemma \ref{lem:VarAbs} in Section \ref{app:ProofsCollected}.

\subsection{Control variates\label{subsec:Control-variates}}
Two different types of control variates $q_k(\theta)$ are used to approximate $\ell_k(\theta)$, both based on a Taylor expansion of the log density $\ell_k(\theta)\coloneqq\log p(y_k|x_k, \theta)$.

The \textit{parameter expanded} control variate suggested by \citet{bardenet2015markov} expands $\ell_k(\theta)$ around some reference value $\theta^{\star}$; it has computational complexity of $O(1)$ \citep{bardenet2015markov} and therefore the overall Computational Cost (CC) for $\widehat{L}_B$ is $m\sum_{l=1}^\lambda\mathcal{X}_l$.

\citet{quiroz2016speeding} present \textit{data expanded} control variates where the expansion of $\ell_k(\theta)$ with respect to $\eta_k = (y_k, x_k)$ is around a reference value of the data $\eta^{\star}$. To make the approximation local, a sparse subset of the data is obtained by clustering the data into $\mathcal{K}$ clusters before the MCMC. For each data point $\eta_k$ belonging to cluster $c$, $q_{k}(\theta)$ is an expansion around the centroid $\eta_c^{\star}$; this control variate has a computational complexity of $O(\mathcal{K})$ \citep{quiroz2016speeding} and an overall computational cost for $\widehat{L}_B$ of $m\sum_{l=1}^\lambda\mathcal{X}_l + O(\mathcal{K})$. See \citet{quiroz2016speeding} for the asymptotic properties of these two types of control variates with respect to $n$.

\subsection{Soft lower bound\label{subsec:On-the-lower}}
While the lower bound of all $\widehat{d}_{m}^{\,\,(h,l)}$ ensures that $\widehat{L}_B$ is positive (Lemma \ref{lem:PoissonEstimator} part (ii)), it is impractical for two reasons. First, for most models it is necessary to evaluate $d_{k}=\ell_{k}-q_{k}$ for all data points to find a lower bound. Second, the optimal implementation outlined in Section \ref{subsec:Unbiased-estimator} requires that $\lambda = d - a$. If the control variates are accurate then $d$ is small relative to $a$, implying that $\lambda \approx -a$. Hence, a large $-a$ implies a large number of products in the block-Poisson estimator, and a large computational cost.

We therefore advocate using a \emph{soft lower bound}, i.e. one that is not necessarily a lower bound for all outcomes of $\widehat{d}_m^{\,\,(h, l)}$ but still gives a $\Pr(\widehat{L}_B\geq 0 )$ close to one. The soft lower bound makes it possible to obtain negative likelihood estimates, and Section \ref{subsec:tuningProductPoisson} shows that $\Pr(\widehat{L}_B \geq 0)$ is a crucial quantity for the efficiency of our method. Lemma \ref{lem:PrNonnegative}
gives an analytically tractable expression for this probability.

\begin{lemma}\label{lem:PrNonnegative}
$$\Pr(\widehat{L}_B \geq 0) = \frac{1}{2}\left(1 + \left(1 - 2\Psi(a, m, \lambda, \gamma)\right)^\lambda\right),$$
where
$$\Psi(a, m, \lambda, \gamma) \coloneqq  \Pr(\xi_l < 0) = \frac{1}{2} \sum_{j = 1}^\infty  \left(1 - \left(1 - 2\Pr\left(A_m < 0\right)\right)^j\right)\Pr(\mathcal{X}_l = j), \quad \mathcal{X}_l \sim \mathrm{Pois}(1),$$
and $A_m \coloneqq (\widehat{d}_m - d)/\lambda + 1$.
\end{lemma}

The probability of a positive estimator $\widehat{L}_B$ can be computed whenever $\Pr(A_m < 0)$ is analytically available. By the CLT, we can expect $A_m$ to be approximately normally distributed when $m$ is sufficiently large as $\widehat{d}_m$ will be normal in this case. When $m$ is small, Section \ref{app:optimalTuningGeneral} of the supplement describes a general approach where $\widehat{d}_m$ is modelled by a mixture of normals. 

\section{The signed block PMMH algorithm\label{sec:Methodology}}
This section outlines the proposed posterior sampling algorithm for the block-Poisson estimator, and derives guidelines for optimally tuning the algorithm.
\subsection{The signed PMMH algorithm\label{subsec:A-pseudo-marginal-algorithm}}
Let $p_{\Theta}(d\theta)$ and $\pi(d\theta)\coloneqq p(d\theta|y) \propto L(\theta)p_{\Theta}(d\theta)$ denote the prior and the posterior measures of $\theta$.
Let $\widehat{L}(\theta, U)$ be an unbiased, but not necessarily positive estimator of the likelihood,
for example the block-Poisson in \eqref{eq:UnbiasedLikelihoodEstimator}; $U$ is the set of all random variables involved when constructing $\widehat L$ and we usually take $U$ to be a set of uniform random numbers with $p_U(du)$ the probability measure of $U$. The unbiasedness of $\widehat{L}(\theta, U)$ means that
\begin{eqnarray}
L(\theta) & = & \int \widehat{L}(\theta, u) p_U(du) , \;\;\text{for any }\; \theta.\label{eq:UnbiasedLikelihood}
\end{eqnarray}
It is invalid to define a target posterior measure
 using the estimator $\widehat{L}(\theta, u)$ because it can be negative.
Instead, we define the joint target   measure
\begin{equation}
\overline{\pi}(d\theta,du) \coloneqq \frac{1}{\overline{C}}|\widehat{L}(\theta, u)|p_{\Theta}(d\theta)p_U(d u), \,\, \overline{C} \coloneqq \int_{\Theta} C(\theta)p_{\Theta}(d\theta), \,\, C(\theta) \coloneqq \int |\widehat{L}(\theta, u)|p_U(du). \label{eq:AugmentedPosterior}
\end{equation}
This is a proper Lebesgue product measure on $\Theta \otimes \mathcal{U}$, and admits the posterior $\pi(d\theta)$ as its marginal measure only if $\widehat{L}(\theta, u) \geq 0$ almost surely. Define
\begin{equation}\label{eq:nu_theta}
    \bar \pi(d\theta) \coloneqq \int  \overline{\pi}(d\theta,du) = \frac{C(\theta)p_{\Theta}(d\theta)}{\overline{C}}.
\end{equation}
Let $S(\theta, u)\coloneqq\mathrm{sign}(\widehat{L}(\theta, u))\in \mathcal{S} \coloneqq \{-1, 1\}$. An important goal in Bayesian inference is to estimate an integral of the form
\[\mathrm{E}_{\pi}[\psi]=\int_\Theta \psi(\theta)\pi(d\theta)\]
for some function $\psi(\theta)$ on $\Theta$. \citet{lyne2015russian} cleverly note that
\begin{eqnarray}
\mathrm{E}_{\pi}[\psi] = \frac{\int_\Theta \psi(\theta)L(\theta)p_\Theta(d\theta)}{\int_\Theta L(\theta) p_\Theta(d\theta)} & = & \frac{\int_\Theta  \int_{\mathcal{U}} \psi(\theta)S(\theta,u)\overline{\pi}(d\theta, du)}{\int_\Theta \int_{\mathcal{U}}S(\theta,u)\overline{\pi}(d\theta, du)}
= \frac{\mathrm{E}_{\ov \pi}[\psi S]}{\mathrm{E}_{\ov \pi}[S]}   .\label{eq:EstimationOfFunctional}
\end{eqnarray}
Hence, we can use a pseudo-marginal scheme \citep{andrieu2009pseudo} to obtain $N$ samples $\{\theta^{(i)},u^{(i)}, i = 1,\dots, N \}$ from $\overline{\pi}(d\theta, du)$ in \eqref{eq:AugmentedPosterior} and then estimate \eqref{eq:EstimationOfFunctional} by
\begin{eqnarray}
\widehat{\mathrm{E}}_{\pi}[\psi] & = & \frac{\sum_{i=1}^{N}\psi(\theta^{(i)})s^{(i)}}{\sum_{i=1}^{N}s^{(i)}}. \label{eq:ISestimator}
\end{eqnarray}
Hence, to compute $\mathrm{E}_{\pi}[\psi]$ it is only necessary to store the $\theta^{(i)}$ and the signs $s^{(i)}=S(\theta^{(i)},u^{(i)})$ in the MCMC.

This elegant observation allows exact
inference in the sense that this estimator is guaranteed to converge $\overline{\pi}$-almost surely to the true value $\mathrm{E}_{\pi}[\psi]$ as $N \rightarrow \infty$, even though the $\theta$ iterates are distributed as $\bar\pi(d\theta) \neq \pi(d\theta)$.

\subsection{The signed block PMMH algorithm\label{subsec:signed-block-algorithm}}

\citet{deligiannidis2015correlated} propose correlating $\log \widehat{L}(\theta,u)$ at the current and proposed MCMC draws using an autoregressive proposal of the underlying random numbers $u$. This correlated pseudo-marginal algorithm is a substantial advance over the standard PMMH because a much larger $\sigma_{\log \widehat{L}}^{2}$ can be tolerated without adversely affecting the sampling efficiency.

We now propose an alternative way to correlate the estimates by partitioning $U$ into $G$ blocks and updating only a single random block together with $\theta$ in each iteration, keeping the other blocks fixed. In the block-Poisson estimator, we use $U_l$ as the set of random numbers to compute $\xi_l$, $l = 1, \dots, \lambda$, and group them as
\begin{eqnarray}
\label{eq:U_for_blocking}
U & = & (U_1, \dots, U_G)\coloneqq U_{1:G},
\end{eqnarray}
so that each $U_i$, $i = 1,\dots, G,$ contains $\lambda/G$ of the $U_l$. Note that we do not require that $G=\lambda$, so the blocks in the block-Poisson estimator do not need to correspond to the blocking of $U$ in \eqref{eq:U_for_blocking}. Section \ref{app:corrFromBlocking} shows that $\mathrm{Corr}\big(\ell(\theta',u'),\ell(\theta,u)\big)\approx1-1/G$, where $(\theta,u)$ and $(\theta',u')$ are the current and proposed draws, respectively. This shows that the block-Poisson estimator in \eqref{eq:UnbiasedLikelihoodEstimator} can produce a pre-specified correlation between the log of the likelihood estimates at successive iterations. Moreover, generating correlation through blocking in a subsampling context is computationally more efficient than using the method in \citet{deligiannidis2015correlated} since it is unnecessary to work with the whole dataset and to generate the full set of underlying random numbers in each MCMC iteration.

Algorithm \ref{alg:implementation} outlines the signed block PMMH with the block-Poisson estimator. The MCMC part of the algorithm targets the absolute measure
\begin{equation}
\overline{\pi}(d\theta,du_{1:G}) \coloneqq \frac{p_{\Theta}(d\theta)}{\overline{C}}\prod_{i = 1}^G| \widehat{L}^{(i)}(\theta, u_i)|p_{U_i}(du_i), \label{eq:AugmentedPosterior_blocking}
\end{equation}
where each $\widehat{L}^{(i)}$ with $\mathrm{E}[\widehat{L}^{(i)}] = L(\theta)^{1/G}$ corresponds to the likelihood estimate when using $U_i$, so that $\mathrm{E}\left[\prod_{i=1}^G \widehat{L}^{(i)}\right] = L(\theta)$. Section \ref{app:convergence} proves the convergence of the MCMC on $\bar\pi$ as well as a central limit theorem for the estimator $\widehat{\mathrm{E}}_{\pi}[\psi]$.

\setcounter{algocf}{0}
\begin{algorithm}
    \caption{The signed block PMMH algorithm}\label{alg:implementation}
    
  \KwIn{Number of blocks $G$, number of MH iterations $N$, and initial values $\theta^{(0)}$, $u^{(0)} = (u^{(0)}_1,\dots,u^{(0)}_G)$.} \vspace{1mm}
  Obtain the optimal tuning parameters $\lambda_{\mathrm{opt}}$,$m_{\mathrm{opt}}$ and $a_{\mathrm{opt}}$ from Algorithm \ref{alg:optTuning}. \\\vspace{1mm}
  \For {$i = 1,\dots, N$}{
 Propose $u' \sim q_U(u^{(i-1)};du')$ by updating a randomly chosen block $u^{(i-1)}_g$ in $u^{(i-1)}$.\\ \vspace{1mm}
 Propose $\theta' \sim q_\Theta(\theta^{(i-1)};d\theta')$.\\ \vspace{1mm}
 Compute $\wh  L (\theta',u')$ by Equation \eqref{eq:UnbiasedLikelihoodEstimator}. \\ \vspace{1mm}
 Set $(\theta^{(i)},u^{(i)}) \gets (\theta',u')$ with probability  $$\alpha_{\Theta,U} (\theta^{(i-1)}, u^{(i-1)}, \theta',u') = 1\wedge\frac{|\wh L (\theta',u')| p_\Theta(\theta') }{ |\wh L (\theta^{(i-1)},u^{(i-1)})| p_\Theta(\theta^{(i-1)})} \frac{q_\Theta(\theta'; d\theta^{(i-1)}) }{q_\Theta(\theta^{(i-1)};d\theta')},$$
else $(\theta^{(i)},u^{(i)}) \gets (\theta^{(i-1)},u^{(i-1)})$. \\ \vspace{1mm}
 Set $s^{(i)} = S(\theta^{(i)},u^{(i)})$. \\ \vspace{1mm}
 }
\KwOut{Return the estimate$$ \widehat{\mathrm{E}}_{\pi}[\psi]  = \frac{\sum_{i=1}^{N}\psi(\theta^{(i)})s^{(i)}}{\sum_{i=1}^{N}s^{(i)}}. $$ }
\end{algorithm}

\subsection{Tuning the signed block PMMH algorithm\label{subsec:tuningProductPoisson}}
This section outlines how to choose the tuning parameters $m$, $\lambda$ and $a$ for the signed block PMMH using the block-Poisson estimator. For brevity and clarity, all assumptions, technical details and derivations are in Section \ref{app:IF}; we focus here on the main ideas and algorithms.
The methodological framework is based on \citet{pitt2012some}, which we extend to the signed block PMMH algorithm. Similarly to \citet{pitt2012some}, the derived guidelines are based on stylized assumptions that make the analysis tractable and transparent. Section \ref{app:testGuidelines} verifies that the guidelines are accurate and robust to deviations from the assumptions. Section \ref{app: block PMMH L geq 0} presents results for the optimal tuning of the block PMMH when the likelihood estimator is always non-negative.

First, the soft lower bound $a$ is set to its optimal value $a=d-\lambda$, with $d$ replaced by an estimate (see Algorithm \ref{alg:optTuning} below), using the simplified estimator in \eqref{eq:SimplifiedLikelihoodEstimator}. We have found that the randomness from replacing $d$ with an estimate does not affect the tuning procedure. 

We now define the objective function which is used to find the optimal $\lambda$ and $m$. A natural objective is the computational time needed to obtain the same precision in the importance sampling estimator as that from a single independent Monte Carlo draw from $\pi(\theta)$. Following \cite{pitt2012some}, we define the Computational Time (CT) as
\begin{equation*}
    \mathrm{CT} \coloneqq \text{Computing cost per iteration}\cdot \frac{\mathrm{V}_{\pi}\widehat{\mathrm{E}}_{\pi}[\psi]}{\mathrm{V}_{\pi}\widehat{\mathrm{E}}_{\mathrm{MC}}[\psi]},
\end{equation*}
where $\mathrm{V}_{\pi}\widehat{\mathrm{E}}_{\mathrm{MC}}[\psi] = \mathrm{V}_{\pi}(\psi)/N$ is the variance of the ideal Monte Carlo estimator from iid sampling from $\pi(\theta)$, and
\begin{equation*}
    \mathrm{V}_{\pi}\widehat{\mathrm{E}}_{\pi}(\psi) = \frac{\Var_{\ov \pi} (\psi S)}{N}\cdot\frac{\mathrm{IF}_{\ov \pi, \psi S}}{(2\tau-1)^2}
\end{equation*}
is the asymptotic variance of the estimator in Algorithm \ref{alg:implementation} given in Theorem \ref{lemma: signed pmmh sampling scheme} in Section \ref{app:convergence}. $\mathrm{IF}_{\ov \pi, \psi S}$ is the usual MCMC inefficiency for samples of $\psi S$ from the MCMC targeting the absolute measure $\ov \pi(d\theta,du_{1:G})$ in \eqref{eq:AugmentedPosterior_blocking}. The factor $(2\tau - 1)^{-2}$, where $\tau \coloneqq \Pr(\widehat L_B > 0)$, comes from the importance sampling correction for the sign. This gives
\begin{equation}\label{eq:realCT}
    \mathrm{CT} = m\lambda \cdot \frac{\Var_{\ov \pi} (\psi S)}{\mathrm{V}_{\pi}(\psi)} \cdot \frac{\mathrm{IF}_{\ov \pi, \psi S}}{(2\tau-1)^2},
\end{equation}
where $m\lambda$ is proportional to the expected cost per iteration (the cost is random when using the block-Poisson estimator).

\begin{figure}
\centering
\includegraphics[width=0.90\linewidth]{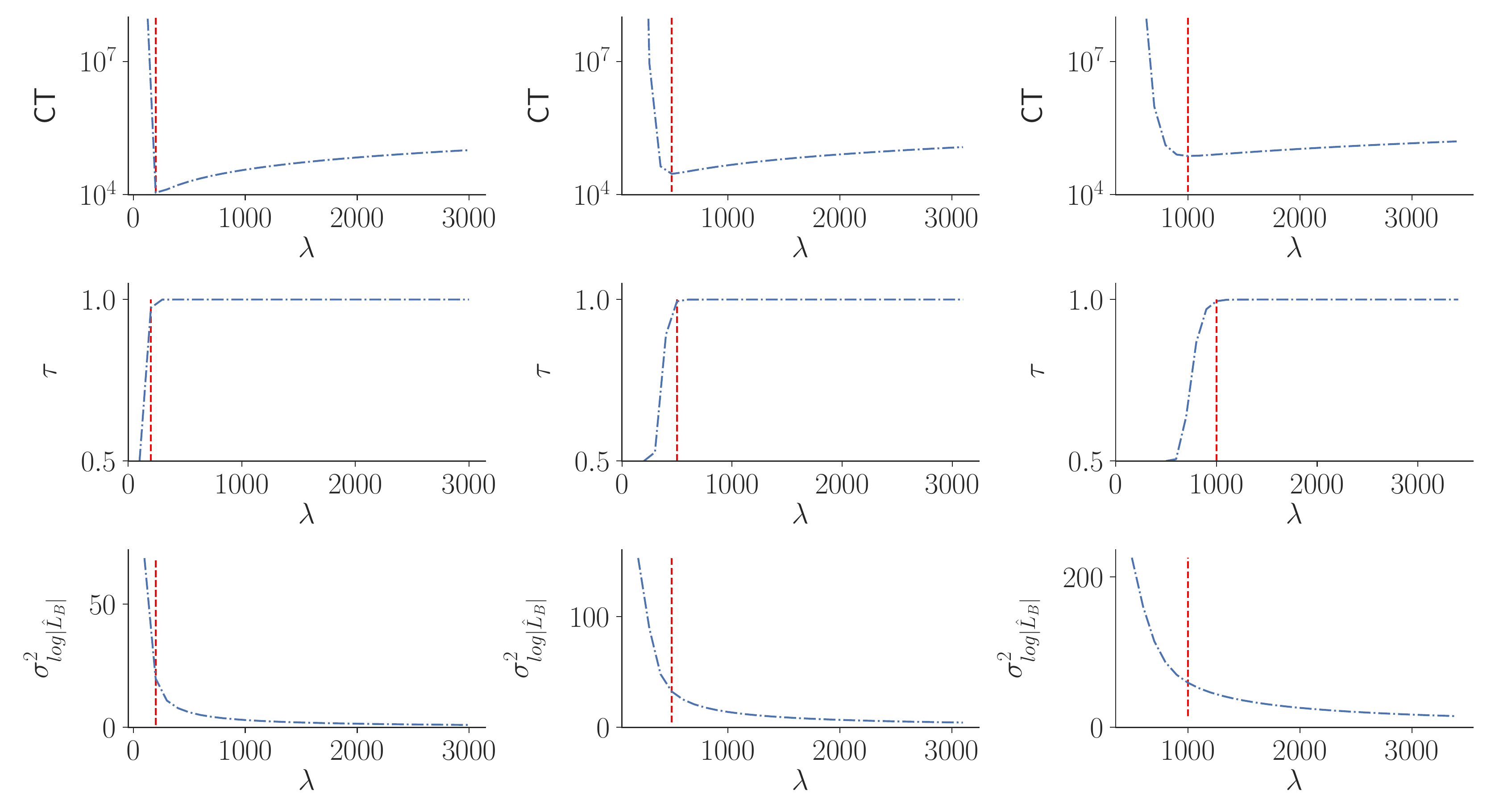}
        %\caption{$\Pr(\widehat{L}>0)$ as a function of $a$.}\label{fig:CT_fig_a}
\caption{Dependence of CT and its parts on $\lambda$ for the settings in Algorithm \ref{alg:optTuning}. The columns correspond to three different $\gamma$ (left $\gamma=90{,}000$, middle $\gamma=400{,}000$, right $\gamma=1,500{,}000$).  The top row shows the Computational Time (CT) in \eqref{eq:CT_Poisson_simplified}. The middle row shows the probability of a positive estimator $\tau$ in Lemma \ref{lem:PrNonnegative}. The bottom rows shows the variance of the log of the absolute value of block-Poisson estimator in \eqref{lem:VarAbs}. The vertical red line marks the optimal value of $\lambda$, i.e. $\lambda_{\mathrm{opt}}$.
}\label{fig:CT_PrPos_sigma2_log_abs_Lhat_func_of_lambda}\end{figure}

The definition of CT in \eqref{eq:realCT} involves $\Var_{\ov \pi} (\psi S)$ which, unlike $\mathrm{IF}_{\ov \pi, \psi S}$ (see Section \ref{app:IF}), depends on the specific functional $\psi$. To make the tuning independent on the choice of $\psi$ we derive tuning guidelines under the assumption $\Var_{\ov \pi} (\psi S) = \mathrm{V}_{\pi}(\psi)$. This approximation is expected to be accurate since negative signs are detrimental for the CT, so the optimal $m$ and $\lambda$ are expected to be in a region where $S=1$ with probability close to one; see Figure \ref{fig:CT_PrPos_sigma2_log_abs_Lhat_func_of_lambda} and the discussion below. 

Hence, $\lambda$ and $m$ in Algorithm \ref{alg:implementation} are chosen by minimizing
\begin{equation}\label{eq:CT_Poisson_simplified}
\mathrm{CT}_B(\lambda, m|\gamma)  \coloneqq m \lambda \frac{\mathrm{IF}_{\ov \pi,  \psi S}\left(\sigma^2_{\log |\widehat{L}_B|}(\lambda, m | \gamma) \right)}{ (2\tau(\lambda, m |\gamma) - 1)^2}.
\end{equation}
The CT in \eqref{eq:CT_Poisson_simplified} is expressed to show explicitly how each factor depends on the tuning parameters $m$ and $\lambda$; the dependence on $\gamma$ is also shown. Note in particular that $\mathrm{IF}_{\ov \pi, \psi S}$ is a function of these arguments only through the variance of the log of the absolute value of the block-Poisson estimator. This is analogous to \cite{pitt2012some}, except that since the MCMC here is run on $|\widehat{L}_B|$, it is the variance of $\log |\widehat{L}_B|$ that enters the IF; see Section \ref{app:IF} for precise definitions of all the terms in \eqref{eq:CT_Poisson_simplified}.

To obtain the CT in \eqref{eq:CT_Poisson_simplified} it is necessary to compute: i) $\sigma^2_{\log |\widehat{L}_B|}$, ii) $\tau$, iii) $\mathrm{IF}(\cdot)$, and iv) $\gamma \coloneqq n^2 \sigma^2_{d_{u_i}}$. The $\mathrm{IF}(\cdot)$ in iii) can be computed by one-dimensional numerical integration; see Lemma \ref{lemma: idealized ineff} in Section \ref{app:IF}.  The variance $\sigma^2_{d_{u_i}}$ in iv) is the intrinsic variability in the population of differences $d_i$ and can be approximated as in Algorithm \ref{alg:optTuning}. Finally, if $\widehat{d}_m$ is assumed normal, then $\sigma^2_{\log |\widehat{L}_B|}$ and $\tau$ can be easily computed from Lemmas \ref{lem:VarAbs} and \ref{lem:PrNonnegative}, respectively. Our recommended strategy is to set $m=30$, where a CLT motivates the normality of $\widehat{d}_m$, and optimize $\mathrm{CT}$ only with respect to $\lambda$. This setting gives the empirical relation $\lambda_{\mathrm{opt}} = \exp\left(-0.1022+   0.4904\log(\gamma_{\mathrm{max}})\right)$ in Algorithm \ref{alg:optTuning} when $G=100$ blocks is used ($\rho=0.99$).

Algorithm \ref{alg:optTuning} summarizes our recommended tuning strategy. The initial optimization in Algorithm \ref{alg:optTuning} is not included in the computational cost since the same optimization initializes all the algorithms in the comparisons. The additional tuning cost in Algorithm \ref{alg:optTuning} is included in our algorithm whenever we compare it against another algorithm. Note that the optimal $\lambda$ is based on $\gamma_{\mathrm{max}}$, resulting in a conservatively low variance of the estimator; this is well known to be a good strategy in pseudo-marginal MCMC \citep{pitt2012some}.

Figure \ref{fig:CT_PrPos_sigma2_log_abs_Lhat_func_of_lambda} plots the CT, $\Pr(\widehat{L} \geq 0)$ and $\sigma^2_{\log |\widehat{L}_B|}$ as functions of $\lambda$ for this strategy; the figure also shows the optimal value $\lambda_{\mathrm{opt}}$. Clearly, the optimal $\lambda_{\mathrm{opt}}$ results in a high probability of a positive estimator, regardless of $\gamma$. 

%\begin{enumerate}[topsep=0pt]
%    \item[1.] Sample $\theta$ values $\mathcal{T}=\{\theta^{(1)},...,\theta^{(M)}\}$ from a multivariate student $t$ approximation of the posterior obtained by optimizing the log posterior based on $\ell(\theta)$ estimated from a single subsample.
 %   \item[2.] Using a subsample of $\widetilde{m}$ observations, estimate $\gamma(\theta)$ by $\widehat{\gamma}(\theta) = n^2 \widehat{\sigma}^2_{d_{u_i}}(\theta)$ for each $\theta \in \mathcal{T}$, and set $\gamma_{\mathrm{max}}\coloneqq \max_{\theta \in \mathcal{T}}  ~\widehat{\gamma}(\theta)$.
 %   \item[3.] With $\rho = 0.99$ and using $m = 30$, set $\lambda$ to its optimal value (see Figure \ref{fig:lambdaopt_vs_gamma}):
%    \begin{equation}\label{eq:relationGammaVsLambda}
 %       \lambda_{\mathrm{opt}} = \exp\left(-0.1022+   0.4904\log(\gamma_{\mathrm{max}})\right).
%    \end{equation}
%    \item[4.]Set the lower bound to $a = \bar d - \lambda$, where $\bar d = M^{-1}\sum_{j=1}^M \widehat d_{\widetilde{m}}(\theta^{(j)})$ using the evaluations from Step 2 to compute the batch means $\widehat{d}_{\widetilde{m}}$.

%\end{enumerate}
\begin{algorithm}
    \caption{Tuning strategy for block PMMH with $m=30$ and $\rho=0.99$}\label{alg:optTuning}
    \KwIn{Number of tuning draws $M$.
    }
    \vspace{2mm}
  Approximate the posterior by a multivariate student $t$ by optimizing a posterior ~\\estimated from a single subsample $\widetilde{u}$ of $\widetilde{m}$ observations.\\
  \vspace{2mm}
  
  \For {$j = 1,\dots, M$}{
  Draw $\theta^{(j)}$ from the approximate Student $t$ posterior \\\vspace{0.3mm}
  Estimate $\gamma(\theta^{(j)})$ by $\widehat{\gamma}(\theta^{(j)}) = n^2 \widehat{\sigma}^2_{d_{u_i}}(\theta^{(j)})$ from the subsample $\widetilde{u}$\\ 
  Compute the batch mean $\widehat{d}_{\widetilde{m}}(\theta^{(j)})$
  \\ \vspace{1mm}
  }
  \vspace{1mm}
  $\gamma_{\mathrm{max}} = \max \{ \widehat{\gamma}(\theta^{(1)}),\ldots,\widehat{\gamma}(\theta^{(M)})\}$. \\ 
  $\bar d = M^{-1}\sum_{j=1}^M \widehat d_{\widetilde{m}}(\theta^{(j)})$. \\ 
 
  \vspace{2mm}
  Set $\lambda_{\mathrm{opt}} = \exp\left(-0.1022+   0.4904\log(\gamma_{\mathrm{max}})\right)$\\
  Set $a_{\mathrm{opt}} = \bar d - \lambda_{\mathrm{opt}}$.\\
    \vspace{1mm}
  \vspace{1mm}
 \KwOut{$\lambda_{\mathrm{opt}}$ and $a_{\mathrm{opt}}$.}
\end{algorithm}

The reason for setting $m=30$ in Algorithm \ref{alg:optTuning} is to obtain an easily implemented tuning strategy that ensures that $\widehat{d}_m$ is approximately normal (CLT). However, when $\widehat{d}_m$ is already normal for smaller $m$, it is beneficial to use a smaller $m$. In fact, we conjecture that if $\widehat{d}_m$ is normal for any $m$, the true minimum of \eqref{eq:CT_Poisson_simplified} is obtained with $m_{\mathrm{opt}} = 1$; see Section \ref{app:ExactVsApprox} for some evidence. This is intuitively understood by noting that the computational cost in the CT is proportional to the product $m\lambda$, so the individual $m$ and $\lambda$ do not matter for the cost. But $\lambda$ has a much larger effect on increasing $\tau$ than $m$, which explains why it is desirable to spend computational resources on more batches (larger $\lambda$) rather than increasing the batch sizes (larger $m$). Fortunately, Section \ref{app:ExactVsApprox} shows that using $m=30$ rather than the optimal $m=1$ only increases CT by a factor of around two, unless $\gamma \coloneqq n^2 \sigma^2_{d_{u_i}}$ is very small, which is unlikely because $n$ is typically very large when subsampling is used. Nevertheless, Section \ref{app:optimalTuningGeneral} proposes a general tuning strategy where the distribution of $\widehat{d}_m$ is approximated by a mixture of normals. All parts of the CT expression are then available in closed form and the CT can be minimized with respect to both $m$ and $\lambda$.
 
Finally, Section \ref{app:ExactVsApprox} documents that the CT of the signed PMMH with the block-Poisson estimator is 2-3 times larger than for the approximate Subsampling MCMC approach in \cite{quiroz2016speeding} for most $\gamma$, and has up to 8 times larger CT when $\gamma$ is very small. Hence, there is a trade-off between exactness and computational time.

\section{Applications\label{sec:Experiments}}
\subsection{Model and data\label{subsec:ModelAndData}}
Our experiments consider the logistic regression for $y_i \in \{0, 1\}$ on covariates $x_i \in \mathbb{R}^p$, with distribution
\begin{eqnarray*}
p(y_{i}|x_{i},\theta) & = & \frac{\exp(y_i x_{i}^{\top}\theta)}{1+\exp(x_{i}^{\top}\theta)},
\end{eqnarray*}
where $\theta$ follows a multivariate normal prior with zero mean and covariance matrix $10I$. We fit the model to the three datasets used in \citet{quiroz2016speeding}: i) the \emph{CovType} data as used in \citet{collobert2002parallel}, with $n=550{,}087$ observations and $p=11$ variables, ii) the firm \emph{Bankruptcy} data as used in \citet{giordani2011taking}, containing $n=4{,}748{,}089$ observations and eight covariates, and iii) the \emph{HIGGS} dataset \citep{baldi2014searching} with $n=1{,}100{,}000$ observations and $21$ covariates as in \cite{quiroz2016speeding}.

Section \ref{app:FireflySupplement} contains some additional empirical results from the experiments.

\subsection{Experiment 1: Comparisons against Firefly Monte Carlo and MCMC\label{subsec:Experiment2}}
The first experiment compares the block-Poisson estimator to both Firefly Monte Carlo \citep{maclaurin2014firefly} and standard MH on the full dataset. Firefly is tuned as in \cite{maclaurin2014firefly} with the optimally tuned lower bound for logistic regression based on a central parameter value $\theta^\star$ and allowing $10\%$ of the observations to change indicator in each iteration. For the block-Poisson estimator we use the parameter expanded control variates, expanded around the same $\theta^\star$, resulting in a small $\gamma(\theta)$. The second experiment explores the effect of larger $\gamma(\theta)$. We check that the normality assumption for $\widehat{d}_m$ when $m=30$ is reasonable and Algorithm \ref{alg:optTuning} returns $\lambda_{\mathrm{opt}} = 100$ after rounding to the nearest allowable $\lambda$. We simulate $N = 55{,}000$ samples from the posterior and discard $5{,}000$ as burn-in. A random walk Metropolis proposal is used, with a scaling factor for the posterior covariance at the mode of $2.38/\sqrt{p}$ for MCMC \citep{roberts1997weak} and $2.5/\sqrt{p}$ for subsampling MCMC \citep{sherlock2013efficiency}. We use the same proposal for Firefly Monte Carlo as for MCMC; \cite{maclaurin2014firefly} use a similar scaling by targeting the acceptance probability $0.234$ \citep{roberts1997weak}, but presumably using the identity matrix instead of the posterior covariance at the mode (since both MCMC and their algorithm have very low effective sample sizes in their Table 1). All algorithms are started at $\theta^\star$.

We measure the performance of our subsampling MCMC using an empirical version of the CT in \eqref{eq:CT_Poisson_simplified}; the $\mathrm{IF}$ for the chain $\{s_i\theta_i,i=1,2,\dots , N\}$ is estimated using the $\mathtt{coda}$ package in $\mathtt{R}$ \citep{coda2006}; the cost is taken as
  the average number of evaluations over the MCMC iterations used when forming the estimator (the number of terms within a factor in the product is random); and $\tau$ is replaced by its empirical estimate. For MCMC we use a similar measure but set $\tau=1$ and the number of evaluations to $n$. We define the estimated Relative Computational Time (RCT) for the block-Poisson algorithm $\mathcal{B}$ against any algorithm $\mathcal{A}$ with $\tau = 1$ as
\begin{equation}\label{eq:RCT_prodPois}
\widehat{\mathrm{RCT}}_{\mathcal{A}} \coloneqq \frac{\overline{\mathrm{CC}}_{\mathcal{A}}\times \widehat{\mathrm{IF}}_{\mathcal{A}}}{\overline{\mathrm{CC}}_{\mathcal{B}}\times\widehat{\mathrm{IF}}_{\mathcal{B}}/(2\widehat{\tau}_{\mathcal{B}} - 1)^2},
\end{equation}
where $\mathrm{CC}$ is the computational cost introduced in Section \ref{subsec:Control-variates} and the bars denote averages over MCMC iterations. This RCT measures how much computational resources 
that algorithm $\mathcal{A}$, compared to the block-Poisson algorithm $\mathcal{B}$, consumes in order to produce a posterior estimate with a given precision; see, e.g., Appendix C of \cite{tran2016block} and \cite{quiroz2016speeding} for more details.

\begin{figure}
\centering

\begin{subfigure}[t]{.30\textwidth}
\centering
\includegraphics[width=\linewidth]{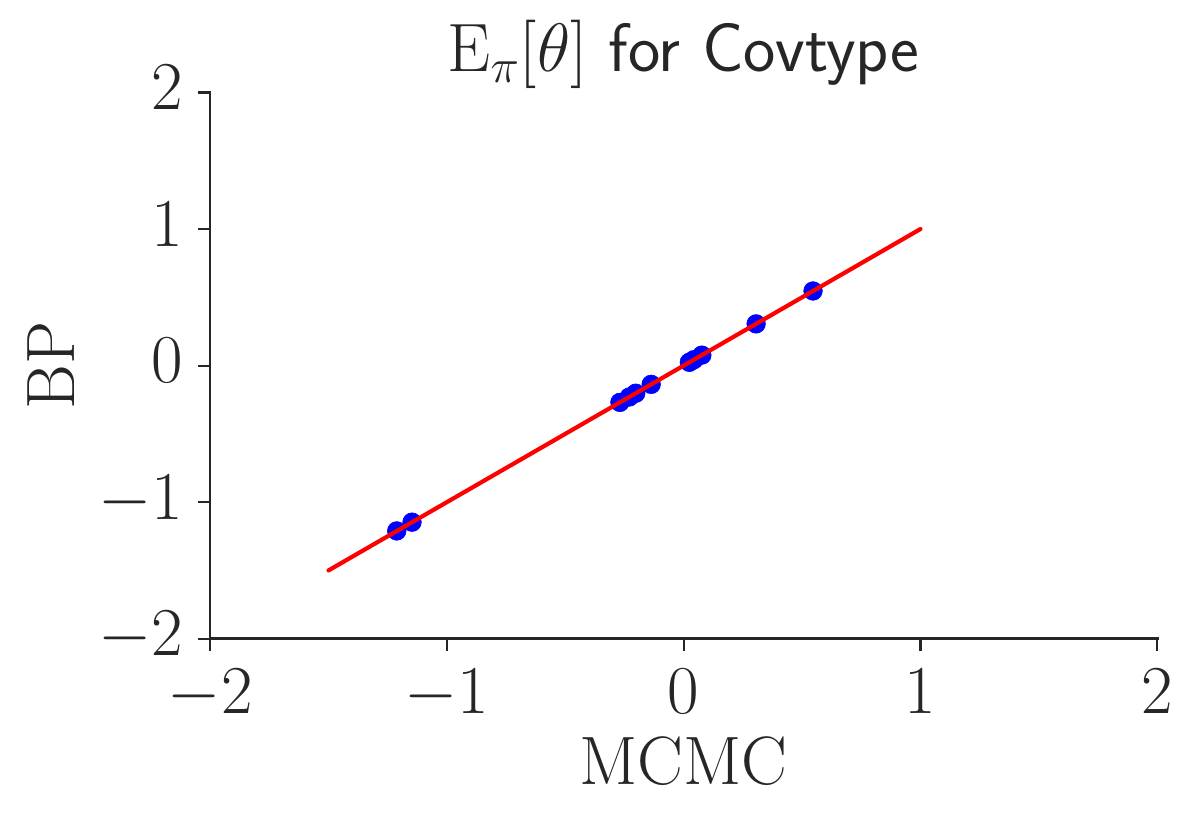}
        %\caption{$\Pr(\widehat{L}>0)$ as a function of $a$.}\label{fig:CT_fig_a}
\end{subfigure}
\begin{subfigure}[t]{.30\textwidth}
\centering
\includegraphics[width=\linewidth]{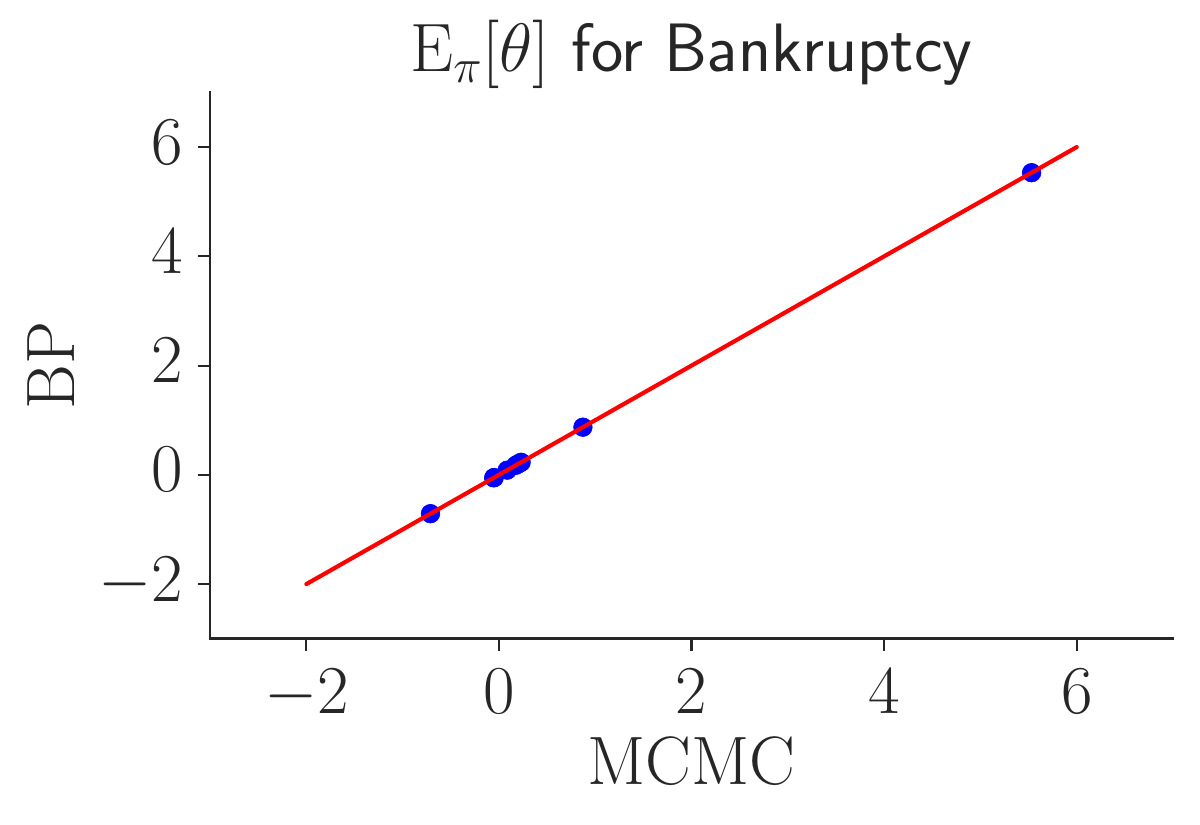}
        %\caption{$\mathrm{CT}$ as a function of $a$}\label{fig:CT_fig_b}
\end{subfigure}
\begin{subfigure}[t]{.30\textwidth}
\centering
\includegraphics[width=\linewidth]{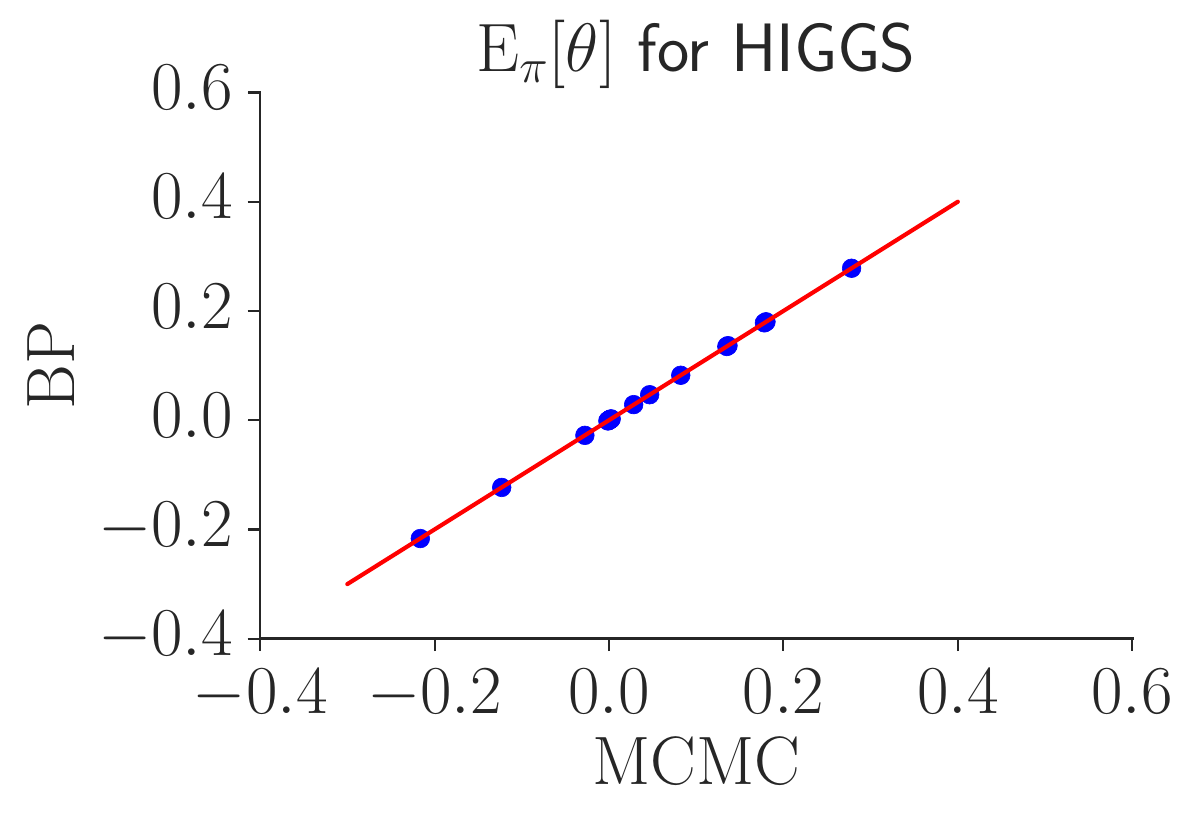}
        %\caption{$\mathrm{CT}$ as a function of $a$}\label{fig:CT_fig_b}
\end{subfigure}

\begin{subfigure}[t]{.30\textwidth}
\centering
\includegraphics[width=\linewidth]{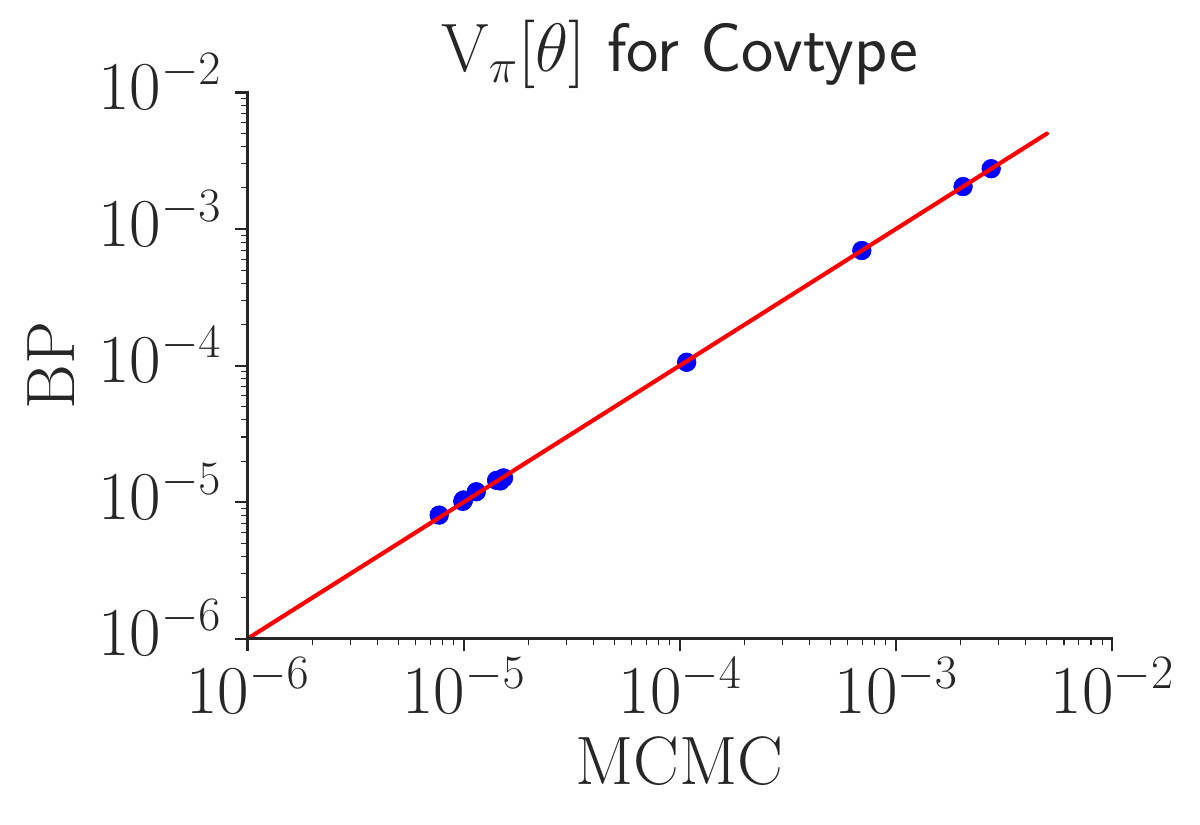}
        %\caption{$\Pr(\widehat{L}>0)$ as a function of $a$.}\label{fig:CT_fig_a}
\end{subfigure}
\begin{subfigure}[t]{.30\textwidth}
\centering
\includegraphics[width=\linewidth]{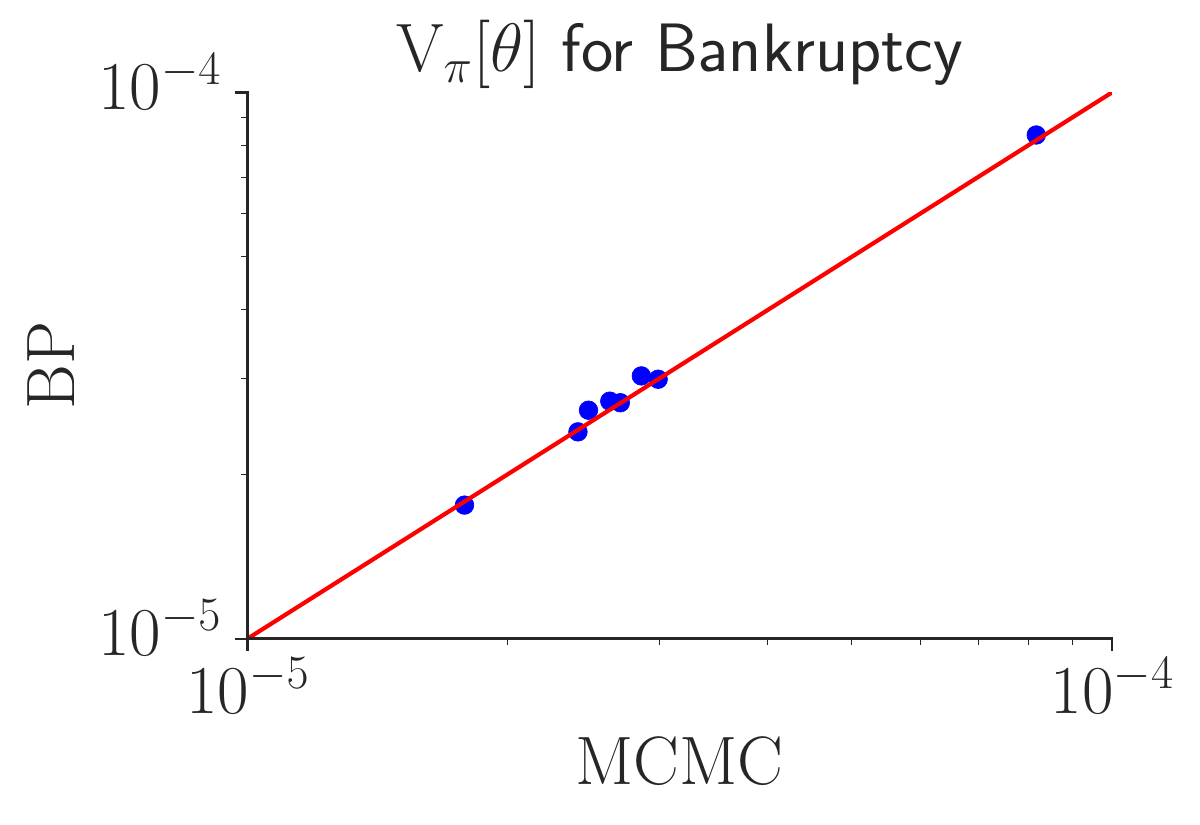}
        %\caption{$\mathrm{CT}$ as a function of $a$}\label{fig:CT_fig_b}
\end{subfigure}
\begin{subfigure}[t]{.30\textwidth}
\centering
\includegraphics[width=\linewidth]{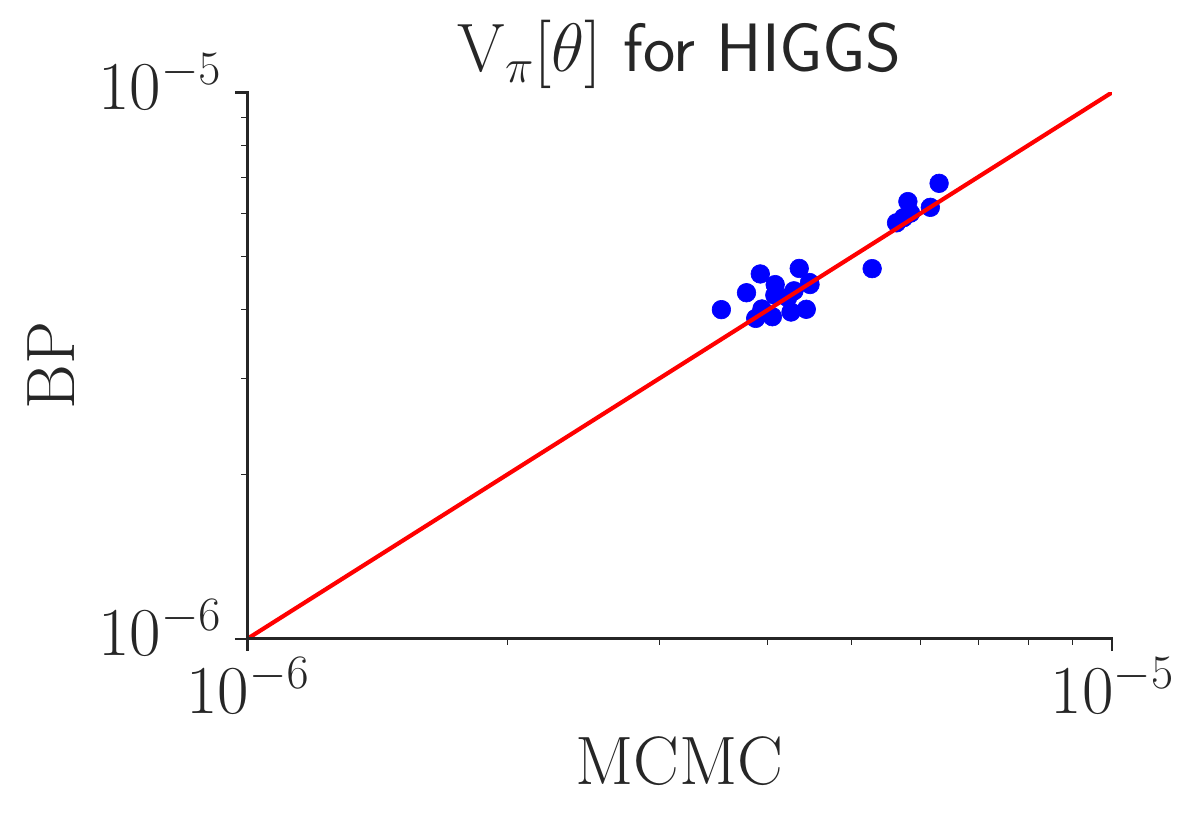}
        %\caption{$\mathrm{CT}$ as a function of $a$}\label{fig:CT_fig_b}
\end{subfigure}

\begin{subfigure}[t]{.30\textwidth}
\centering
\includegraphics[width=\linewidth]{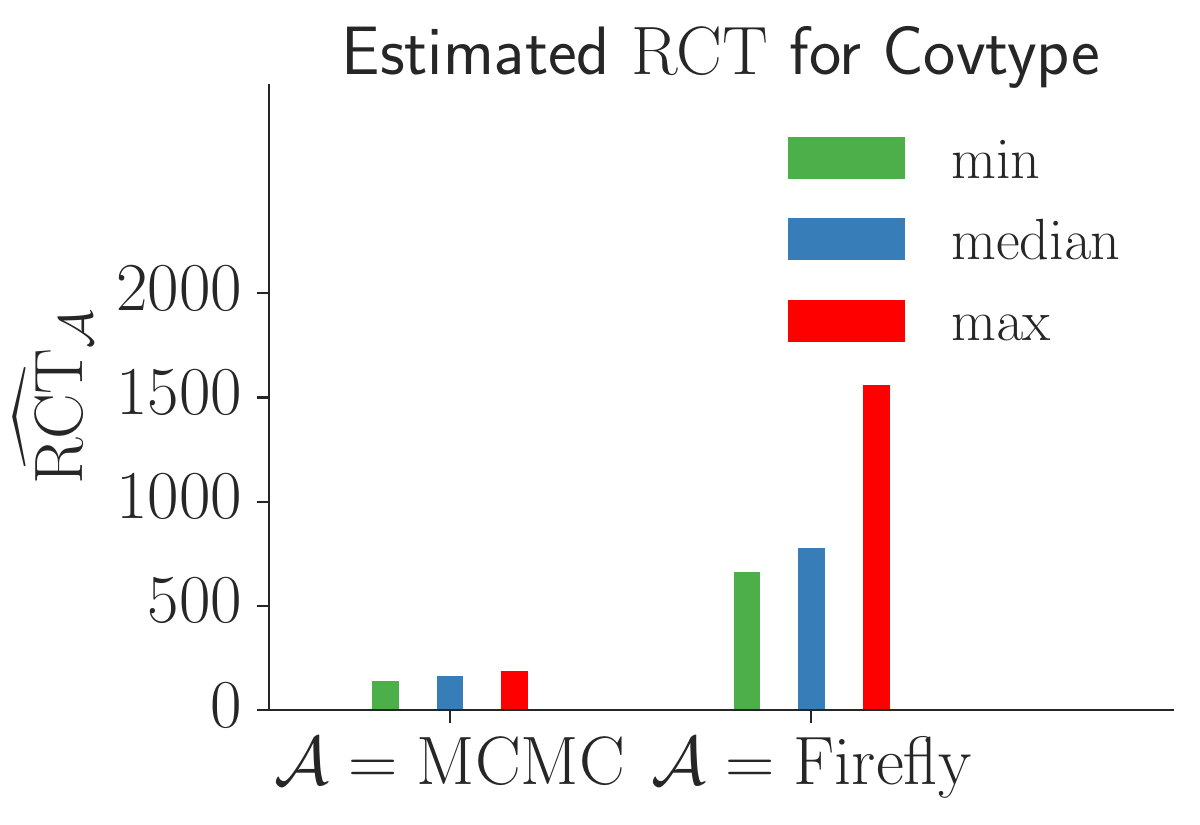}
        \caption{Covtype data.}%\label{fig:CT_lambda1_fig}
\end{subfigure}
\begin{subfigure}[t]{.30\textwidth}
\centering
\includegraphics[width=\linewidth]{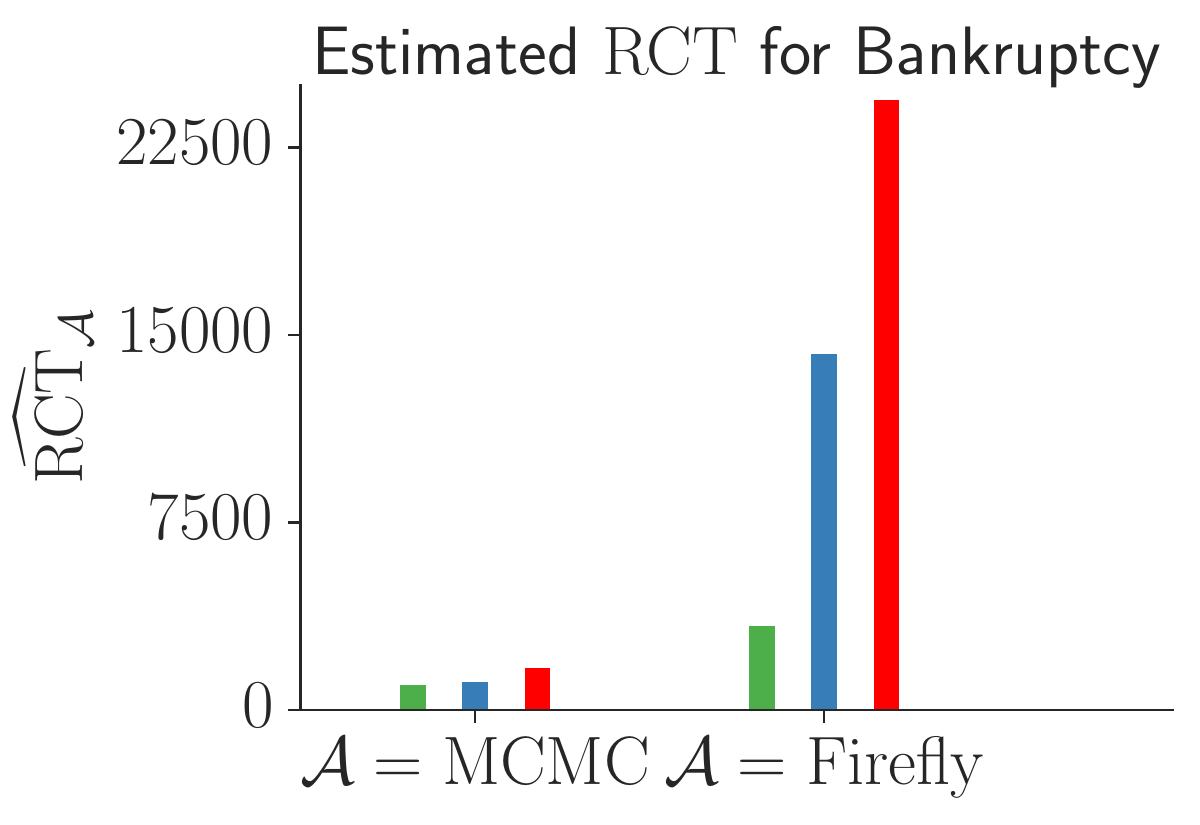}
        \caption{Bankruptcy data.}%\label{fig:CT_lambda100_fig}
\end{subfigure}
\begin{subfigure}[t]{.30\textwidth}
\centering
\includegraphics[width=\linewidth]{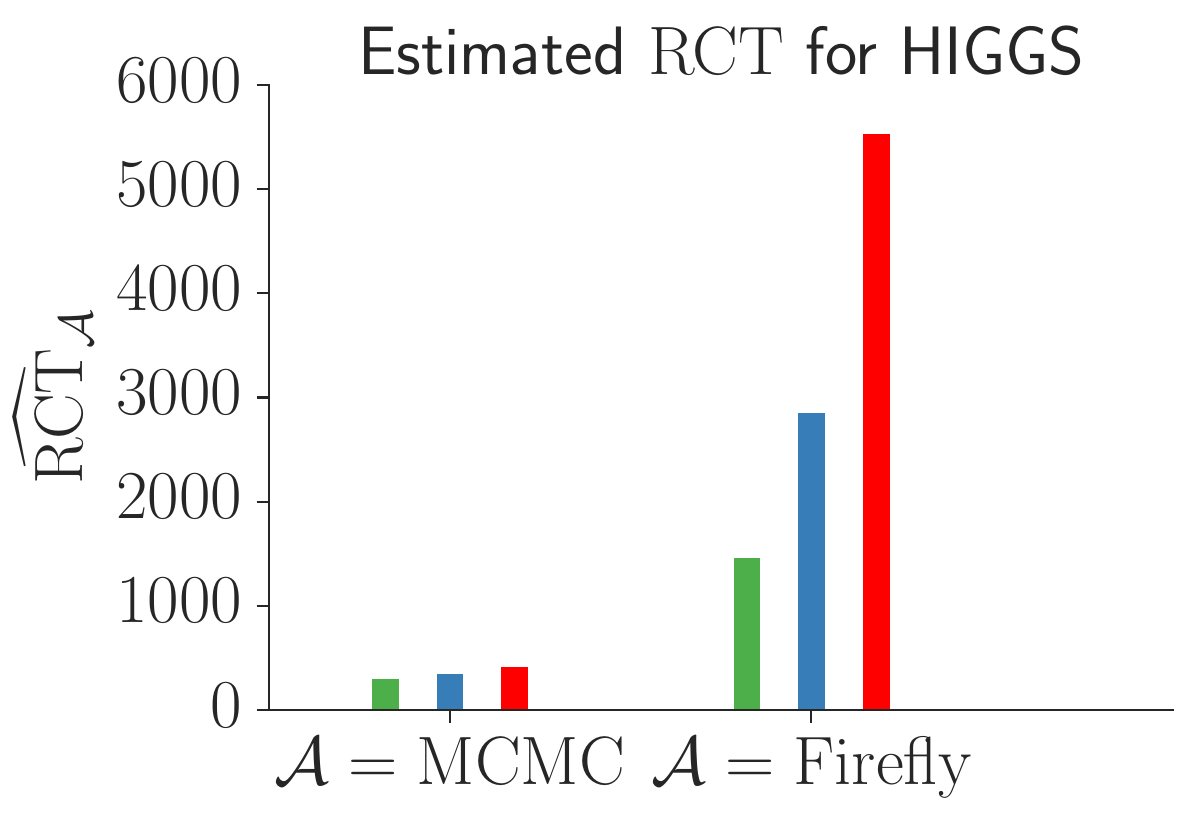}
        \caption{HIGGS data.}%\label{fig:CT_lambda100_fig}
\end{subfigure}

\caption{Results from Experiment 1 using parameter expanded control variates. Estimates of posterior expectation (upper), variance (middle) and Relative Computational Time (RCT) in \eqref{eq:RCT_prodPois} (lower) of correlated product-Poisson with $\lambda=100$ relative to MCMC and Firefly Monte Carlo. The results are shown for three datasets, Covtype (A), Bankruptcy (B) and HIGGS (C). }\label{fig:Expectation_Variance_CT_experiment1}
\end{figure}

Figure \ref{fig:Expectation_Variance_CT_experiment1} shows the accuracy of estimating the posterior expectations and variances using our method on the three datasets. Although some estimates are visually off the 45-degree line, all deviations are within the usual Monte Carlo error, and we conclude that the estimates are very accurate. Figure \ref{fig:Expectation_Variance_CT_experiment1} also shows the relative computational times in \eqref{eq:RCT_prodPois} compared to MCMC and Firefly Monte Carlo, respectively; the block-Poisson approach has a two orders of magnitude improvement against MCMC and three or even four orders of magnitude against Firefly Monte Carlo. Although Firefly Monte Carlo has a lower computational cost than MCMC (less density evaluations), its extreme autocorrelation (documented in Section \ref{app:FireflySupplement}) grossly inflates the CT. 

\subsection{Experiment 2: Performance when the control variates are poor.}

The next experiment is a serious test of the methodology when the control variates are poor so that $\gamma$ is large and there are outliers in the differences $d_k$. To obtain poor covariates, we use the data expanded control variates in \citet{quiroz2016speeding}, with a very small number of clusters in relation to $n$. We follow \cite{quiroz2016speeding} and choose $K = 1042, 16374, 355$ for the Covtype, Bankruptcy and HIGGS datasets. The deliberately poor control variates generate severe outliers in the $d_k$ population and normality is therefore not guaranteed for $m = 30$. We found that $m=100,100, 600$ are large enough to obtain normality in the Covtype, Bankruptcy and HIGGS datasets, respectively, and the corresponding optimal $\lambda$ are $\lambda_{\mathrm{opt}} = 500,  1100, 300$. The other settings are the same as in Experiment 1.

\begin{figure}
\centering

\begin{subfigure}[t]{.30\textwidth}
\centering
\includegraphics[width=\linewidth]{Expectation_MCMC_vs_ProdPois_Covtype.pdf}
        %\caption{$\Pr(\widehat{L}>0)$ as a function of $a$.}\label{fig:CT_fig_a}
\end{subfigure}
\begin{subfigure}[t]{.30\textwidth}
\centering
\includegraphics[width=\linewidth]{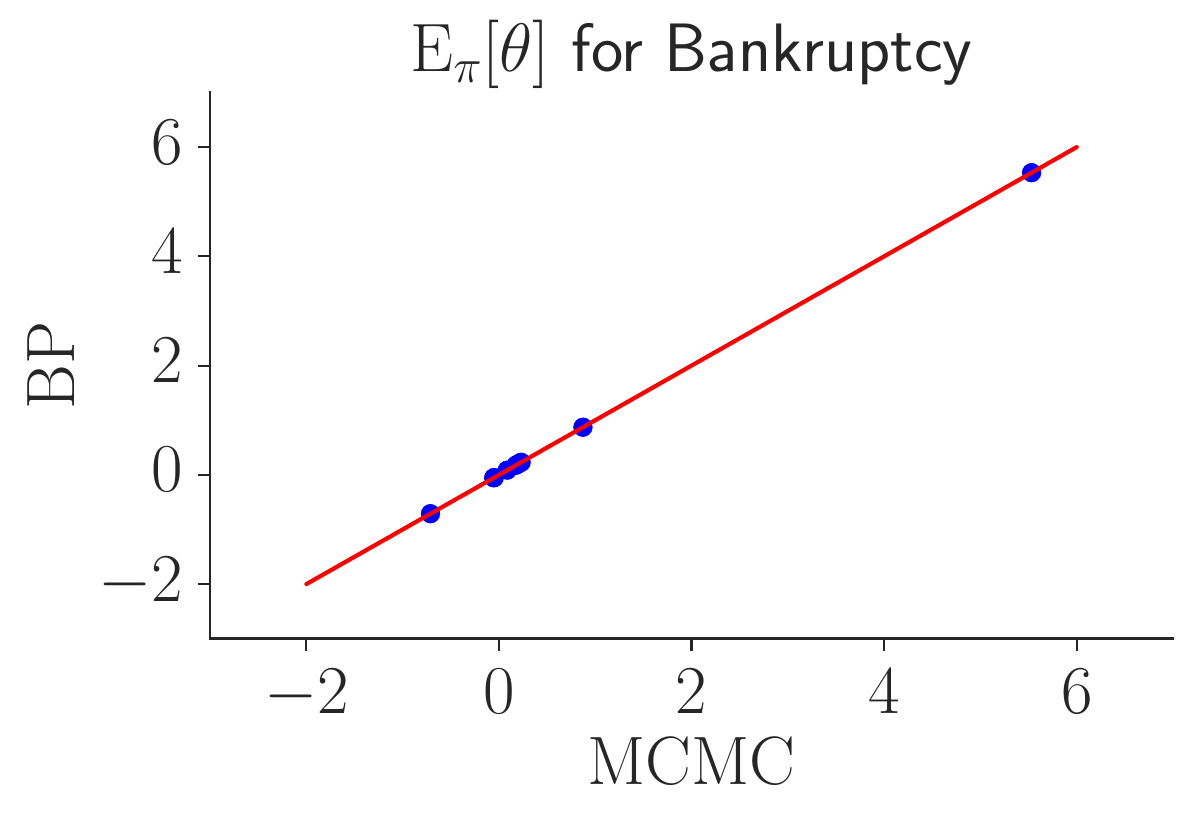}
        %\caption{$\mathrm{CT}$ as a function of $a$}\label{fig:CT_fig_b}
\end{subfigure}
\begin{subfigure}[t]{.30\textwidth}
\centering
\includegraphics[width=\linewidth]{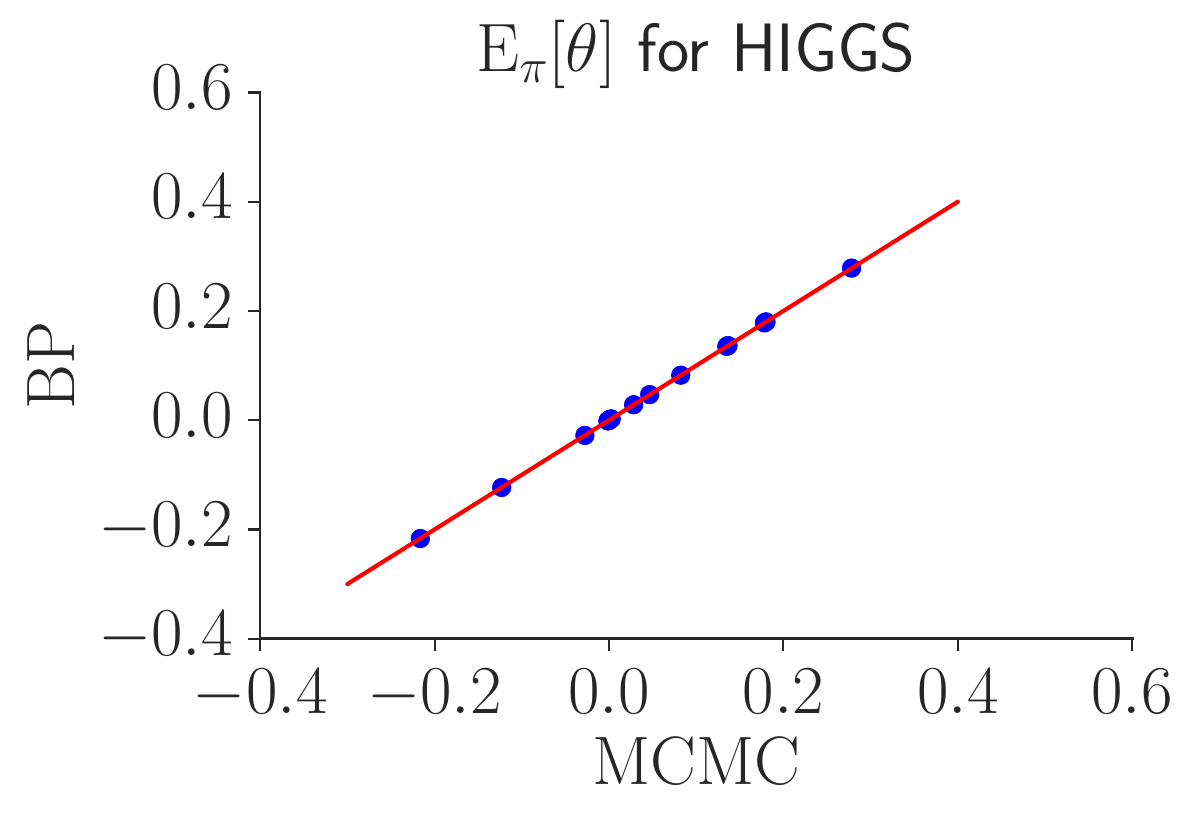}
        %\caption{$\mathrm{CT}$ as a function of $a$}\label{fig:CT_fig_b}
\end{subfigure}

\begin{subfigure}[t]{.30\textwidth}
\centering
\includegraphics[width=\linewidth]{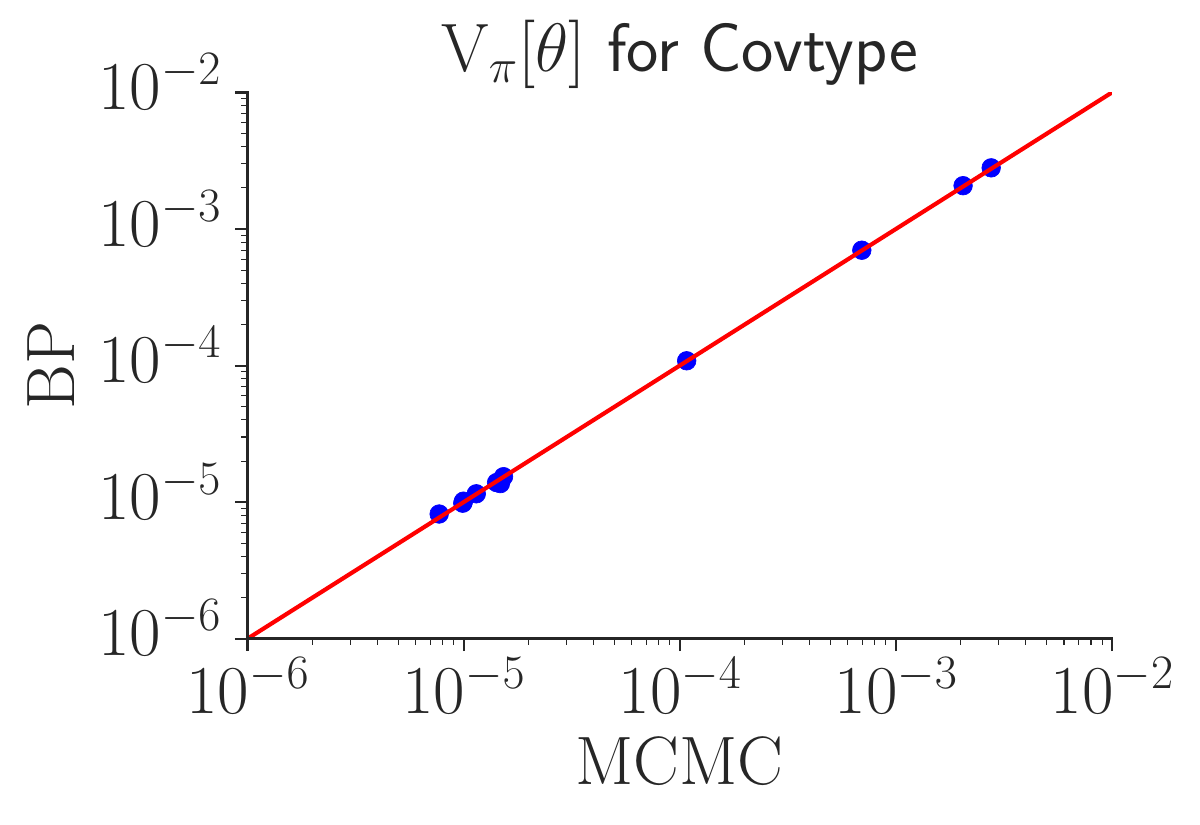}
        %\caption{$\Pr(\widehat{L}>0)$ as a function of $a$.}\label{fig:CT_fig_a}
\end{subfigure}
\begin{subfigure}[t]{.30\textwidth}
\centering
\includegraphics[width=\linewidth]{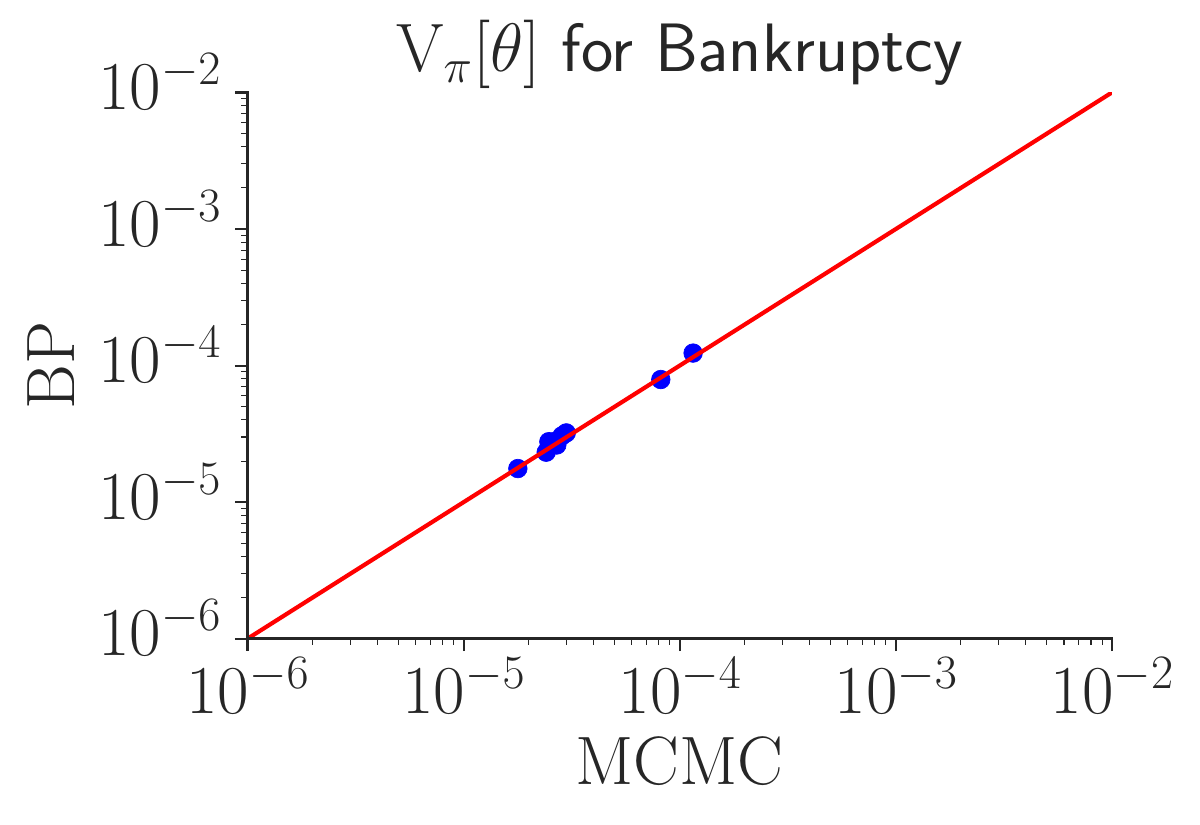}
        %\caption{$\mathrm{CT}$ as a function of $a$}\label{fig:CT_fig_b}
\end{subfigure}
\begin{subfigure}[t]{.30\textwidth}
\centering
\includegraphics[width=\linewidth]{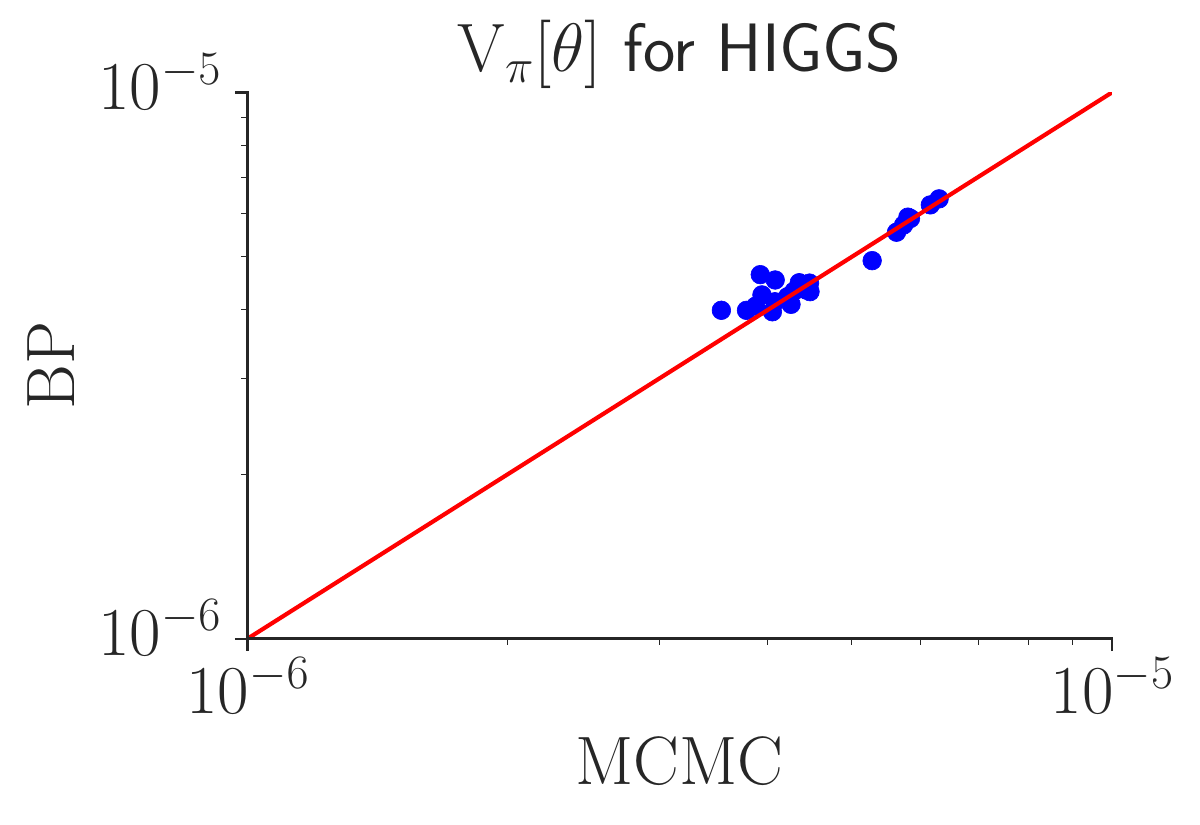}
        %\caption{$\mathrm{CT}$ as a function of $a$}\label{fig:CT_fig_b}
\end{subfigure}

\begin{subfigure}[t]{.30\textwidth}
\centering
\includegraphics[width=\linewidth]{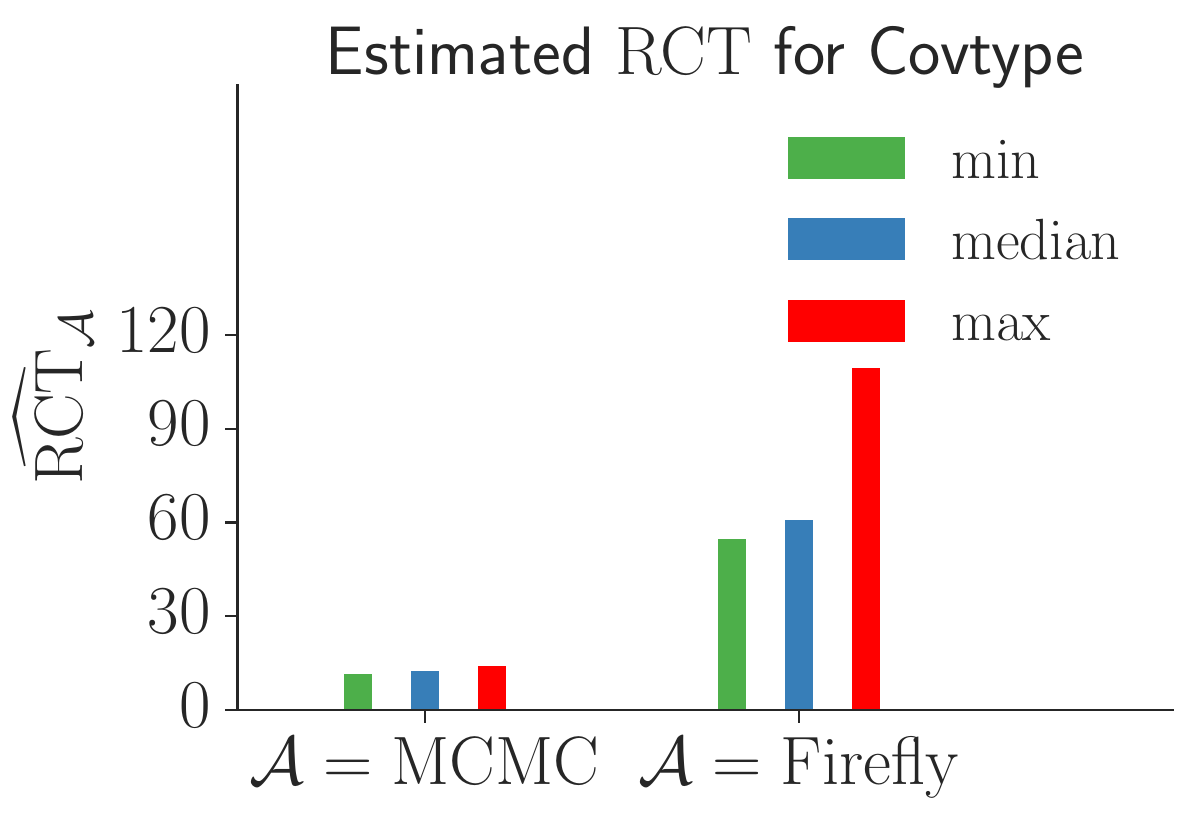}
        \caption{Covtype data.}%\label{fig:CT_lambda1_fig}
\end{subfigure}
\begin{subfigure}[t]{.30\textwidth}
\centering
\includegraphics[width=\linewidth]{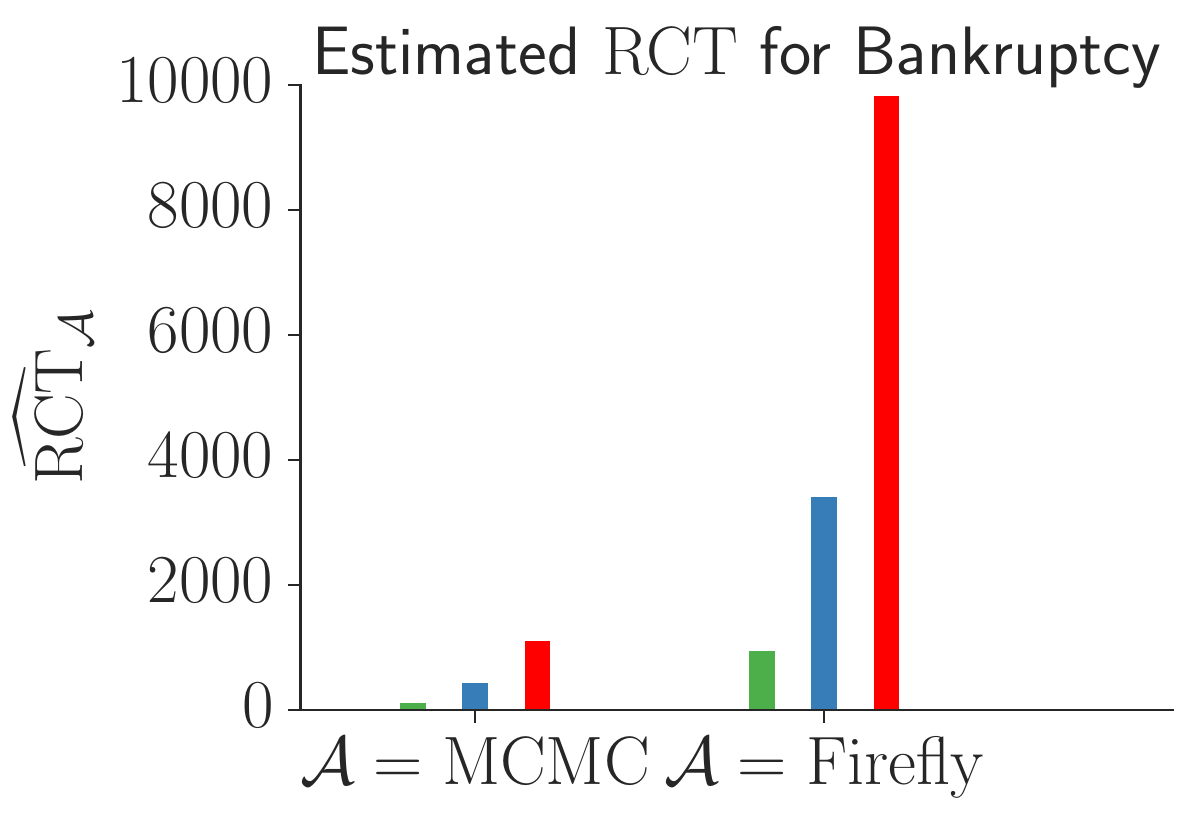}
        \caption{Bankruptcy data.}%\label{fig:CT_lambda100_fig}
\end{subfigure}
\begin{subfigure}[t]{.30\textwidth}
\centering
\includegraphics[width=\linewidth]{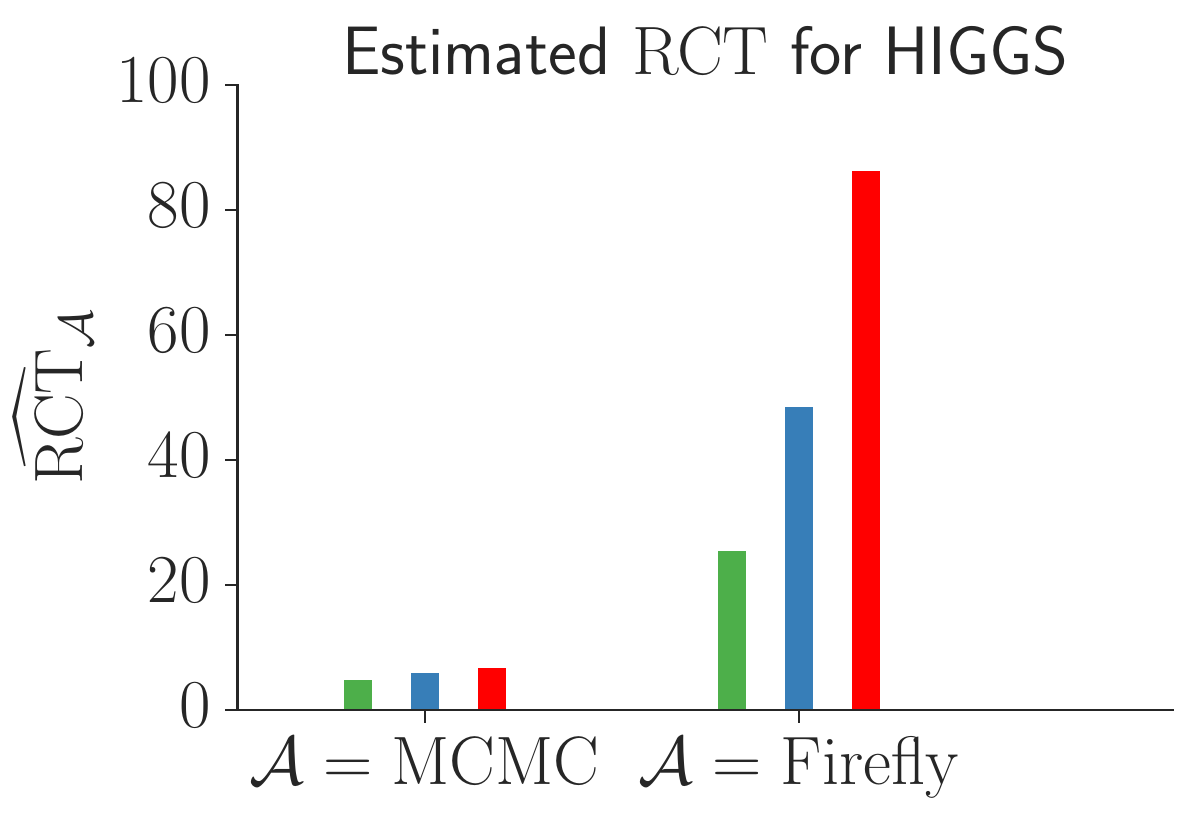}
        \caption{HIGGS data.}%\label{fig:CT_lambda100_fig}
\end{subfigure}

\caption{Results from Experiment 2 using data expanded control variates with large $\gamma$. Estimates of posterior expectation (upper), variance (middle) and Relative Computational Time (RCT) in \eqref{eq:RCT_prodPois} (lower) of block-Poisson with $\lambda = 500,  1100, 300$ (Covtype, Bankruptcy, HIGGS) relative to MCMC and Firefly Monte Carlo. The results are shown for three datasets, Covtype (A), Bankruptcy (B) and HIGGS (C). }\label{fig:Expectation_Variance_CT_experiment2}
\end{figure}

Figure \ref{fig:Expectation_Variance_CT_experiment2} summarizes the results. Although the control variates are now significantly less accurate than in Experiment 1, our algorithm is still more efficient than MCMC and dramatically more efficient that Firefly Monte Carlo. Note that Firefly Monte Carlo is based on the same tight lower bound as in Experiment 1, yet our algorithm with poor control variates still performs much better. This example illustrates that even when $\gamma$ is large and varies a lot over the parameter space, our guidelines suggest a $\lambda$ which results in an efficient MCMC chain that still outperforms MH on the full dataset and Firefly Monte Carlo.

\subsection{Experiment 3: Comparisons against zig-zag --- scaling with respect to $p$\label{subsec:zigzag_experiment}}

The third experiment explores how the performance of MCMC with the block-Poisson estimator scales with the number of parameters, $p$, and how it compares with the recently proposed zig-zag sampler with subsampling. 

The zig-zag process in \cite{bierkens2016zig} is a \textit{continuous time} piecewise deterministic multivariate Markov process set up so that it has the posterior of interest as its invariant distribution. The process evolves deterministically in time in a linear fashion until one of its components changes direction (sign). The time to this event, i.e.\ the switching time, is a random variable that is controlled by \textit{switching rates}, one for each component. A rejection sampling step determines if the component with lowest switching rate makes a switch. This rejection step requires upper bounds of the switching rates for each dimension; the looser the bound, the lower is the acceptance probability. Importantly, the analytical bounds need to be derived for every new model, making it hard to automate the zig-zag sampler, whereas in our algorithm the gradient and Hessian of the log density needed for the control variates are readily computed using automatic differentiation. 

\cite{bierkens2016zig} also propose a subsampling variant of this algorithm where, in each zig-zag iteration, \textit{a single data point} is required to determine the probability of a switch. When used with control variates, the performance of zig-zag with subsampling is shown by \cite{bierkens2016zig} to be super-efficient and they demonstrate impressive performance for logistic regression with a small number of parameters. 

We now compare our method with the zig-zag subsampling algorithm in a logistic regression setup, using their code at \url{https://github.com/jbierkens/zigzag-experiments}. To make the comparison with zig-zag fair, we implement our exact subsampling algorithm using a proposal for $\theta$ based on Hamiltonian dynamics \citep{neal2011mcmc}, following \cite{dang2017hamiltonian} who propose an efficient way to use data subsampling in Hamiltonian Monte Carlo (HMC). A Hamiltonian proposal uses a continuous time dynamics to explore the posterior more efficiently and this time-continuity makes it more similar to the zig-zag sampler; see \cite{nishimura2017discontinuous} for a connection between their discontinuous Hamiltonian Monte Carlo (HMC) algorithm and the zig-zag sampler. This proposal also has the added benefit of demonstrating that our subsampling approach applies to HMC samplers. We follow \cite{dang2017hamiltonian} and set the stepsize $\epsilon=0.2$ for the discretization of the Hamiltonian dynamics and use  $L = 6$ leap-frog steps for solving the dynamics. The Poisson estimator is implemented with $m=30$ and $\lambda=100$, using $3{,}000$ subsamples used on average.

We follow \cite{bierkens2016zig} and use epochs as the computational budget. One epoch corresponds to the number of iterations required to evaluate the gradient of all $n$ data points. Since zig-zag evaluates the gradient of the full dataset in each iteration, one epoch is $1$ iteration. For zig-zag with subsampling one epoch is $n$ iterations since it evaluates the gradient of a single data point in each iteration. Our HMC based method uses $L$ evaluations of the gradient of the log-density and a log-density evaluation for each of the $m\lambda$ subsamples for updating $\theta$ given $u$. Since we only change one block of $u$ in the update step for $u|\theta$, this step uses $m$ log-density evaluations; see \cite{dang2017hamiltonian} for details. Thus the number of iterations per epoch for our method is therefore $$\frac{n}{(L+1)\cdot \lambda m + m}=\frac{1{,}100{,}000}{(6+1)\cdot 1{,}000\cdot 30 + 30}\approx 52,306,$$ for our settings.

\begin{figure}
\centering
\includegraphics[width=0.55\linewidth]{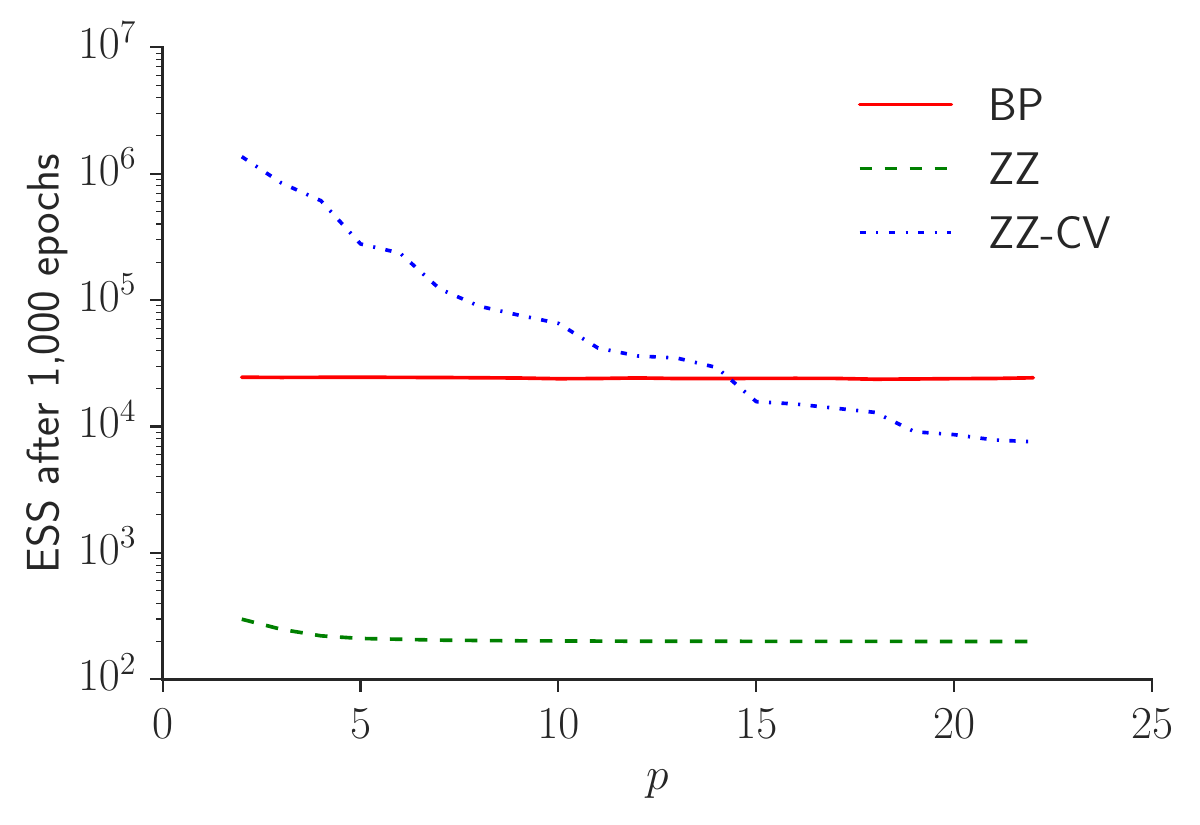}

\caption{Effective sample size (ESS) after $1{,}000$ epochs vs the number of parameters $p$; ESS is the median over parameters. The results are shown for the block-Poisson sampler with a Hamiltonian proposal (BP), zig-zag sampler (ZZ), and zig-zag sampler with subsampling and control variates (ZZ-CV). The vertical axis is in logarithmic scale.}\label{fig:zigzag_results}
\end{figure}

We use the HIGGS dataset for this experiment since it has the largest number of covariates among the three datasets, $p=22$, including an intercept. For each $j=2,\dots, 22$, we run the algorithms on a dataset consisting of the first $p=j$ covariates (including the intercept) and compare the effective sample sizes obtained. Figure \ref{fig:zigzag_results} shows, for each $p$, the median effective sample size of the parameters when using a computational budget of $1{,}000$ epochs for both zig-zag algorithms. The effective sample size is defined as the number of samples $N$ divided by the inefficiency factor. Clearly, the subsampling zig-zag sampler is extremely efficient for small $p$; for $p=2$ it is approximately a factor $56$ times more efficient than our algorithm. However, the efficiency of subsampling zig-zag deteriorates when $p$ increases, since the acceptance probability that determines the switch decreases, indicating that the bounds lose their tightness as discussed above. Note that the efficiency of our method remains essentially unchanged with dimension. With $p=15$, our method is $1.6$ times more efficient than zig-zag, and for $p=22$ our approach is $3.2$ times more efficient. The performance of zig-zag depends on the tightness of the upper bound, and it is unclear that equally tight bounds are readily available for more complex models.

\section{Conclusions\label{sec:Conclusions}}

We propose an algorithm for fast exact simulation-based inference where the likelihood is estimated cheaply
by efficient data subsampling. At the core of the algorithm is a novel and unbiased block-Poisson estimator of the likelihood targeting a given correlation between estimates at successive MCMC iterations, which is known to increase the efficiency \citep{deligiannidis2015correlated}.

For computational reasons, the block-Poisson estimator is set up to allow for occasional negative likelihood estimates. The proposed algorithm is therefore a signed PMMH algorithm \citep{lyne2015russian} where the pseudo-marginal MCMC is based on the absolute value of the likelihood estimate followed by a sign-correcting importance sampling step.

A major contribution of the article is the derivation of practical guidelines for tuning the parameters in the signed PMMH algorithm by minimizing the asymptotic variance of the importance sampling estimator per unit of computing time. The guidelines are based on idealized assumptions, but we demonstrate that the guidelines are accurate, effective and robust to features of the data. The framework proposed here applies to a larger set of problems, in particular models with intractable normalizing constants.

We demonstrate the performance of our algorithm by applying it to three commonly used datasets that are modelled by logistic regression. The proposed algorithm dramatically outperforms MH on the full dataset and  Firefly Monte Carlo, and is shown to  scale better with respect to the dimension of the parameter space than the zig-zag subsampling algorithm.

\if0\blind
{
\section{Acknowledgments}
We thank the Editor, Associate Editor and two reviewers for helping improve both the content and presentation of the article. Matias Quiroz, Robert Kohn and Khue-Dung Dang were partially financially supported by Australian
Research Council Center of Excellence grant CE140100049.
Mattias Villani was partially financially supported by Swedish Foundation for Strategic Research (Smart Systems: RIT 15-0097).
}\fi

\bibliographystyle{apalike}
\addcontentsline{toc}{section}{\refname}\bibliography{ref}

\newpage

\renewcommand{\thesection}{S\arabic{section}}
\renewcommand{\theproposition}{S\arabic{proposition}}
\renewcommand{\theassumption}{S\arabic{assumption}}

\renewcommand{\thethm}{S\arabic{thm}}

\renewcommand{\thelemma}{S\arabic{lemma}}
\renewcommand{\thealgocf}{S\arabic{algocf}}
\renewcommand{\thefigure}{S\arabic{figure}}
\renewcommand{\thetable}{S\arabic{table}}
\renewcommand{\thepage}{S\arabic{page}}
\renewcommand{\thetable}{S\arabic{table}}
\renewcommand{\thepage}{S\arabic{page}}
\setcounter{page}{1}
\setcounter{section}{0}
\setcounter{algocf}{0}
\setcounter{lemma}{0}
\setcounter{assumption}{0}
\setcounter{table}{0}
\setcounter{thm}{0}

We refer to equations, sections, lemmas in the main paper as Equation~1, Section~1, Lemma~1 etc., and to equations, sections and lemmas, etc in the supplement as Equation~(S1), Section~S1 and Lemma~S1, etc.

The supplement is organized as follows. Section \ref{app:propertiesSignedBlockPMMH} presents some properties of the signed block PMMH algorithm.
Section \ref{app:IF} derives a tractable formula for the inefficiency factor of the algorithm. Section \ref{app:optimalTuningGeneral} shows a more general approach to obtain the optimal tuning for the signed block PMMH algorithm. Section \ref{app:ExactVsApprox} compares the computational time between the block-Poisson estimator and the approximate method in \cite{quiroz2016speeding}. Section \ref{app:testGuidelines} evaluates the robustness of the tuning guideline to deviations from its assumption.  Section \ref{app: block PMMH L geq 0} analyses a special case of the signed block PMMH algorithm when the estimator is almost surely positive. Section \ref{app:FireflySupplement} provides some additional details for the experiments in Section \ref{sec:Experiments}. Section 
\ref{app:ProofsCollected} provides the proofs of some results in the paper and the supplement.

\section{Properties of the signed block PMMH algorithm}\label{app:propertiesSignedBlockPMMH}
\subsection{Correlation from blocking}\label{app:corrFromBlocking}
We first show that the correlation induced by block PMMH has a simple form. Let $u \coloneqq (u_1, \dots, u_G)$ and  $u^\prime \coloneqq (u^\prime_1, \dots, u^\prime_G)$ denote the current and proposed values of $U$, respectively. We  will write $u_i \in \mathcal{U}$ for $i=1, \dots, G,$ so that $u_{1:G} \in \mathcal{U}^{\,G}$. Then
\begin{lemma}\label{lemma: u_i are indep pi bar}
$U_1, \dots, U_G$ are independent ($\ov \pi$) conditional on $\theta$,
i.e., $\ov \pi(u_{1:G}|\theta) = \prod_{i=1}^G \pi(u_i|\theta)$.
\end{lemma}
Let $\ell (\theta',u') = \log |\wh L_B(\theta',u') | $ and $\ell (\theta,u) = \log |\wh L_B(\theta,u) |$. Then,
\begin{align*}
\ell (\theta',u') & = \sum_{i=1}^G \log |L^{(i)} (\theta',u_i')| \quad \text{and} \quad \ell (\theta,u) = \sum_{i=1}^G \log |L^{(i)} (\theta,u_i)|;
\end{align*}
$\ell (\theta',u')$ and $\ell (\theta,u)$ have $G-1$ out of $G$ terms of the $u_i$ in common, and the $u_i$ are independent by
Lemma~\ref{lemma: u_i are indep pi bar}. Hence, for $G$ large and $\theta$ close to $\theta'$,
we argue that  $\ell (\theta',u')$ and $\ell (\theta,u)$ are approximately normal
with a correlation close to $\rho = 1-\frac1G$. Furthermore, if the sample size is large, then
$\theta$ and $\theta'$ are likely to be close by the  Bernstein von Mises theorem \citep[Chapter 10.2]{van1998asymptotic}.
In our case $G = 100$ and the sample size $n$ is large.
Hence, $\mathrm{Corr}\big(\ell(\theta',u'),\ell(\theta,u)\big)\approx1-1/G$, which demonstrates why the block-Poisson estimator in \eqref{eq:UnbiasedLikelihoodEstimator} allows us
to target a given correlation between the log of the likelihood estimator at the current and proposed draws.

\subsection{Convergence of the signed block PMMH algorithm}\label{app:convergence}
The proposal distribution updating a single block at random is
\begin{align}\label{eq: proposal q for u}
q_U(u; du')&\coloneqq
\frac1G \sum_{i=1}^G p_{U_{i}} (du_{i}^\prime) \prod_{j \neq i} \delta_{u_{j}} (d u_{j}^\prime),
\end{align}
where $\delta_a(\cdot)$ is the Dirac delta measure centered at $a$. Let  $q_\Theta(\theta; d\theta') $ be the proposal  for $\theta$ so that the joint proposal for $\theta$ and $u$ is $q_\Theta(\theta; d\theta')q_U(u; du')$. Consider now the signed block Metropolis-Hastings sampling scheme for $\theta$ and $u$.

\begin{algorithm} %% Algorithm 1
\caption{Block PMMH sampling for the absolute measure}\label{alg: signed block pmmh}
  \KwIn{Current values of $\theta$,$u$.} \vspace{1mm}

Generate $\theta^\prime, u^\prime $ using the proposal $q_\Theta(\theta; d\theta')q_U(u; du')$. \vspace{1mm}

Set $\theta \gets \theta^\prime$ and $u \gets u^\prime$ with probability
%$\alpha_{\Theta,U} (\theta, u, \theta',u') = 1\wedge r_{\Theta,U} (\theta, u; \theta',u')$, where
\begin{align*}
\alpha_{\Theta,U} (\theta, u, \theta',u') = 1\wedge \frac{\ov \pi(d\theta',du')}{\ov \pi(d\theta,du)} \times \frac{q_\Theta(\theta'; d\theta )
q_U(u'; d u )}{q_\Theta(\theta; d\theta')q_U(u; du') }.
\end{align*} \vspace{1mm}

\KwOut{New values of $\theta$, $u$.}

\end{algorithm}
\noindent The next lemma gives a workable expression for the acceptance probability of Algorithm \ref{alg: signed block pmmh}.
\begin{lemma}\label{lemma: reversible u}
\begin{align}\label{eq: simplified accept proposal theta u}
\frac {\ov \pi(d\theta',du')}{\ov \pi(d\theta,du)} \times \frac{q_\Theta(\theta'; d\theta)q_U(u';du)}{q_\Theta(\theta; d\theta')q_U(u; du') }
& = \frac{|\wh L (\theta',u')| p_\Theta(d\theta') }{ |\wh L (\theta,u)| p_\Theta(d\theta)} \frac{q_\Theta(\theta'; d\theta) }{q_\Theta(\theta; d\theta')}.
\end{align}
\end{lemma}

Theorem \ref{lemma: signed pmmh sampling scheme} below proves the ergodic properties of Algorithm~\ref{alg: signed block pmmh} based on assumptions about the following sampling scheme targeting the intractable $\bar \pi(d\theta)$ and some additional conditions stated in Assumption \ref{ass: ergodic assumpy signed pmmh}.

\begin{algorithm}
\caption{MH sampling for the $\theta$-marginal absolute measure}\label{alg: theta based sampling scheme}
\KwIn{Current value of $\theta$.} \vspace{1mm}

Generate $\theta^\prime$ using the proposal $q_\Theta(\theta; d\theta')$. \\ \vspace{1mm}

Set $\theta  \gets \theta^\prime$ with probability %$\alpha_\Theta(\theta, \theta') = 1\wedge r_\Theta(\theta, \theta')$, where
\begin{align*}
\alpha_\Theta(\theta, \theta') = 1\wedge \frac{ \bar \pi(d\theta') q(\theta'; d\theta)} { \bar \pi(d\theta) q(\theta; d\theta') }.
\end{align*} \vspace{1mm}

\KwOut{New value of $\theta$.}
\end{algorithm}

%\begin{ass} \label{ass: ergodic assumpy signed pmmh}% Assumption 1
%$~$
%\begin{enumerate}[topsep=0pt, label={\emph{(\roman*)}}]
%\item For all $\theta \in \Theta$ and $ u \in {\mathcal{U}}$, $0<|\wh L(\theta, u %)|<\infty$.
%\item For all $\theta \in \Theta$ and $ u \in {\mathcal{U}}$,
%\[\rho(\theta,u):=1-\int\alpha_{\Theta,U}(\theta,u;\theta',u')q_\Theta(\theta;d\theta')q_U(u;du')>0.\]
%\end{enumerate}
%\end{ass}

\begin{assumption} \label{ass: ergodic assumpy signed pmmh}% Assumption 1
$~$
\begin{enumerate}[topsep=0pt, label={\emph{(\roman*)}}]
\item For all $\theta \in \Theta$ and $ u \in {\mathcal{U}}^{\,G}$, $-\infty < \wh L(\theta, u ) < \infty $ and
$C(\theta) > 0 $,  where $C(\theta)$ is defined in \eqref{eq:AugmentedPosterior}.
\item Let $P_\Theta(\theta; d\theta') = K_\Theta(\theta; d\theta') + \delta_\theta(d\theta') \left(1-\int  K_\Theta(\theta; d\theta')\right)  $ and
$K_\Theta(\theta; d\theta') = \alpha_{\Theta}(\theta, \theta') q_\Theta(\theta; d\theta')$ be the Markov transition kernel and sub-stochastic kernel of Algorithm~\ref{alg: theta based sampling scheme}. We assume that if $P_\Theta(\theta;B)>0$ for some $\theta$ and $B$ with $\bar \pi(B)>0$, then $K_\Theta(\theta;B)>0$.
\end{enumerate}
\end{assumption}
%Note that $\rho(\theta,u)$ is the probability of staying at $(\theta,u)$
%and Assumption \ref{ass: ergodic assumpy signed pmmh}(ii) says that Algorithm \ref{alg: %signed block pmmh} has a non-zero probability of not moving in each iteration.  
If $\theta' \not=\theta$, then $K_\Theta(\theta; d\theta) = q_\Theta(\theta; d\theta)$ and $P_\Theta(\theta; d\theta') = K_\Theta(\theta; d\theta')$; a sufficient condition for Assumption \ref{ass: ergodic assumpy signed pmmh}(ii) is that if $P_\Theta(\theta; d\theta)>0$ then $q_\Theta(\theta; d\theta)>0$; i.e. the proposal kernel $q_\Theta(\theta;\cdot) >0$ whenever the transition probability $P_\Theta(\theta; d\theta)>0$.

Theorem \ref{lemma: signed pmmh sampling scheme} gives some convergence properties of the block PMMH sampling scheme for $\overline{\pi}(\cdot,\cdot)$; Part~(iii) and (iv) of Theorem \ref{lemma: signed pmmh sampling scheme}  were also obtained by \cite{lyne2015russian}.
\begin{thm}\label{lemma: signed pmmh sampling scheme}
Suppose that the Markov chain in Algorithm~\ref{alg: theta based sampling scheme} is irreducible and aperiodic and that Assumption~\ref{ass: ergodic assumpy signed pmmh} holds. Then,
\begin{enumerate}[topsep=0pt, label={\emph{(\roman*)}}]
\item
 Algorithm~\ref{alg: signed block pmmh} is reversible.
\item
Samples from Algorithm~\ref{alg: signed block pmmh} converges to $\ov \pi$ in total variation norm.
\item
Suppose also that $\E_\pi[|\psi|]< \infty$ and $\E_{\ov \pi} (S) \neq 0 $. Then, $\wh \E_\pi[\psi] \rightarrow \E_\pi[\psi] $ $\ov \pi$-almost surely.
\item
Define the Inefficiency Factor (IF)
\begin{align}\label{eq: ineff psi S}
\mathrm{IF}_{\ov \pi, \psi S} & = \frac{\Var_{\ov \pi} (\psi S) + 2 \sum_{j=1}^\infty \Gamma_j}{\Var_{\ov \pi} (\psi S)},
\end{align}
where $\Gamma_j = \Cov_{\ov \pi} \Big ( S_0\psi(\theta_0),S_j\psi(\theta_j) \Big)$
and $\Big \{ (\theta_0,S_0), \dots, (\theta_j,S_j),\dots \Big \}$ are the MCMC iterates generated by the sampling scheme in Algorithm \ref{alg: signed block pmmh}. If $\Var_{\ov \pi} [\psi S]\mathrm{IF}_{\ov \pi, \psi S} < \infty$ and $\E_{\ov \pi} (S) \neq 0 $, then
\begin{align} \label{eq: CLT}
\sqrt N \Big ( \wh \E_\pi(\psi) - \E_\pi(\psi) \Big) & \stackrel{d}{\rightarrow} \mathcal{N}\Big (0, \frac{\Var_{\ov \pi} (\psi S)\mathrm{IF}_{\ov \pi, \psi S}}{\E_{\ov \pi} (S)^2} \Big), \textit{and } {\E_{\ov \pi} (S)^2}=(2\tau-1)^2.
\end{align}
\end{enumerate}
\end{thm}

%We can show that the variance of the estimator $\wh \E_\pi(\psi)$ %does converge to the limiting variance in \eqref{eq: CLT}. 
%Write the estimator $\wh \E_\pi(\psi)$ in \eqref{eq:ISestimator} as %$\wh \E_\pi(\psi)=X_N/S_N$ with 
%$X_N:=(1/N)\sum_{i=1}^{N}\psi(\thet%a^{(i)})s^{(i)}$ and %$S_N:=(1/N)\sum_{i=1}^{N}s^{(i)}$. Then, it can be shown that
%\[V\Big(\frac{\sqrt N X_N}{S_N}\Big)\to \frac{\Var_{\ov \pi} (\psi %S)\mathrm{IF}_{\ov \pi, \psi S}}{\E_{\ov \pi} (S)^2}.\]
%Indeed, without loss of generality, consider  $\E_\pi(\psi)=0$. Under %some regularity conditions, we can assume that $\sqrt N %X_N\stackrel{d}{\rightarrow} \mathcal{N}(0,A)$ with $A:=\Var_{\ov %\pi} (\psi S)\mathrm{IF}_{\ov \pi, \psi S}$ and that %$\E|S_N-\mu_S|\to0$ with $\mu_S:=2\tau-1$.
%By Taylor's expansion for the function $g(x,y)=x/y$ at $x=0$ and %$y=\mu_S$, we have that
%\[\frac{x}{y}=\frac{x}{\mu_S}-\frac{x(y-\mu_S)}{\mu_S^2}+o(|y-\mu_S|)%\]

\section{The inefficiency factor for signed block PMMH\label{app:IF}}
This section presents the framework for obtaining heuristic guidelines for choosing $\lambda$ and $m$
in Section \ref{sec:Methodology}. In particular, we obtain a tractable expression for the inefficiency factor
for the signed block PMMH sampler, which is used for minimizing the computational time $\mathrm{CT}_B$. The analysis
follows \cite{pitt2012some}, and our simplifying assumptions are in the same spirit.

Define,
\begin{equation}
\begin{aligned}
C_i(\theta) & = \int_{\mathcal{U}} | \wh L^{(i)} (\theta, u_i) |p_{U}(d u_i) , \wt Z_i(\theta,u_i)  = \log | \wh L^{(i)} (\theta, u_i)| - \log C_i(\theta), \quad i=1, \dots, G, \\
\label{eq:Z_defined}\\
Z(\theta, u) & = \sum_{i=1}^G \wt Z_i(\theta,u_i)\quad {\rm and} \quad  \wt Z_{1:G} = \Big(\wt Z_1, \dots , \wt Z_G  \Big) .
\end{aligned}
\end{equation}
%We then have the following lemma whose proof is straightforward and is omitted.
Lemma \ref{eq: indep z i stuff} justifies the mapping of $(\theta,u)$ into a more tractable set of variables. Its proof is straightforward and omitted. 
\begin{lemma}\label{eq: indep z i stuff}
\begin{equation}
\begin{aligned}\label{eq: various things about pi bar}
\ov \pi (d\theta,d u) & = \exp(z) \bar \pi(d\theta) \prod_{i=1}^G p_{U_i}(du_i) , \quad \,
\ov \pi ( d \wt z_{1:G}|\theta )  = \prod_{i=1}^G \exp( \wt z_i) p_{Z_i}(d\wt z_i), \\
& \E_{u_i \sim p_{U_i}}  \Big(\exp\left(\wt Z_i\right)|\theta \Big) = 1 \quad \text{and} \quad
\E_{u \sim p_{U}}  \Big(\exp(Z)|\theta \Big) = 1,
\end{aligned}
\end{equation}
with $\bar \pi(d\theta)$ in \eqref{eq:nu_theta}.
\end{lemma}
Equation \eqref{eq: various things about pi bar} shows that the $\wt z_i$ are $\ov \pi$-independent conditional on $\theta$. Let $ v = \wh L (\theta, u) $ and take $w$ such that $(\theta, v, z, w)$ is a diffeomorphism  of $(\theta, u)$.
Then, $p(d\theta, dv, d z, d w) = p(d\theta, du)$. This allows the transformation of the measure $\ov \pi (\theta,u)$ to a workable expression in Lemma \ref{lem:change_of_measure}. Next, we
assume that,
\begin{assumption}\label{ass:change_of_measure}
%\begin{enumerate}[topsep=0pt, label={\emph{(\roman*)}}]
%    \item
$S \coloneqq \mathrm{sign}(\widehat L(\theta, U))$ and $Z(\theta,U)$ are $\ov\pi$-independent given $\theta$.
%\end{enumerate}
\end{assumption}
Assumption \ref{ass:change_of_measure} is reasonable as $Z$ is defined in terms of the absolute value of the estimator and therefore ignores the sign.
\begin{lemma}\label{lem:change_of_measure}
Suppose that Assumption \ref{ass:change_of_measure} holds. Then,
\begin{align}\label{eq:hypothetical_target}
\ov \pi (d\theta,ds, dz) & = \exp(z) \bar \pi(d \theta) p(ds|\theta)p(dz|\theta).
\end{align}
\end{lemma}
Similarly to \cite{pitt2012some}, we consider a hypothetical chain targeting \eqref{eq:hypothetical_target}.
The following assumption presents the idealized proposal which makes the derivation of the inefficiency tractable.

\begin{assumption}\label{Assumption1}$~$
\begin{enumerate}[topsep=0pt, label={\emph{(\roman*)}}]
    \item If $U_i \sim p_{U_i}(\cdot) , i=1, \dots, G$, then
    the distribution of $Z = \sum_{i=1}^G \wt Z_i(\theta,U_i)$ conditional on $\theta$
    is $\mathcal{N}(-\sigma^2/2, \sigma^2)$ with $\sigma^2 \coloneqq \mathrm{V}_{u \sim p_U}[Z]$, which is independent of $\theta$.
    \item $q(\theta,s,z; d\theta',ds',dz')=\bar \pi(d\theta')p(ds'|\theta')q(z; dz' | \sigma^2, \rho)$, with $\rho = 1-\frac1G$ and
    \begin{equation}\label{eq:idealized_proposal_zprime_given_z}
    q(z; dz' | \rho,\sigma^2)=\mathcal{N}\left(dz'\bigg|-\frac{\sigma^2}{2}(1-\rho)+\rho z,\sigma^2(1-\rho^2)\right) \text{ for } \rho = 1 - \frac{1}{G}.
    \end{equation}
\end{enumerate}
\end{assumption}
The mean of $z$ in Part (i) of Assumption \ref{Assumption1} is $-1/2$ of the variance, which
is now consistent with the fact that $\E_{u \sim p_U}[e^Z]=1$. This
 implies that, if $Z \sim \overline{\pi}$, then $Z \sim \mathcal{N}(\sigma^2/2, \sigma^2)$.
 The proposal $q(z; dz'|\rho,\sigma^2)$ in Part (ii) of Assumption \ref{Assumption1}
 implies that the correlation between the current $z$ and proposal $z'$ is $\rho= 1- 1/G$.
 The assumption that $\rho = 1 - 1/G$ is plausible because the current $Z$ and proposed $Z^\prime$ differ only by one block;
 see also the discussion at the end of Section~\ref{subsec:signed-block-algorithm}. The inefficiency factor for the independent pseudo-marginal method in \cite{pitt2012some} is derived using $\rho=0$. The next lemma uses their proof, but with the proposal in \eqref{eq:idealized_proposal_zprime_given_z}.
\begin{lemma} \label{lemma: idealized ineff} Suppose that Assumption \ref{Assumption1} holds. Then,
\begin{enumerate}[topsep=0pt, label={\emph{(\roman*)}}]
\item The MH acceptance probability of the proposal  $q (\theta, s, z; d\theta',ds',d z')$ is
\begin{align*}
 \min \{1, \exp(z'-z) \}.
\end{align*}
\item
The acceptance probability conditional on $z$ of the idealized sampling scheme is
\begin{align*}
k(z|\sigma^2, \rho) & = \int  \min \{1, \exp(z'-z) \} q(z; dz'|\sigma^2, \rho).
\end{align*}
\item
The inefficiency of the sampling scheme is
\begin{align*}
\IF_{\ov \pi} (\sigma^2, \rho) & = 1 + 2 \mathrm{E}_{\ov \pi(z|\theta)}  \Big( \frac{1-k(z| \sigma^2, \rho)}{k(z| \sigma^2, \rho)} \Big),
\end{align*}
where
\beq\label{conditional_acceptance_prob}
k(z|\sigma^2, \rho) =\exp(-x+w^2/2)\Phi\left(\frac{x}{w}-w\right)+\Phi\left(\frac{-x}{w}\right),
\eeq
with $x:=\bigg (z + \frac{\sigma^2}{2}\bigg ) (1-\rho) $, $w:=\sigma\sqrt{ 1- \rho^2}$ and $\Phi$ denotes the standard normal cumulative density function.
\end{enumerate}
\end{lemma}
The inefficiency $\IF_{\ov \pi} (\sigma^2, \rho)$ can be computed accurately using one-dimensional numerical integration
 as $\ov\pi(z|\theta)\sim \mathcal{N}(\sigma^2/2,\sigma^2)$. We end this section by presenting an alternative set of more restrictive assumptions implying Part (i) of Assumption \ref{Assumption1} and the proposal in Part (ii) of Assumption \ref{Assumption1}, i.e. \eqref{eq:idealized_proposal_zprime_given_z}. This assumption provides greater understanding of the results above.
\begin{assumption}\label{ass:more_restrictive_assumptions}If $U_i \sim p_{U}(\cdot) , i=1, \dots, G$, then
    the distribution of $\wt Z_i(\theta,U_i)$ conditional on $\theta$
    is $\mathcal{N}(-\sigma^2/2G, \sigma^2/G)$ with $\sigma^2 \coloneqq \mathrm{V}_{u \sim p_U}[Z]$ for $Z = \sum_{i=1}^G \wt Z_i(\theta,U_i)$, which is independent of $\theta$.
\end{assumption}
Lemma \ref{lem:bivariate_dist_z} then implies the desired result.
\begin{lemma}\label{lem:bivariate_dist_z} Suppose that Assumption \ref{ass:more_restrictive_assumptions} holds and let $$Z = \sum_{i=1}^G \wt Z_i(\theta,U_i) \text{ and }  Z'=\sum_{i\not=j,i=1}^G \wt Z_i(\theta',U_i)+\wt Z_j(\theta',U_j'),$$ with $U_i \sim \overline{\pi}$ and $U_j'\sim p_{U_j}(\cdot).$ Then, \begin{equation}
    \label{eq:Joint_dist_z_zprime}
    \begin{bmatrix}z\\
    z^\prime
    \end{bmatrix} \sim \mathcal{N}\left( \begin{bmatrix}\frac{\sigma^{2}}{2}\\
    -\frac{\sigma^{2}}{2}(1-2\rho)
    \end{bmatrix}, \sigma^{2}\begin{bmatrix}1 & \rho\\
    \rho & 1
    \end{bmatrix} \right), \text{ with } \rho = 1-\frac{1}{G}.
    \end{equation}
\end{lemma}

\section{Optimal tuning - general case}\label{app:optimalTuningGeneral}

Algorithm \ref{alg:optTuning} assumes $m=30$ so that $\widehat{d}_m$ can be assumed to be normal and the optimal $\lambda$ can be computed easily. Section \ref{subsec:tuningProductPoisson} argues that this can be suboptimal. We therefore consider an alternative approach, based on assuming that $\widehat{d}_m$ is a mixture of normals, which makes the tuning general since a mixture of normals with enough mixture components can approximate any distribution. We propose fitting a mixture of normals to $\widehat{d}_m$ using characteristic functions.

The following lemma generalizes Lemma \ref{lem:VarAbs}; its proof is in Section \ref{app:ProofsCollected}. The notation $\mathrm{Mix-}\mathcal{N}(\mu,\sigma^{2},\omega)$ means that we have a standardized finite mixture with $C$ components. Its parameters are $\mu \coloneqq \mu_{1:C}$, $\sigma \coloneqq \sigma^2_{1:C}$, $\omega \coloneqq \omega_{1:C}$ with $\sum_j \omega_j = 1$ and, by standardization, its mean and variance are $\sum_j \omega_j \mu_j  = 0$ and $\sum_j \omega_j (\sigma_j^2 + \mu_j^2) -1 = 1$.
\begin{lemma}
\label{lem:VarAbsL_finite_mixture}Let $\bar{d}_{m}^{\,\,(h,l)}=\sqrt{\frac{m}{\gamma}}(\widehat{d}_{m}^{\,\,(h,l)}-d)\overset{iid}{\sim}\mathrm{Mix-}\mathcal{N}(\mu,\sigma^{2},\omega)$
follow mixture of normals for all $h$ and $l$ such that $\mathrm{E}\left[\bar{d}_{m}^{\,\,(h,l)}\right]=0$
and $\mathrm{V}\left[\bar{d}_{m}^{\,\,(h,l)}\right]=1$. The variance of $\log\left|\widehat{L}_B\right|$
when $a=d-\lambda$ is then
\begin{align*}
\mathrm{V}\left[\log\left|\widehat{L}_B\right|\right] & =\lambda\sum_{c=1}^{C}\omega_{c}(\nu_{c}^{2}+(\eta_{c}-\eta)^{2})+\lambda\eta^{2},
\end{align*}
where
\[
\eta_{c}=\log\left(\frac{\sigma_{c}}{\lambda}\sqrt{\frac{\gamma}{m}}\right)+\frac{1}{2}\left(\log2+\mathrm{E}_{J_{c}}\left[\psi^{(0)}(1/2+J_{c})\right]\right),
\]
$\eta=\sum_{c=1}^{C}\omega_{c}\eta_{c}$ and
\[
\nu_{c}^{2}\coloneqq\frac{1}{4}\left(\mathrm{E}_{J_{c}}\left[\psi^{(1)}(1/2+J_{c})\right]+\mathrm{V}_{J_{c}}\left[\psi^{(0)}(1/2+J_{c})\right]\right)
\]
with $J_{c}\sim\mathrm{Pois}\left(\frac{(\mu_{c}+\sqrt{\frac{m}{\gamma}}\lambda)^{2}}{2\sigma_{c}^{2}}\right)$
and $\psi^{(q)}$ is the polygamma function of order $q$. Furthermore,
$\mathrm{V}\left[\log\left|\widehat{L}_B\right|\right]<\infty$ for all
$m>0$, $\lambda>0$ and $\mu,\sigma,\omega$.
\end{lemma}

\begin{figure}
\centering
\includegraphics[width=0.90\linewidth]{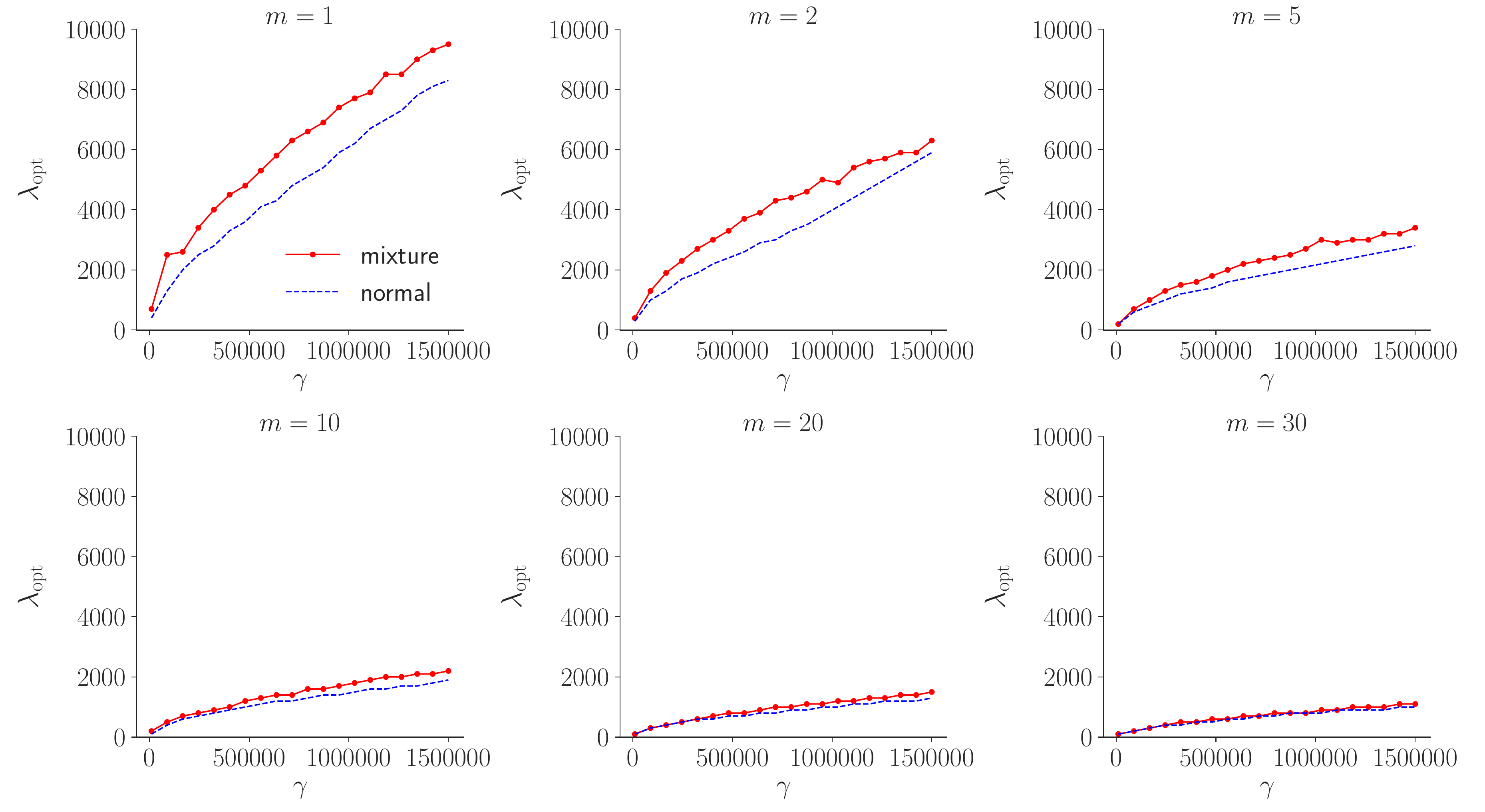}
        %\caption{$\Pr(\widehat{L}>0)$ as a function of $a$.}\label{fig:CT_fig_a}
\caption{Optimal $\lambda$ as a function of $\gamma$ for the normal approach vs the finite mixture of normal approach. Each figure shows, for a specific value of $m$ (see title of each panel), the $\lambda$ obtained by minimizing the CT in \eqref{eq:CT_Poisson_simplified} under a normal assumption for $\widehat{d}_m$ (blue dashed line) and a finite mixture of normal assumption for $\widehat{d}_m$ (red solid line).
}\label{fig:lambda_opt_normal_vs_mixture}\end{figure}

Given the finite mixture distribution for $\widehat{d}_m$, it is straightforward to compute the probability that the estimator is positive, as this becomes a mixture of normal cumulative distribution functions. We can numerically evaluate the CT in \eqref{eq:CT_Poisson_simplified} for the finite mixture of normals case, and can optimize $\lambda$ for a given $m$. Figure \ref{fig:lambda_opt_normal_vs_mixture} compares $\lambda_{\mathrm{opt}}$ from the mixture of normal approach to the simpler approach in Algorithm \ref{alg:optTuning} where the $d_k$ follow a Student $t$ distribution with 4 degrees of freedom. The figure illustrates that when $m=30$ (and above) the two approaches result in nearly indistinguishable optimal values for all $\gamma$; when $m$ is small, the optimal guidelines can differ substantially, especially for larger values of $\gamma$.

We now outline how to find the parameters $(\omega_{1:G}, \sigma^2_{1:G}, \mu_{1:G})$
 in the finite mixture distribution. Let $X_{1},...,X_{n}\vert\theta\overset{iid}{\sim}f_{X}(x\vert\theta)$
with a finite mean $\mu$ and variance $\sigma^{2}$. We are interested
in approximating the distribution of the sample mean $\bar{X}_{n}=n^{-1}\sum_{i=1}^{n}X_{i}$
by a mixture of normals
\[
f_{\bar{X}_{n}}(x)\approx g_{\bar{X}_{n}}(x\vert\eta,\psi^{2},\omega)=\sum_{c=1}^{C}\omega_{c}\mathcal{N}(x\vert\eta_{c},\psi_{c}^{2}),
\]
where $\mathcal{N}(x\vert\eta,\psi^{2})$ denotes a normal density with mean
$\eta$ and variance $\psi^{2}$. Let $\beta=\{\omega,\eta,\psi\}$
denote all the parameters of the mixture, and let $g_{\bar{X}_{n}}^{(\beta)}(x)$
denote the density for the sample mean from the mixture with parameters
$\beta$.

The aim is  to find the necessary number of mixture components
$C$ and the parameters $\beta$ of the mixture that approximates
$f_{\bar{X}_{n}}(x)$ well. We will do so by minimizing the $L_{2}$
distance between $f_{\bar{X}_{n}}(x)$ and $g(x\vert\theta,\psi^{2},\omega)$
\[
d\left(f_{\bar{X}_{n}},g_{\bar{X}n}^{(\beta)}\right)=\int\left(f_{\bar{X}_{n}}(x)-g_{\bar{X}_{n}}^{(\beta)}(x)\right)^{2}dx.
\]
The $L_{2}$ distance is very convenient since by Plancherel's theorem
we can turn the density matching problem into a Characteristic Function (CF) matching problem,
\[
d\left(f_{\bar{X}_{n}},g_{\bar{X}n}^{(\beta)}\right)=\int\left(f_{\bar{X}_{n}}(x)-g_{\bar{X}_n}^{(\beta)}(x)\right)^{2}dx=\int\left(\varphi_{\bar{X}_{n}}(t)-\varphi_{\bar{X}_{n}}^{(\beta)}(t)\right)^{2}dt=d\left(\varphi_{\bar{X}_{n}}(t),\varphi_{\bar{X}_{n}}^{(\beta)}(t)\right),
\]
where $\varphi_{X}(t)$ is the characteristic function $\varphi_{X}(t)\coloneqq\mathrm{E}\left[e^{itX}\right]$
for a random variable $X$. Matching CFs is especially attractive here since the density of the sample mean $f_{\bar{X}_{n}}$ may be
intractable, but its CF is straightforward to obtain,
\[
\varphi_{\bar{X}_{n}}(t)=\left(\varphi_{X}(t/n)\right)^{n},
\]
with $\varphi_{X}(t)$ the CF of $f_{X}(x\vert\theta)$. We
will match CFs for the standardized mean $\bar{Z}_{n}=(\sqrt{n}/\sigma)(\bar{X}_{n}-\mu)$.
Using the property $\varphi_{a+bX}(t)=e^{ita}\varphi_{X}(bt)$, we
obtain the CF for the standardized mean as
\[
\varphi_{\bar{Z}_{n}}(t)=e^{-it\mu\sqrt{n}/\sigma}\left(\varphi_{X}(t/(\sqrt{n}\sigma))\right)^{n}.
\]
The CF for a normal mixture $\sum_{c=1}^{C}\omega_{c}\mathcal{N}(x\vert\eta_{c},\psi_{c}^{2})$
is $\varphi_{X}(t)=\sum_{c=1}^{C}\omega_{c}\varphi_{X_{c}}(t)$, where
$\varphi_{X_{c}}(t)=\exp(i\eta_{c}t-\psi_{c}^{2}t^{2}/2)$ is the
CF of the $c$th mixture component.

We minimize $d\left(\varphi_{\bar{Z}_{n}}(t),\varphi_{\bar{Z}_{n}}^{(\beta)}(t)\right)$
with respect to $\beta$ for a given $C$ by reparameterizing the
standard deviations in the mixture components in exponential form
and the weights $\omega$ using the softmax function. The minimization
is subject to the restrictions that the mixture has zero mean and
unit variance: $\sum_{c=1}^{C}\omega_{c}\eta_{c}=0$ and $\sum_{c=1}^{C}\omega_{c}\left(\kappa_{c}^{2}+(\eta_{c}-\eta)^{2}\right)=1$.
These restrictions can be enforced directly or indirecly via penalties.

\section{Exact vs approximate subsampling MCMC}\label{app:ExactVsApprox}
\begin{figure}
\centering
\includegraphics[width=0.6\linewidth]{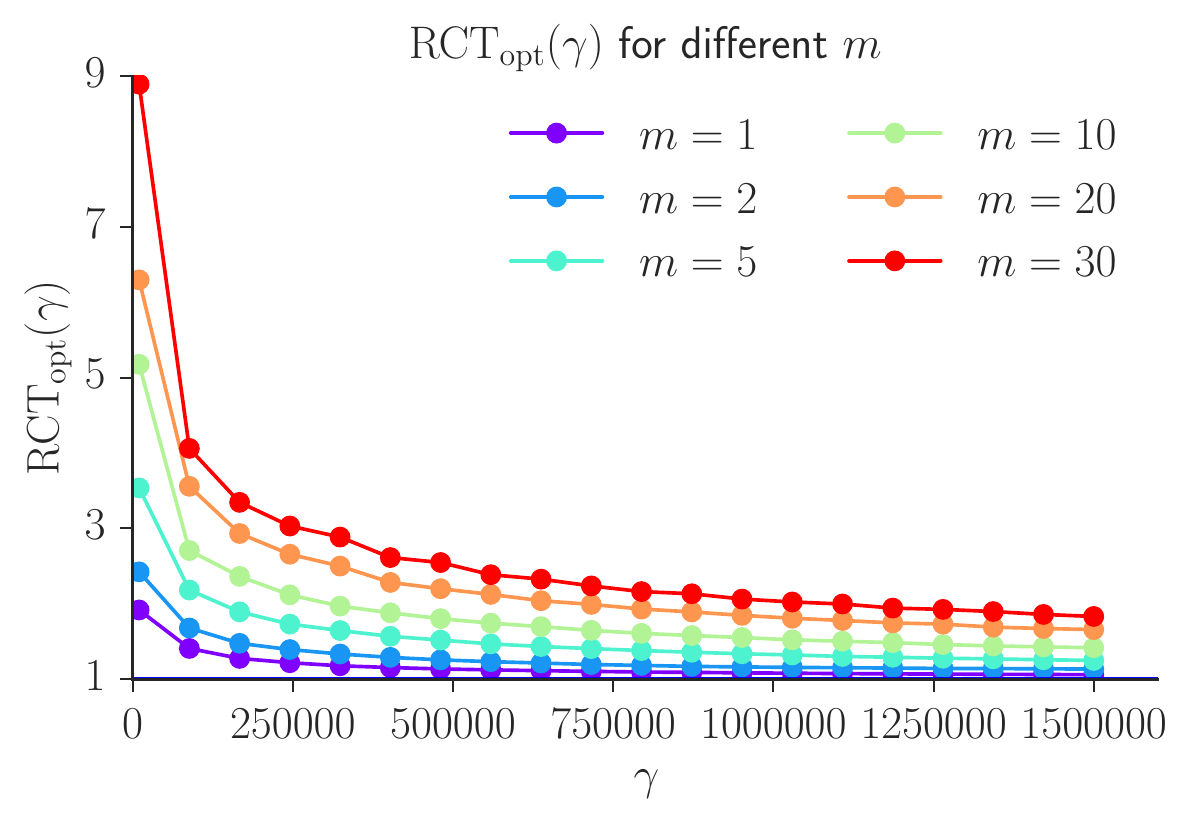}
        %\caption{$\Pr(\widehat{L}>0)$ as a function of $a$.}\label{fig:CT_fig_a}
\caption{RCT in \eqref{eq:RCT_approx_vs_exact} as a function of $\lambda$ for different values of $m$.}\label{fig:RCT_approx_vs_exact}
\end{figure}

\cite{quiroz2016speeding} propose a framework for subsampling MCMC using the bias-corrected likelihood estimator,
\begin{eqnarray*}
\widehat{L}_A & \coloneqq &  \exp\left(q + \widehat{d}_{M}
 - \frac{n^2}{2M}  \widehat{\sigma}^2_{d_{u_i}} \right),
\end{eqnarray*}
based on a subsample of size $M$. The bias-correction is only approximate so the pseudo-marginal sampler in \cite{quiroz2016speeding} targets a perturbed posterior which is within $O(n^{-1}M^-{2})$ from the true posterior in total variation distance. Thus, unlike the current exact approach, the posterior moments from the sampler in \cite{quiroz2016speeding} are only approximate. We can show that if the $d_k$ are normally distributed, then
\begin{eqnarray*}
\sigma^2_{\log \widehat{L}_A} \coloneqq \mathrm{V}[\log \widehat{L}_A] & = &  \frac{\gamma}{M} + \frac{\gamma^2}{2M^3},
\end{eqnarray*}
and we can then define the computational time
\begin{eqnarray*}
\mathrm{CT}_{\mathrm{A}}(M | \gamma, \rho = 0.99) & = & M\cdot \mathrm{IF}\left(\sigma^2_{\log \widehat{L}_A}(M | \gamma, \rho = 0.99)\right).
\end{eqnarray*}
Similarly to the block-Poisson estimator in Section \ref{subsec:tuningProductPoisson}, we can minimize this $\mathrm{CT}$ with respect to $M$, for each $\gamma$. The Relative Computational Time (RCT) of the optimal exact method proposed here against the approx in \cite{quiroz2016speeding} is
\begin{equation}\label{eq:RCT_approx_vs_exact}
    \mathrm{RCT}_{\mathrm{opt}}(\gamma) = \frac{\mathrm{CT}_{\mathrm{E}}(\lambda_{\mathrm{opt}} | \gamma, m, \rho = 0.99)}{\mathrm{CT}_{\mathrm{A}}(M_{\mathrm{opt}} | \gamma, \rho = 0.99)}.
\end{equation}
Figure \ref{fig:RCT_approx_vs_exact} plots $\mathrm{RCT}_{\mathrm{opt}}(\gamma)$ as a function of $\gamma$ for several values of $m$ and shows that the approximate approach has lower CT, by
 a factor between 2 and 9 for $m=30$, depending of the value of $\gamma$; thus there is a trade-off between exactness and computational cost.

\section{Testing the robustness of the tuning guidelines}\label{app:testGuidelines}
We now test the accuracy of the guidelines for tuning $\lambda$ when $m = 30$. In particular, we will test the robustness of the guidelines to deviations from the main assumption behind the derivation of the guidelines, i.e. $\gamma = n^2 \sigma^2_d$ is independent of $\theta$. To do so, we construct the differences $d_k = \ell_k - q_k$ in the following two ways:
\begin{enumerate}[topsep = 0pt]
\item \emph{Correct assumption}: We sample a population $d_k$ that does not depend on $\theta$ and obtain $\sigma^2_d = V[d_k] = 1/n\sum_{k=1}^n (d_k - \overline{d})^2$.
\item \emph{Incorrect assumption}: For each $\theta$, we compute $d_k(\theta) = \ell_k(\theta) - q_k(\theta)$, taking $q_k(\theta)$ to be the data expanded control variate in Section \ref{subsec:Control-variates}. We then compute $\gamma(\theta) = n^2 \sigma^2_d(\theta)$; in this setting, $\gamma$ depends on $\theta$ and therefore violates the main assumption in the optimal tuning derivation.
\end{enumerate}

The guidelines are tested as follows. We consider a simple Poisson regression model with one covariate and no intercept, $y_k \sim \mathrm{Poisson}\left(\exp(\theta x_k)\right)$. To make the experiment as clean as possible we choose a perfect proposal by computing $\bar \pi(\theta) \propto \int |\widehat{L}(\theta, u)|p_U(u)du$ on a grid of $\theta$ values (by Monte Carlo simulation). Since $\theta$ is one dimensional, this distribution is easily sampled by the inverse cdf method. For each $\lambda$, we can now compute the empirical version of the CT in Section \ref{sec:Experiments} by estimating the integrated auto-correlation time of $s\theta$ from a long MCMC chain for $\theta$. Repeating this procedure for each $\lambda$ on a grid we can select the $\lambda$ that minimizes the empirical CT and compare it to $\lambda_{\mathrm{opt}}$ from Algorithm \ref{alg:optTuning}. 

\begin{figure}
\centering

\begin{subfigure}[t]{.45\textwidth}
\centering
\includegraphics[width=\linewidth]{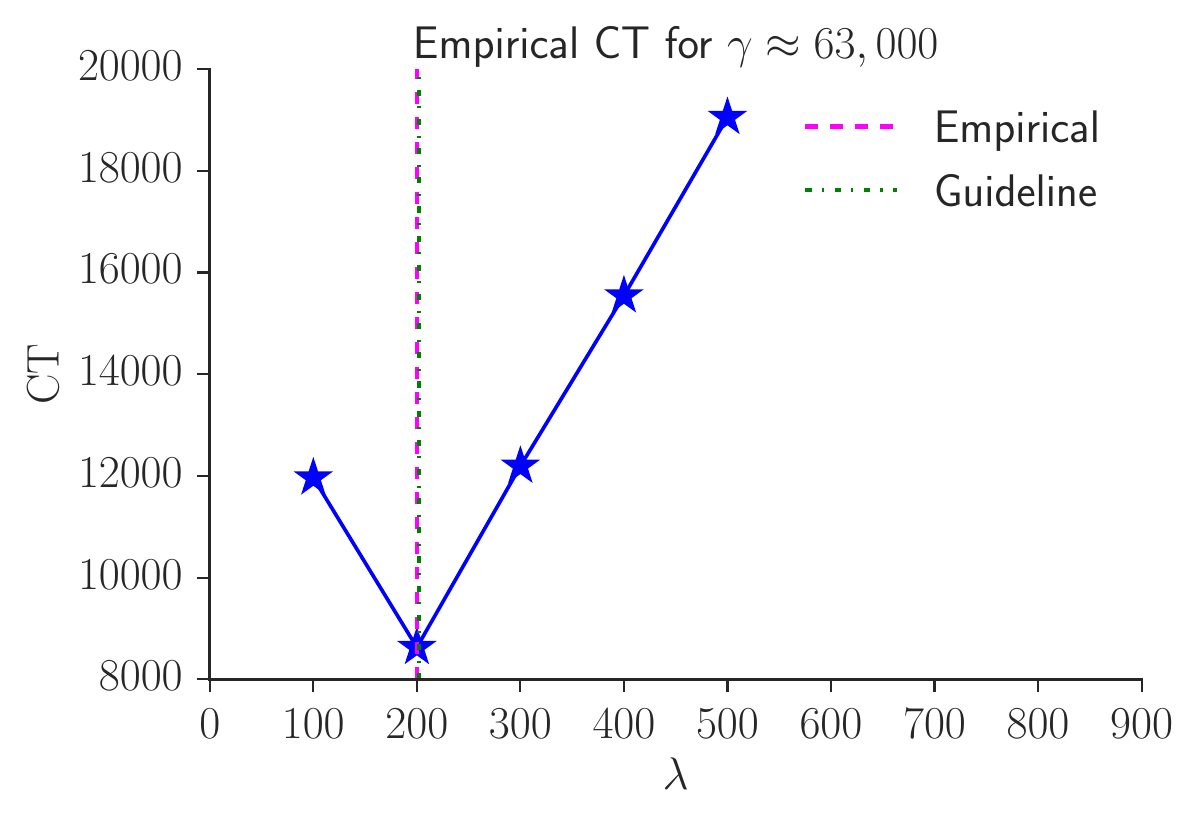}
        %\caption{$\Pr(\widehat{L}>0)$ as a function of $a$.}\label{fig:CT_fig_a}
\end{subfigure}
\begin{subfigure}[t]{.45\textwidth}
\centering
\includegraphics[width=\linewidth]{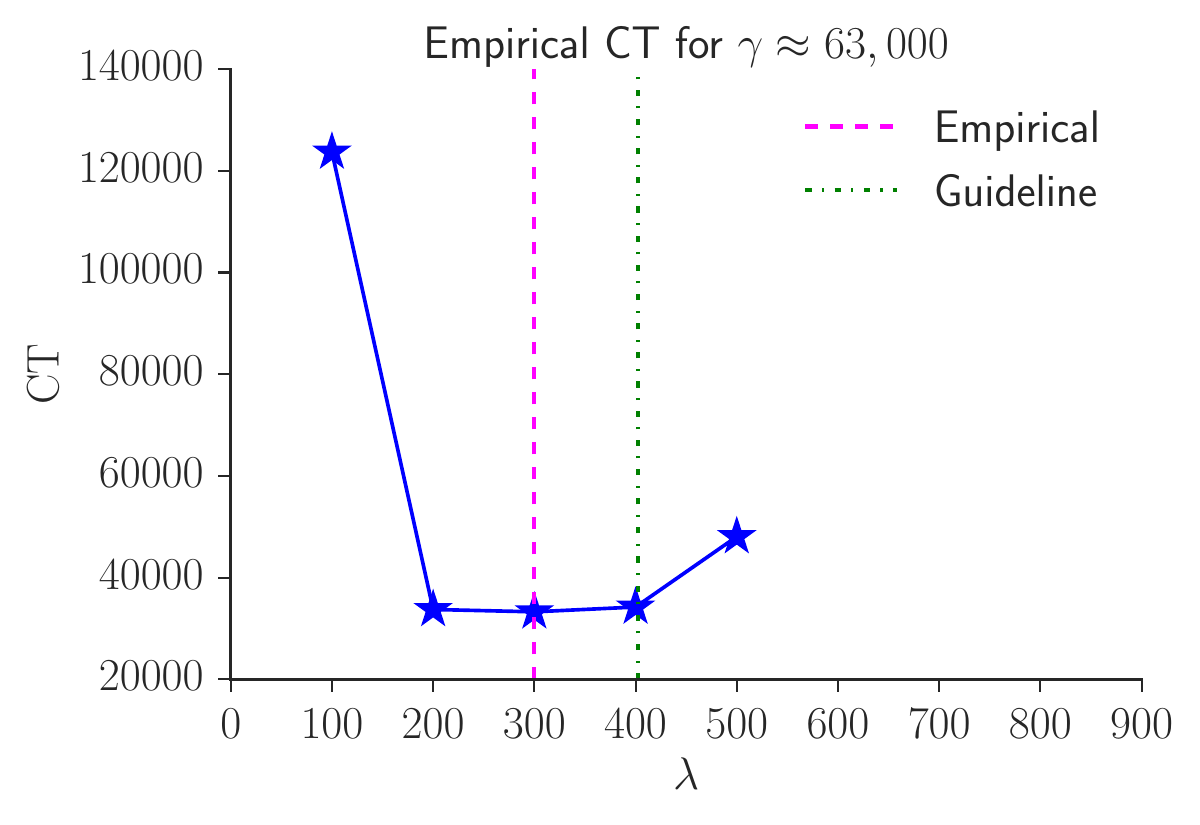}
        %\caption{$\mathrm{CT}$ as a function of $a$}\label{fig:CT_fig_b}
\end{subfigure}

\begin{subfigure}[t]{.45\textwidth}
\centering
\includegraphics[width=\linewidth]{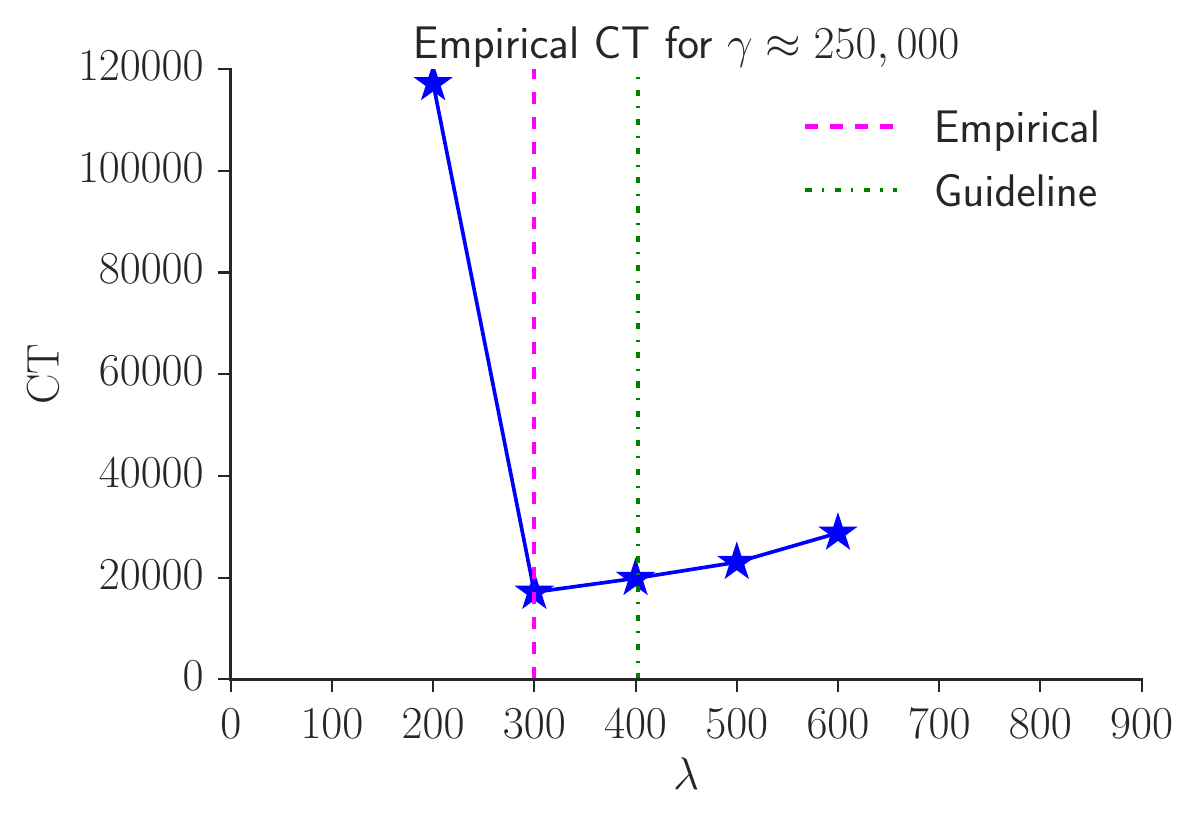}
        \caption{$\gamma$ does not depend on $\theta$.}%\label{fig:CT_fig_b}
\end{subfigure}
\begin{subfigure}[t]{0.45\textwidth}
\centering
\includegraphics[width=\linewidth]{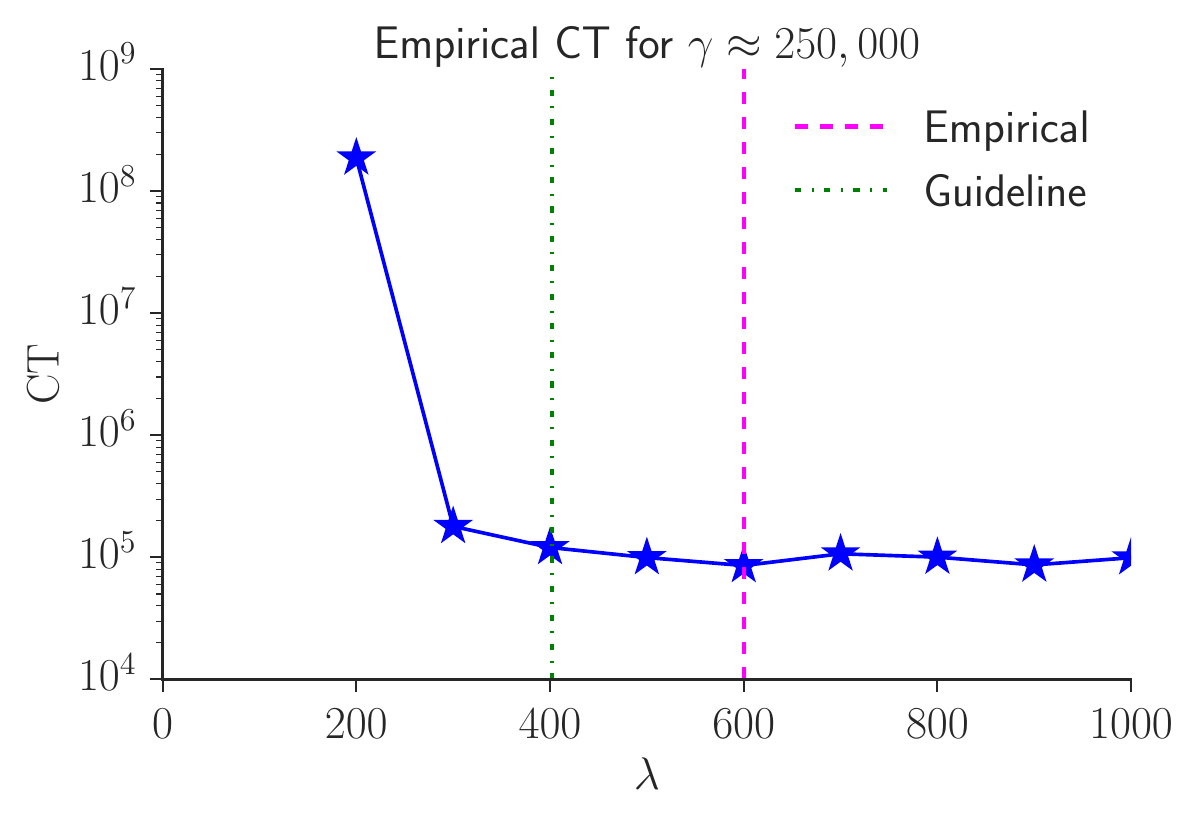}
        \caption{$\gamma$ depends on $\theta$.}%\label{fig:CT_fig_a}
\end{subfigure}

\caption{Testing the robustness of the tuning guidelines in Algorithm \ref{alg:optTuning}. Panels (A) show the empirical CT for two different $\gamma$ values, when the assumption that $\gamma$ is independent of $\theta$ is correct. Panels (B) show the empirical CT for two different $\gamma$ values, but in a setting that violates the assumption that $\gamma$ depends not depend on $\theta$. The vertical lines correspond to the optimal $\lambda$ according to the guidelines in Algorithm \ref{alg:optTuning} (dotted green) and the empirically observed optimum (dashed magenta).}\label{fig:Empirical_vs_Guidelines}
\end{figure}

Figure \ref{fig:Empirical_vs_Guidelines} shows that when the $\lambda_{\mathrm{opt}}$ provided by the guidelines does not agree with the empirically obtained one, the difference in computational time between the different $\lambda$ is very small. We therefore conclude that the guidelines are sensible, and more importantly, they never suggest a $\lambda$ which is too small resulting in a catastrophically large CT.

\section{The block PMMH with $\wh L \geq 0 $\label{app: block PMMH L geq 0}}
This section derives a special case of our algorithm in which the sign is 1 $\overline{\pi}$-almost surely, which we refer to as the block PMMH. We provide guidelines for its optimal tuning.

\subsection{Convergence of the block PMMH\label{app:convergenceBlockPMMH}}
Suppose that the $\widehat{L}^{(i)}(\theta,u) \geq  0, i=1, \dots, G$ for all $\theta \in \Theta, u_{1:G} \in {\mathcal{  U}}^{\, G}$,
so that the likelihood estimator is $\overline{\pi}$-almost surely positive, i.e. $\Pr_{\overline{\pi}}(S = 1) = 1$.
Then, $\bar \pi (d\theta) = \pi(d\theta)$ (the posterior), $C(\theta) = L(\theta)$ (the likelihood),
$\ov C = p(y)$ (the marginal likelihood), \eqref{eq:AugmentedPosterior_blocking} becomes
\begin{eqnarray*}
\overline{\pi}(d\theta,du_{1:G}) & =   \frac{p_{\Theta}(d\theta)}{\overline{C}} \prod_{i = 1}^G \widehat{L}^{(i)}(\theta, u_i) p_{U}(du_i), \label{eq:AugmentedPosterior_blocking_pos_estimator}
\end{eqnarray*}
and  \eqref{eq:ISestimator} becomes
\begin{eqnarray}
\widehat{\mathrm{E}}_{\pi}[\psi] & = & \frac{1}{N}\sum_{i=1}^{N}\psi(\theta^{(i)}). \label{eq:ISestimator_pos_estimator}
\end{eqnarray}
Then the following convergence result holds. Its proof follows from the proof of Theorem \ref{lemma: signed pmmh sampling scheme}.
\begin{cor}\label{cor:block_pmmh} Suppose that $\widehat{L}(\theta,u) \geq 0$, and that the Markov chain in Algorithm~\ref{alg: theta based sampling scheme} is irreducible and aperiodic. Then,
\begin{enumerate}[topsep=0pt, label={\emph{(\roman*)}}]
\item
 Algorithm~\ref{alg: signed block pmmh} is reversible.
\item
 Algorithm~\ref{alg: signed block pmmh} converges to $\overline \pi$ in total variation norm, with marginal $\pi(d\theta) = \int_{\mathcal{U}}\overline \pi (d\theta, du)$.
\item  Suppose that $\E_\pi[|\psi|]< \infty$. Then, $\wh \E_\pi[\psi] \rightarrow \E_\pi[\psi] $ a.s. ($\,\ov \pi$).
\item
Define
\begin{align*}%\label{eq: ineff psi block}
\mathrm{IF}_{\overline \pi, \psi} & = \frac{\Var_{\overline \pi} (\psi) + 2 \sum_{j=1}^\infty \overline{\Gamma}_j}{\Var_{\overline \pi} (\psi) } ,
\end{align*}
where $\overline{\Gamma} = \Cov_{\overline \pi} \Big(\psi(\theta_0),\psi(\theta_j)\Big)$
and $(\theta_0, \dots, \theta_j,\dots )$ are the MCMC iterates of $\theta$. If $\Var_{\overline \pi} [\psi]\mathrm{IF}_{\overline \pi, \psi}=\Var_{\pi} [\psi]\mathrm{IF}_{\overline \pi, \psi} < \infty$, then
\begin{align*} \label{eq: CLT}
\sqrt N \Big( \wh \E_\pi(\psi) - \E_\pi(\psi) \Big) & \rightarrow \mathcal{N}\Big(0, \Var_{\pi} (\psi )\mathrm{IF}_{\overline \pi, \psi} \Big).
\end{align*}
\end{enumerate}
\end{cor}
We call the algorithm with a nonnegative likelihood estimator in Corollary \ref{cor:block_pmmh} the block PMMH.
By Part (ii) of Corollary \ref{cor:block_pmmh}, unlike the signed block PMMH, the iterates of $\theta$ for the block PMMH are distributed according to $\pi(d\theta)$.

\subsection{Tuning the block PMMH\label{SS: tunning block PMMH}}
We now consider tuning the block PMMH, which is a special case of the signed block PMMH with only positive signs and therefore $\tau \coloneqq 1$; see Section \ref{subsec:A-pseudo-marginal-algorithm}.

\citet{pitt2012some} show that the optimal number of particles $m$, when $\sigma_{\log \widehat{L}}^{2} \propto 1/m$, is given implicitly by the $\sigma_{\log \widehat{L}}^{2}$ that minimizes the computational time
\begin{equation}\label{eq:CT_pos_estimator}
    \mathrm{CT}(\sigma_{\log \widehat{L}}^2) \coloneqq m\cdot \mathrm{IF}(\sigma_{\log \widehat{L}}^2) \propto \frac{\mathrm{IF}(\sigma_{\log \widehat{L}}^2)}{\sigma_{\log \widehat{L}}^2},
\end{equation}
where $\mathrm{IF}$ denotes the inefficiency factor of the pseudo-marginal chain; see also \citet{doucet2012efficient}.
This computational time only depends on $\sigma^2_{\log \widehat{L}}$, which in turn is proportional to $1/m$,
hence simplifying the expression compared to \eqref{eq:CT_Poisson_simplified}. \cite{pitt2012some} derive, based on several assumptions which we also invoke in Section \ref{app:IF}, an analytic expression for the IF and shows that  $\mathrm{CT}(\sigma_{\log \widehat{L}}^2)$
 in \eqref{eq:CT_pos_estimator} is minimized when $\sigma_{\log \widehat{L}}^2 \approx 1$. Under less restrictive assumptions
\citet{doucet2012efficient,sherlock2013efficiency} show that the optimal value of $\sigma_{\log \widehat{L}}^2$ ranges between approximately $1$ and $3.3$.

The next lemma obtains the optimal $\sigma_{\log \widehat{L}}^2$ for the block PMMH and is proved using the IF derived in \cite{pitt2012some}, but incorporating our block scheme. The lemma shows that the block PMMH with $G=100$ speeds up the independent pseudo-marginal significantly, as the optimal value of $\sigma_{\log \widehat{L}}^{2}$ is $234$, which is much larger than the recommended range $1$ to $3.3$ in \cite{pitt2012some}.
\begin{lemma}\label{lem:Optimal_sigma2_pos_estimator} Given the assumptions in Section \ref{app:IF}, $\sigma_{\log \widehat{L}}^2 \approx 2.16^2/(1-\rho^2)$ minimizes \eqref{eq:CT_pos_estimator} when $\rho = 1 - 1/G$ is close to $1$.
\end{lemma}

\begin{figure}
\centering

\begin{subfigure}[t]{.45\textwidth}
\centering
\includegraphics[width=\linewidth]{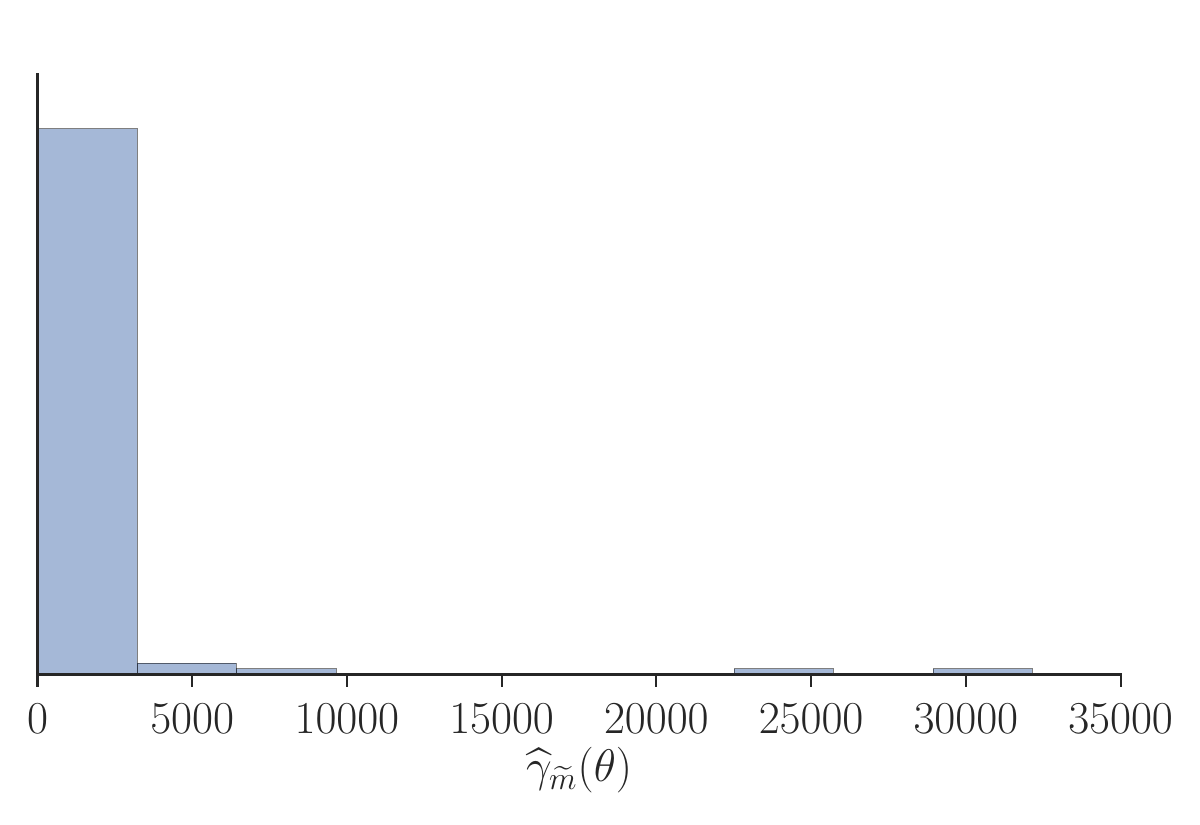}
\end{subfigure}
\begin{subfigure}[t]{.45\textwidth}
\centering
\includegraphics[width=\linewidth]{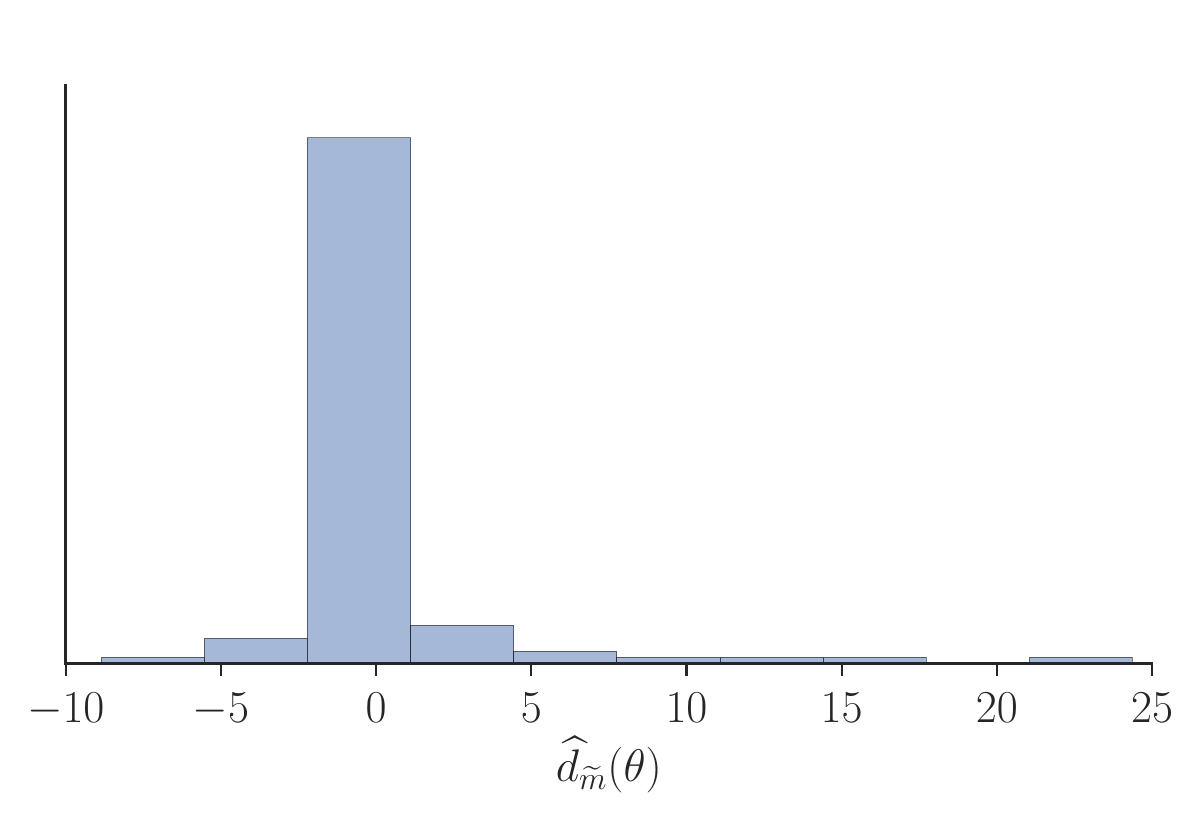}
\end{subfigure}

\begin{subfigure}[t]{.45\textwidth}
\centering
\includegraphics[width=\linewidth]{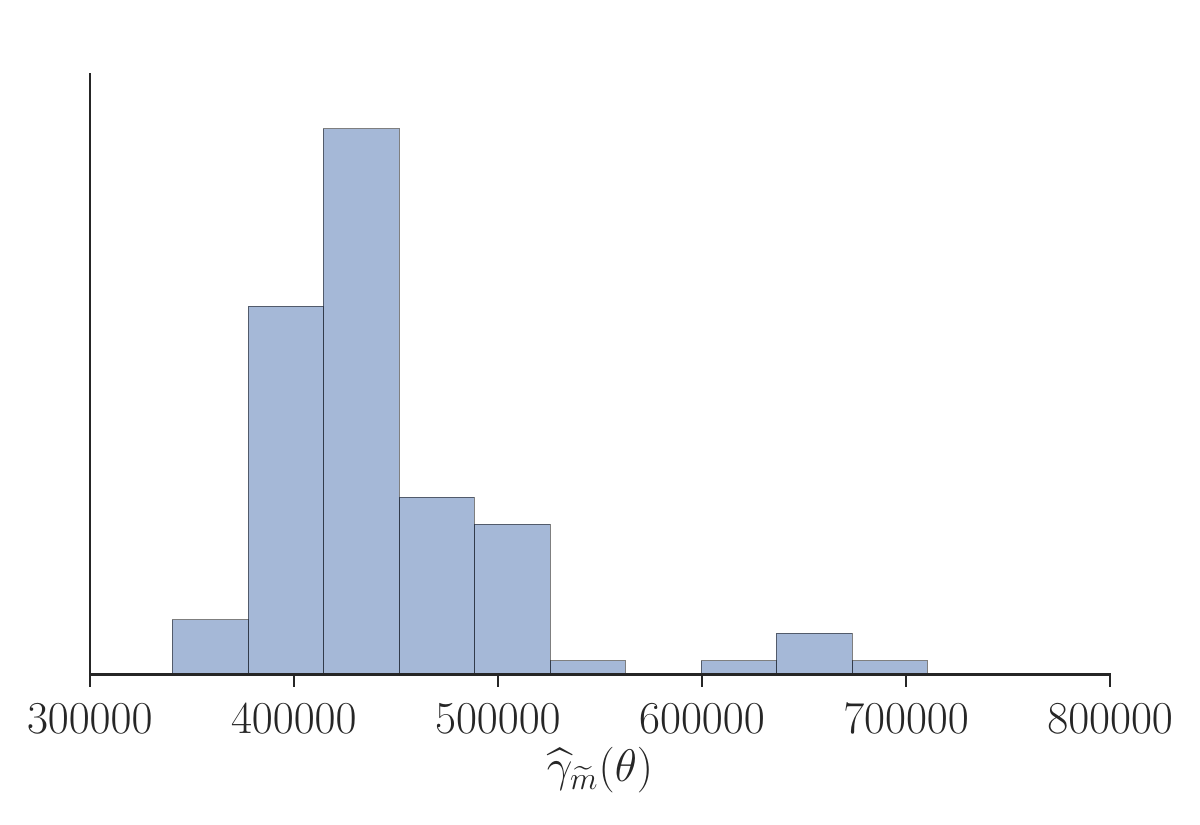}
        \caption{Estimated $\gamma(\theta)$'s.}%\label{fig:CT_lambda100_fig}
\end{subfigure}
\begin{subfigure}[t]{.45\textwidth}
\centering
\includegraphics[width=\linewidth]{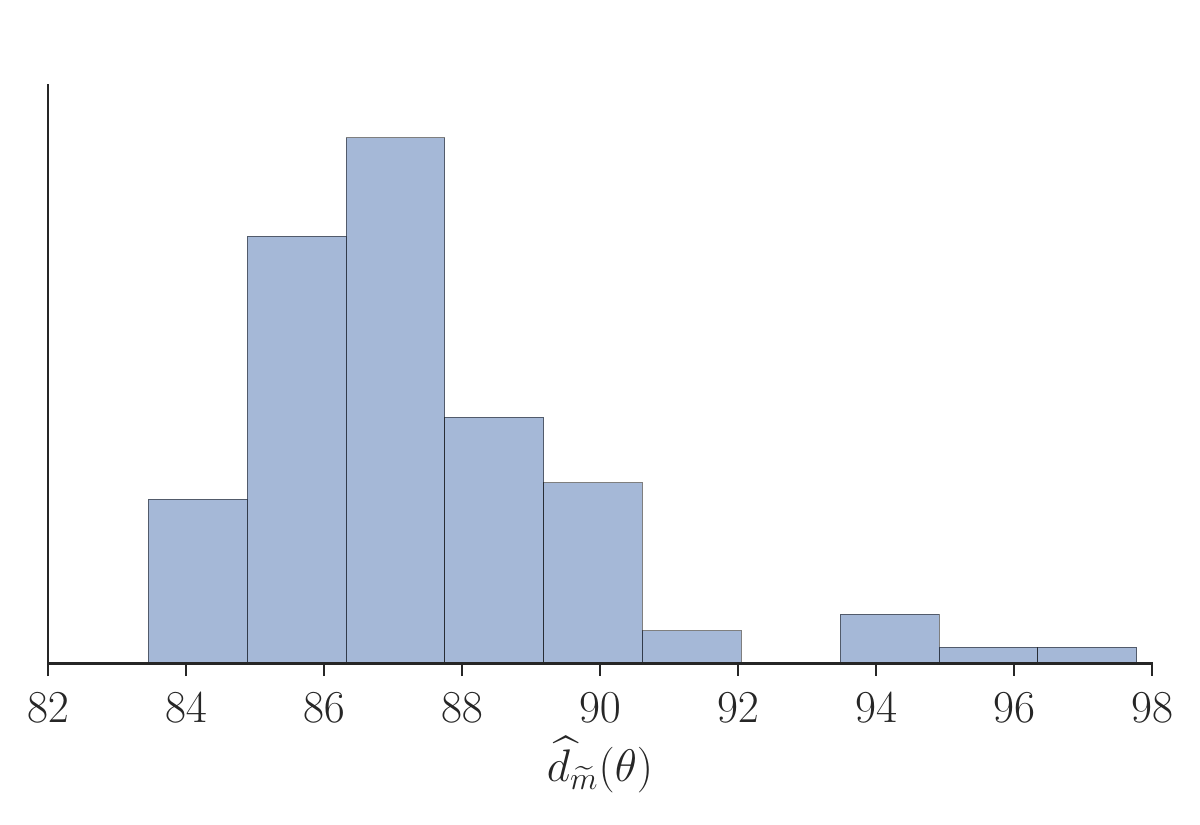}
        \caption{Estimated $d(\theta)$'s.}%\label{fig:CT_lambda100_fig}
\end{subfigure}

\caption{Estimating $\gamma(\theta)$ (left panel) and $d(\theta)$ (right panel) for different $\theta$ on the Covtype dataset. Estimates  of $\gamma(\theta) = n^2\sigma^2_{d_{u_i}}$ and $d(\theta)$ based on $\widetilde{m} = 0.10n$ with parameter expanded control variates (top row) and data expanded control variates (bottom row). The histograms are obtained by (for a fixed $u$) estimating the quantities unbiasedly for 100 values of $\theta$ sampled from a multivariate Student t approximation of the posterior with 5 degrees of freedom. This generates over-dispersed $\theta$ values and hence a $\gamma_{\mathrm{max}}$ which is conservative. Note that this figure is not useful for evaluating the normality assumption of $\widehat{d}_{\widetilde{m}}$, because $u$ is fixed.}\label{fig:gammahat_and_dhat}
\end{figure}

\begin{figure}
\centering

\begin{subfigure}[t]{.45\textwidth}
\centering
\includegraphics[width=\linewidth]{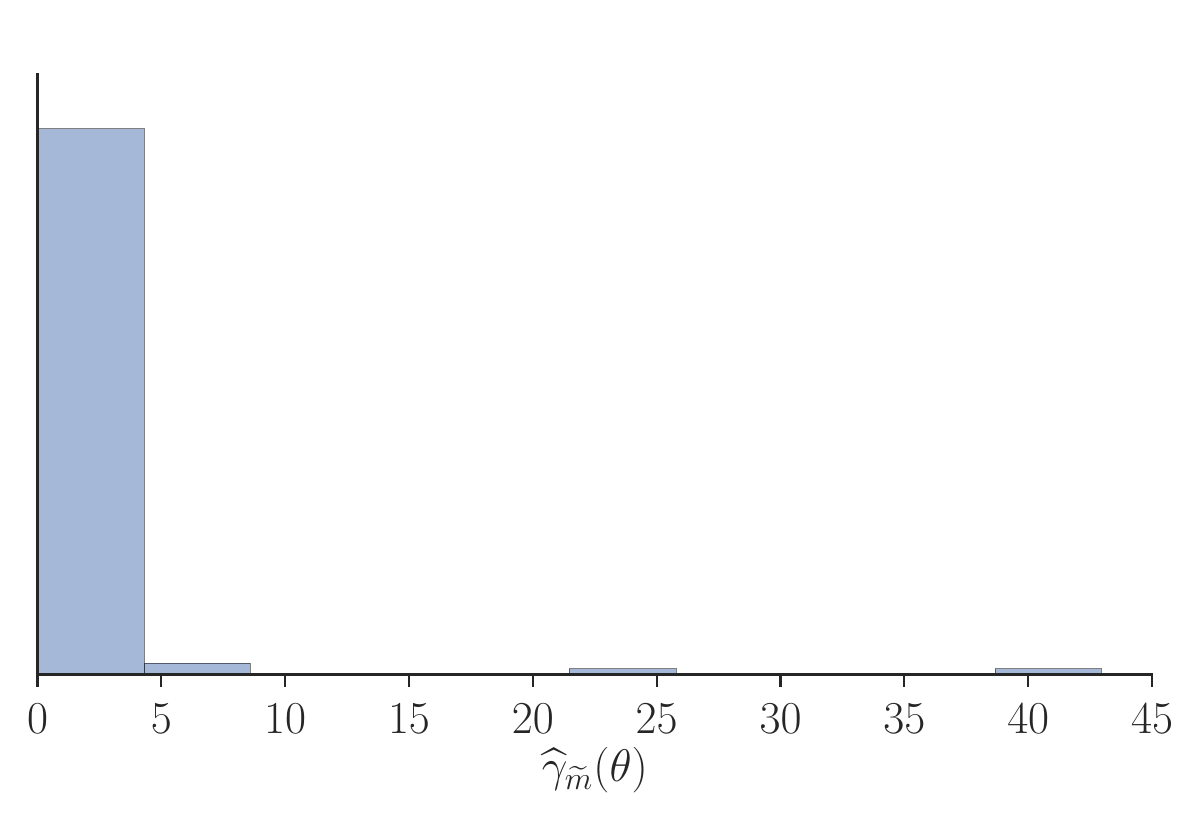}
\end{subfigure}
\begin{subfigure}[t]{.45\textwidth}
\centering
\includegraphics[width=\linewidth]{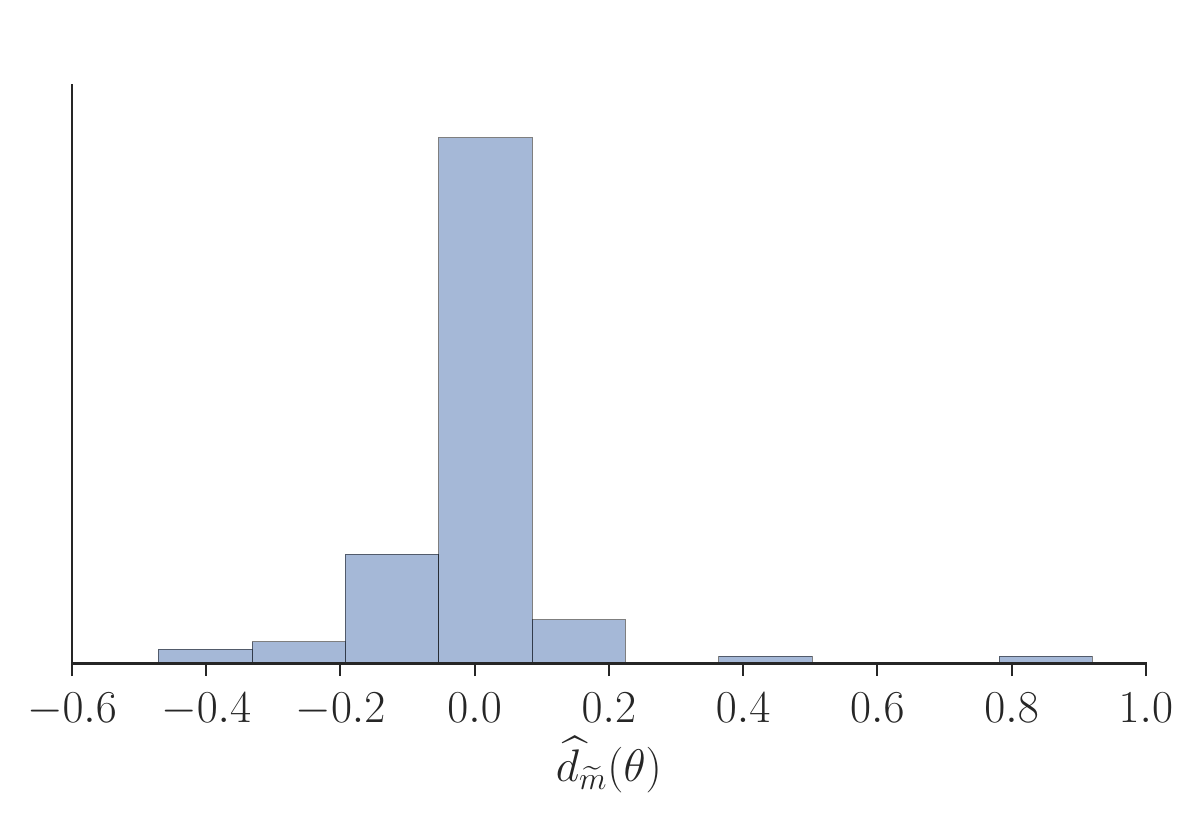}
\end{subfigure} 

\begin{subfigure}[t]{.45\textwidth}
\centering
\includegraphics[width=\linewidth]{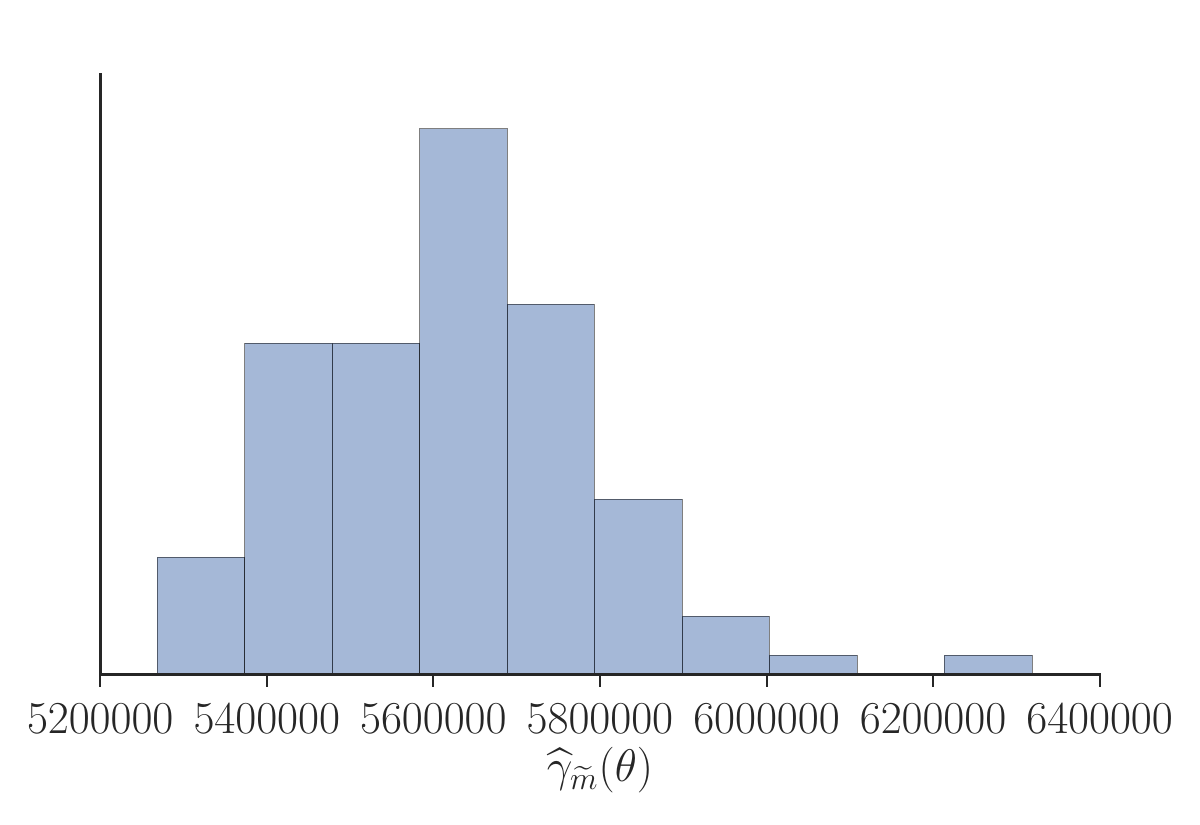} 
        \caption{Estimated $\gamma(\theta)$'s.}%\label{fig:CT_lambda100_fig}
\end{subfigure}
\begin{subfigure}[t]{.45\textwidth}
\centering
\includegraphics[width=\linewidth]{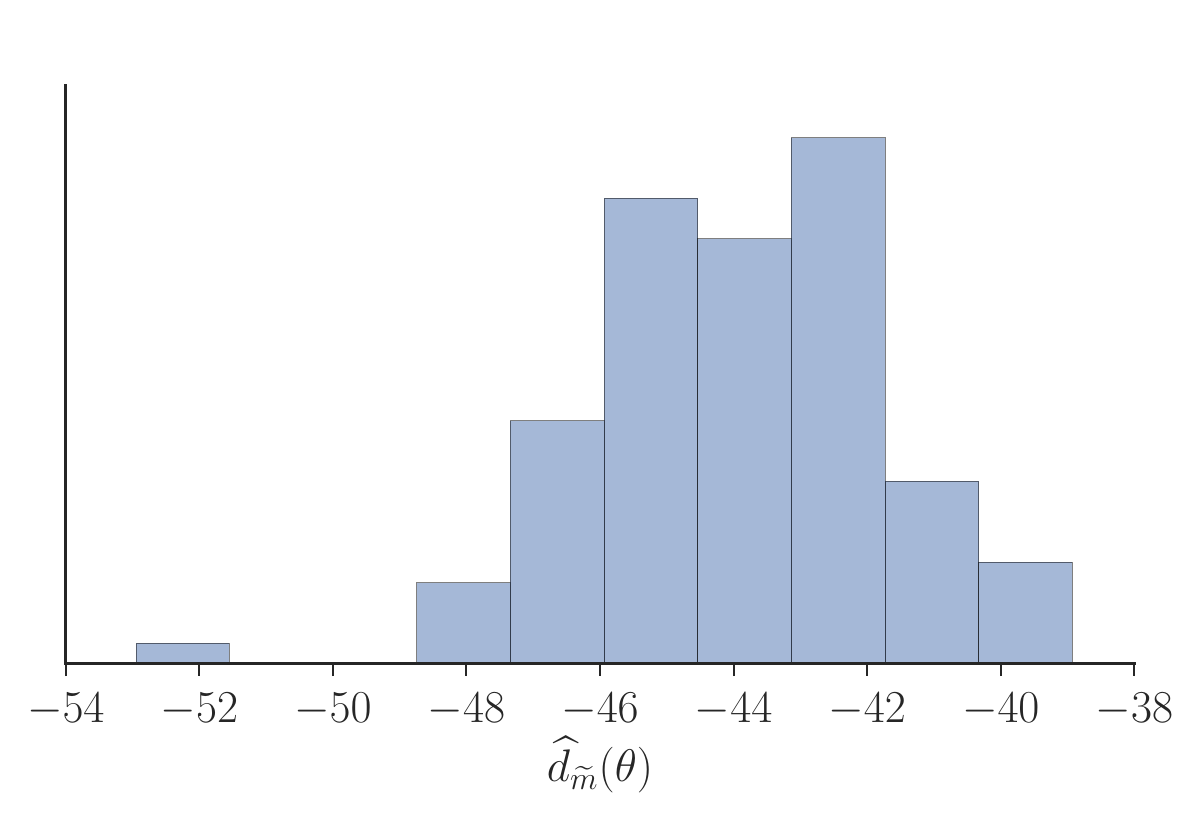}
        \caption{Estimated $d(\theta)$'s.}%\label{fig:CT_lambda100_fig}
\end{subfigure}

\caption{Estimating $\gamma(\theta)$ (left panel) and $d(\theta)$ (right panel) for different $\theta$ on the bankruptcy dataset. Estimates  of $\gamma(\theta) = n^2\sigma^2_{d_{u_i}}$ and $d(\theta)$ based on $\widetilde{m} = 0.10n$ using parameter expanded control variates (top row) and data expanded control variates (bottom row). The histograms are obtained by, for a fixed $u$, estimating the quantities unbiasedly for 100 values of $\theta$ sampled from a multivariate Student t approximation of the posterior with 5 degrees of freedom. This generates over-dispersed $\theta$ values and hence we obtain a conservative $\gamma_{\mathrm{max}}$. This figure is not useful for evaluating the normality assumption of $\widehat{d}_{\widetilde{m}}$, because $u$ is fixed.}\label{fig:gammahat_and_dhat_bankruptcy}
\end{figure}

\begin{figure}
\centering

\begin{subfigure}[t]{.45\textwidth}
\centering
\includegraphics[width=\linewidth]{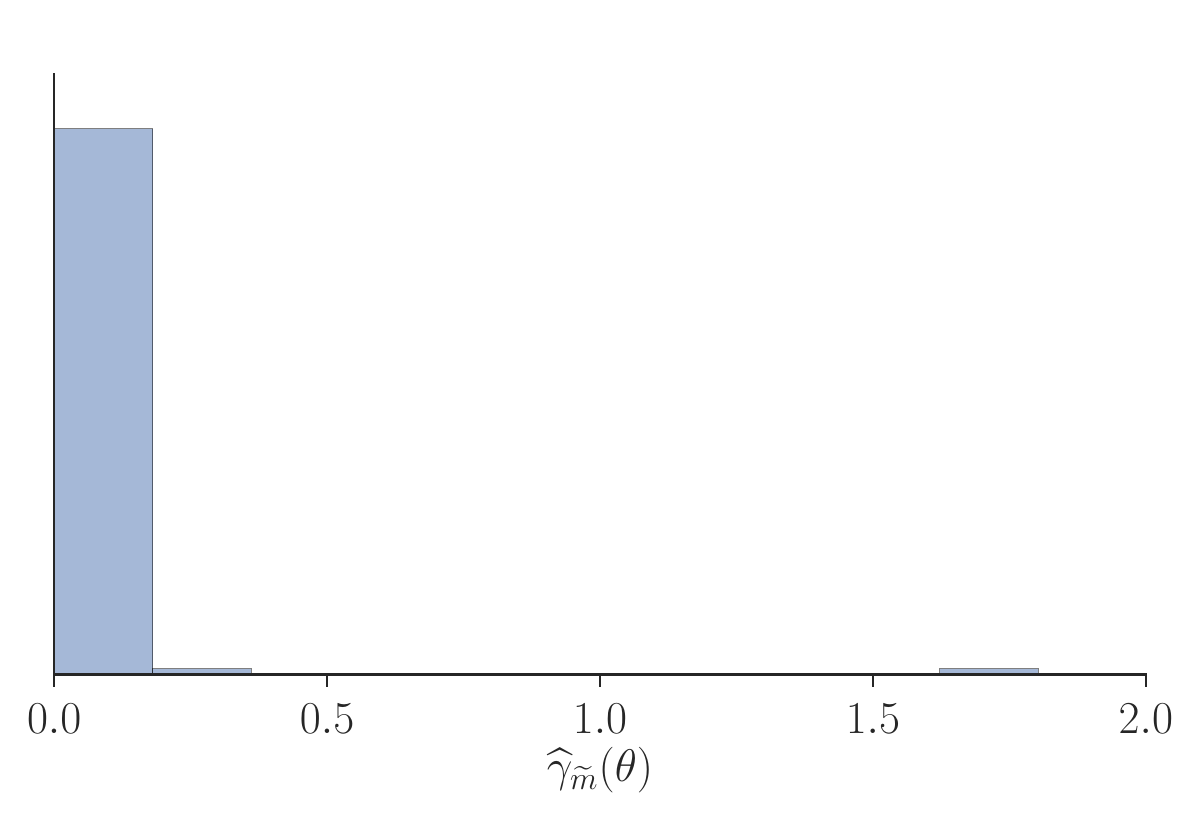}
\end{subfigure}
\begin{subfigure}[t]{.45\textwidth}
\centering
\includegraphics[width=\linewidth]{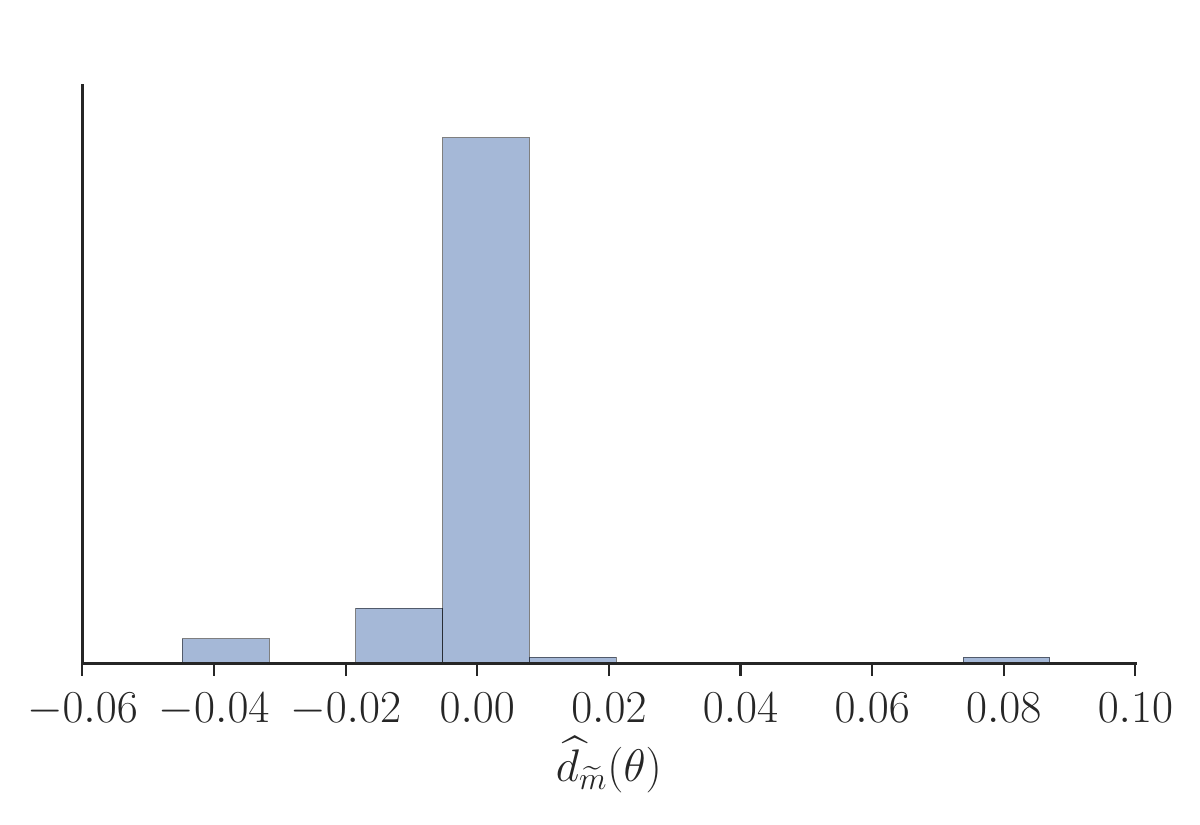}
\end{subfigure} 

\begin{subfigure}[t]{.45\textwidth}
\centering
\includegraphics[width=\linewidth]{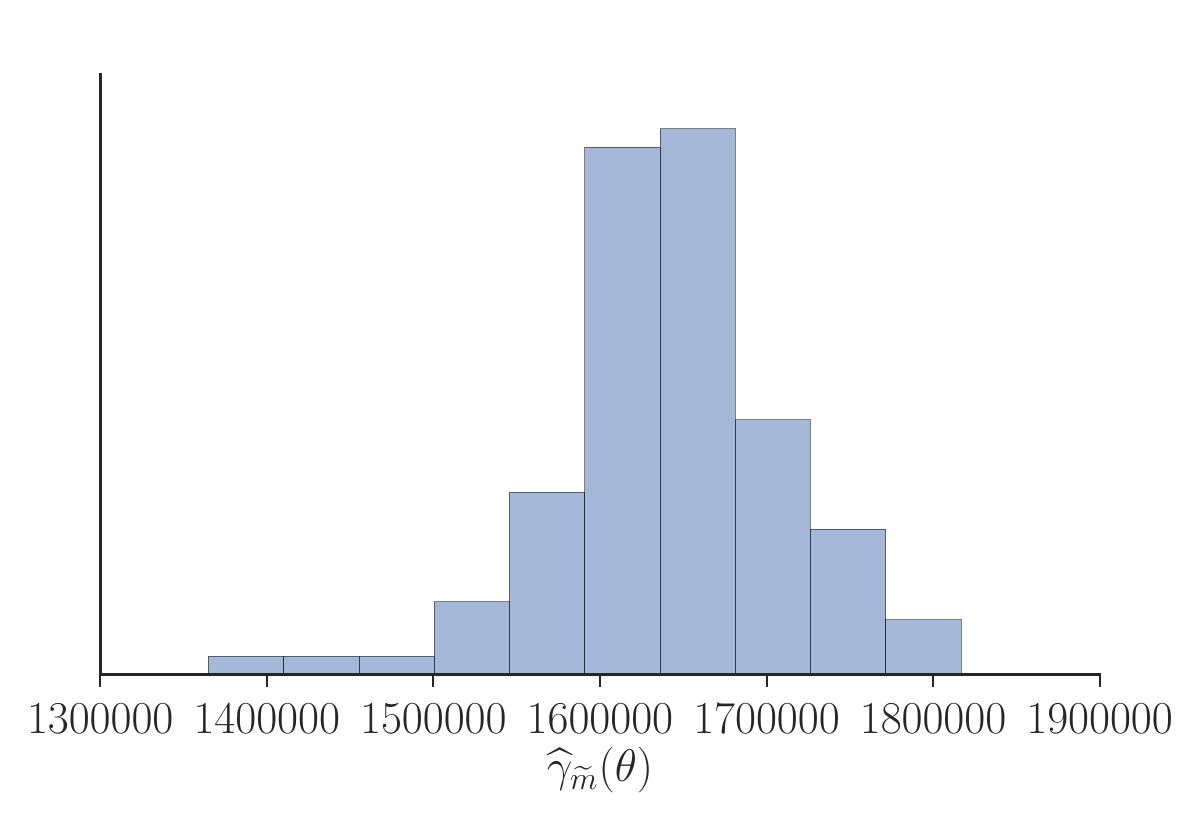} 
        \caption{Estimated $\gamma(\theta)$'s.}%\label{fig:CT_lambda100_fig}
\end{subfigure}
\begin{subfigure}[t]{.45\textwidth}
\centering
\includegraphics[width=\linewidth]{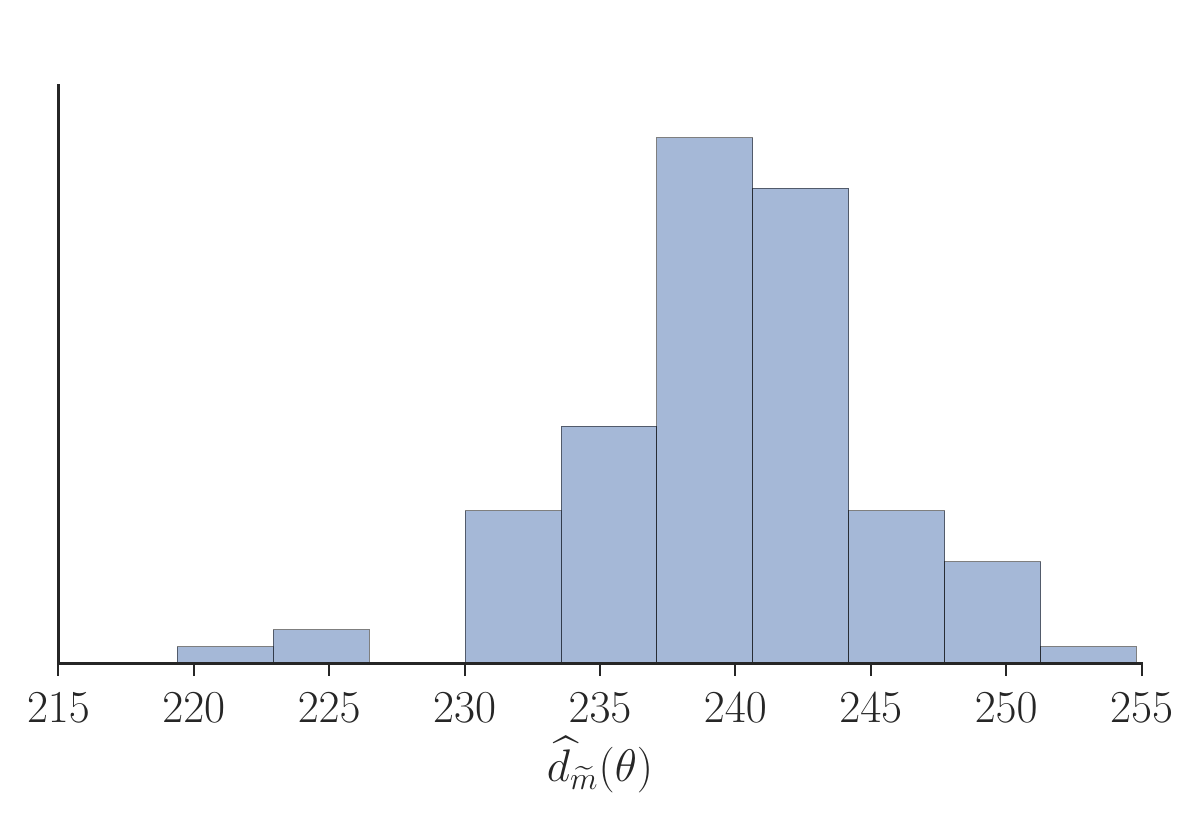} 
        \caption{Estimated $d(\theta)$'s.}
\end{subfigure}

\caption{Estimating $\gamma(\theta)$ (left panel) and $d(\theta)$ (right panel) for different $\theta$ on the HIGGS dataset. Estimates  of $\gamma(\theta) = n^2\sigma^2_{d_{u_i}}$ and $d(\theta)$ based on $\widetilde{m} = 0.10n$ for parameter expanded control variates (top row) and data expanded control variates (bottom row). The histograms are obtained by (for a fixed $u$) estimating the quantities unbiasedly for 100 values of $\theta$ sampled from a multivariate Student t approximation of the posterior with 5 degrees of freedom. This generates over-dispersed $\theta$ values and hence we obtain a $\gamma_{\mathrm{max}}$ which is conservative. Note that this figure is not useful for evaluating the normality assumption of $\widehat{d}_{\widetilde{m}}$, because $u$ is fixed.}\label{fig:gammahat_and_dhat_HIGGS}
\end{figure}

\begin{figure}
\centering

\begin{subfigure}[t]{.50\textwidth}
\centering
\includegraphics[width=\linewidth]{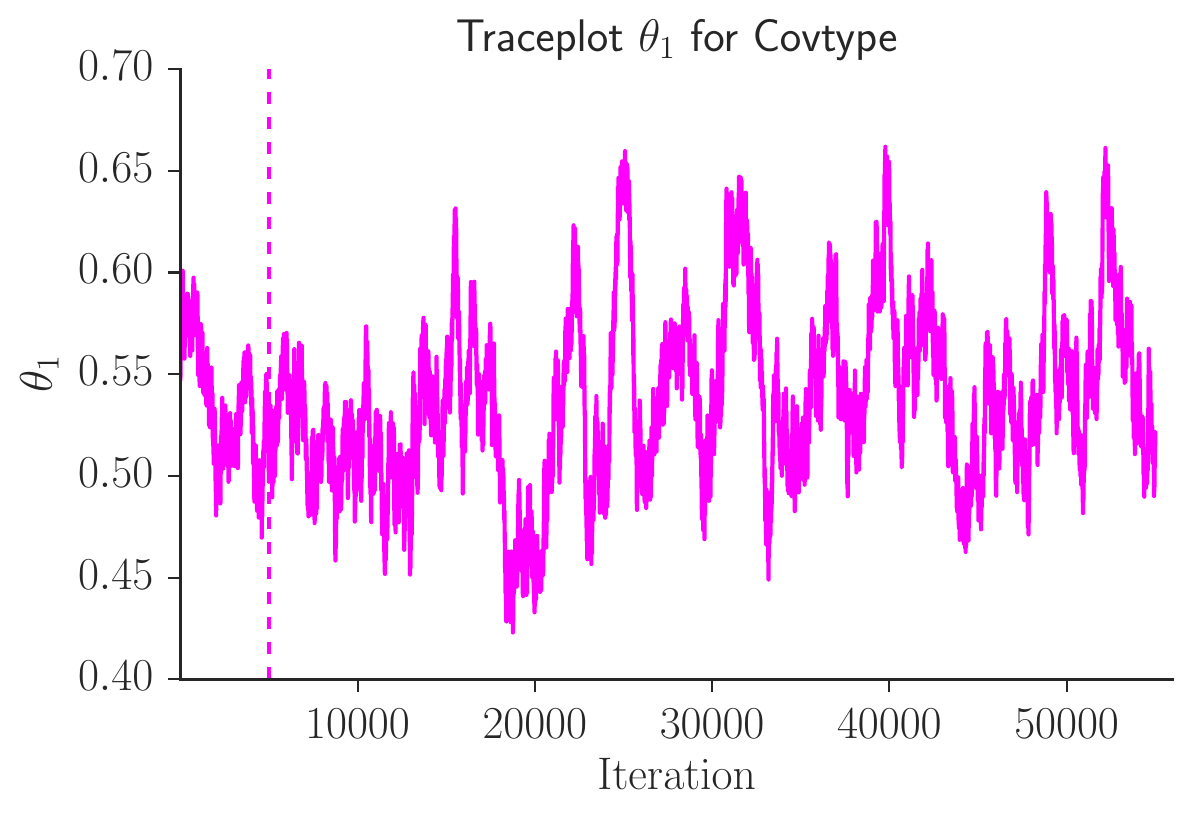}
        %\caption{$\Pr(\widehat{L}>0)$ as a function of $a$.}\label{fig:CT_fig_a}
\end{subfigure}

\begin{subfigure}[t]{.50\textwidth}
\centering
\includegraphics[width=\linewidth]{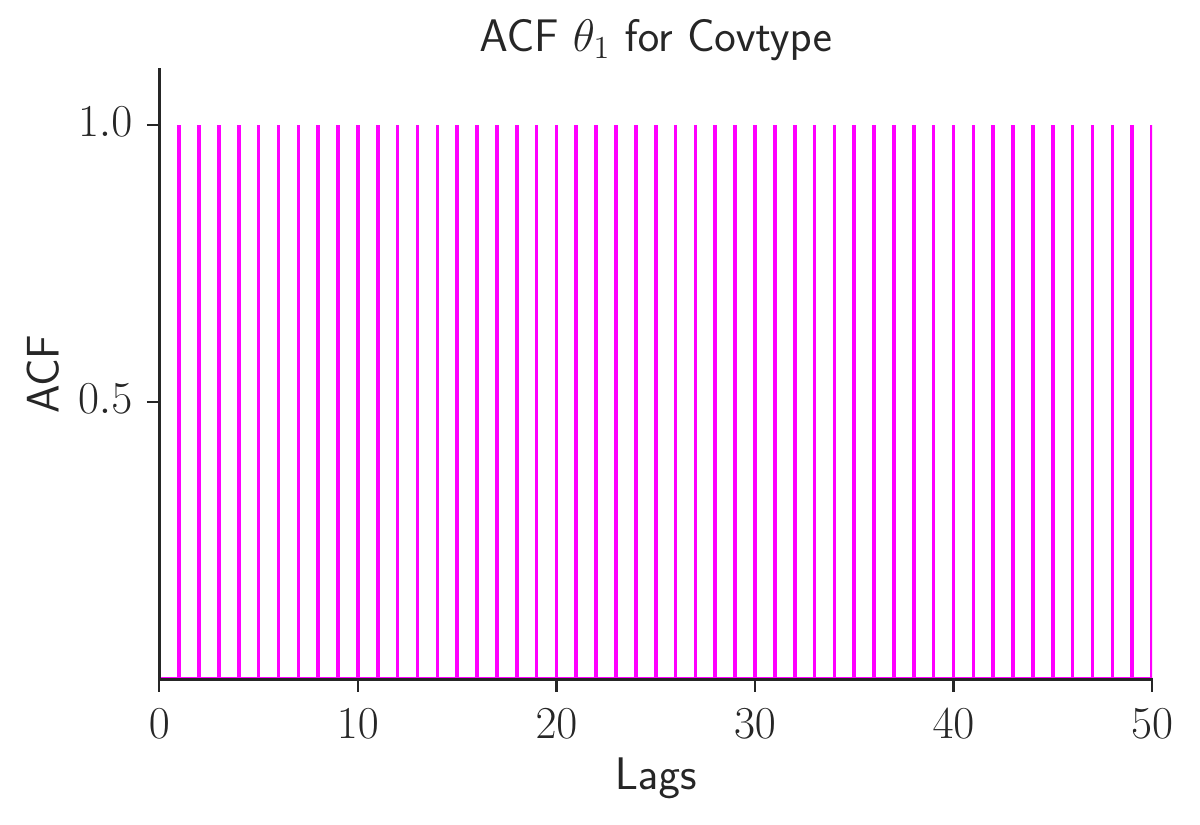}
        %\caption{$\Pr(\widehat{L}>0)$ as a function of $a$.}\label{fig:CT_fig_a}
\end{subfigure}

\begin{subfigure}[t]{.50\textwidth}
\centering
\includegraphics[width=\linewidth]{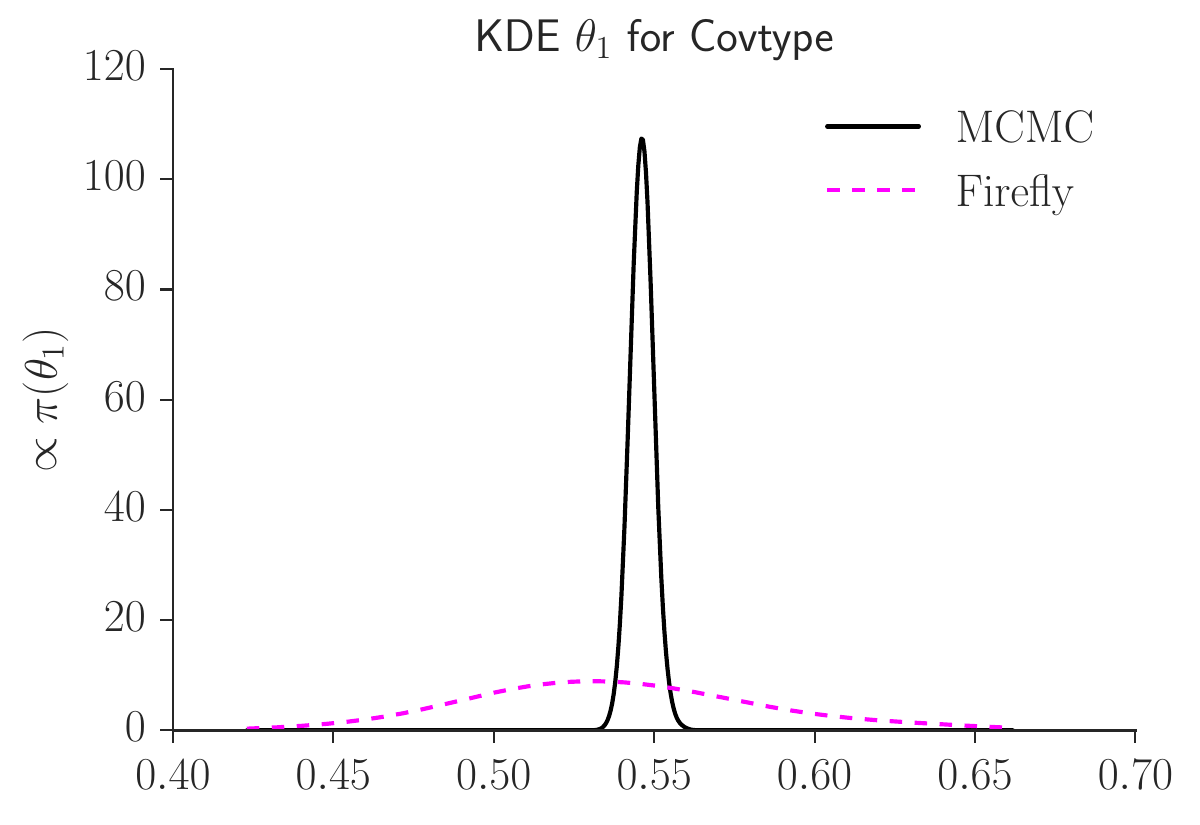}
\end{subfigure}

\caption{Results for Firefly Monte Carlo. Trace plots (upper panel), estimate of autocorrelation function (middle panel) and kernel density estimate (lower panel) of the posterior samples using FireFly Monte Carlo for the Covtype dataset. The vertical line in the upper panel marks the burn-in at iterate $5{,}000$.}\label{fig:FireFly_Draws_KDE_ACF}
\end{figure}

\begin{figure}
\centering

\begin{subfigure}[t]{.50\textwidth}
\centering
\includegraphics[width=\linewidth]{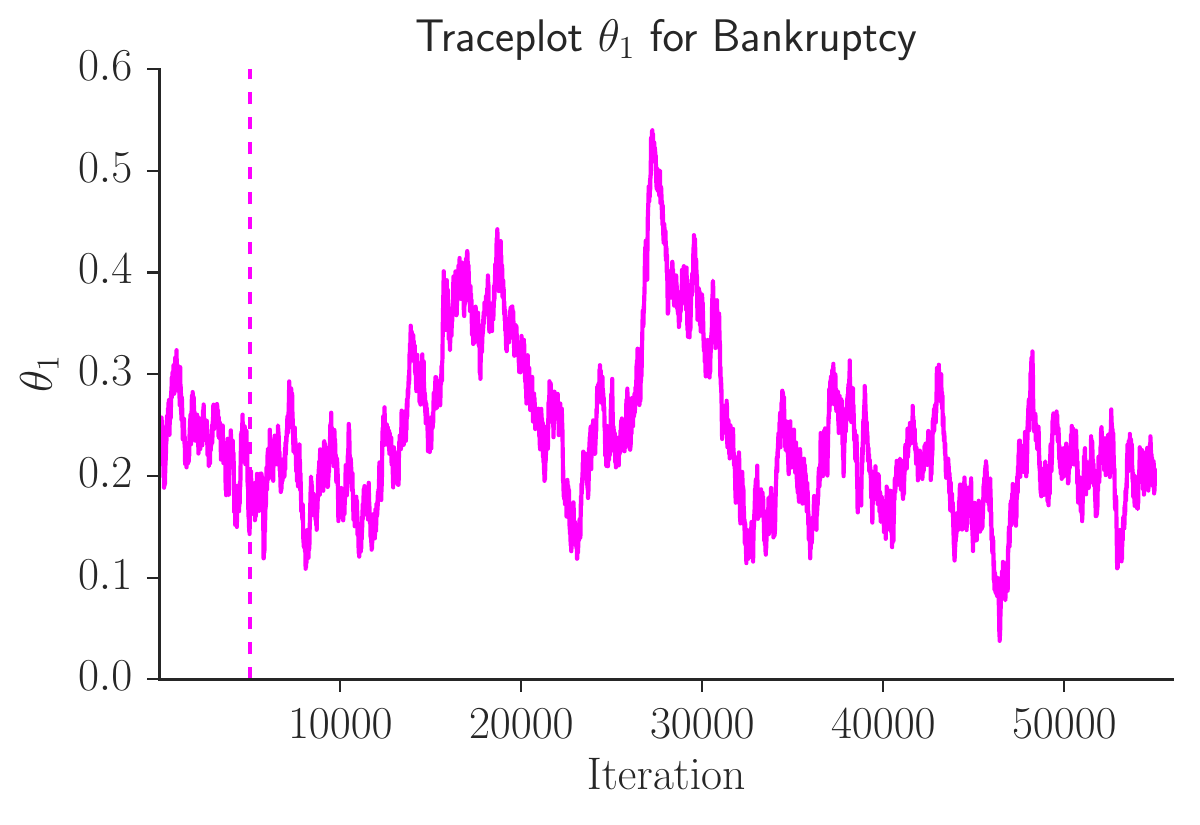}
        %\caption{$\Pr(\widehat{L}>0)$ as a function of $a$.}\label{fig:CT_fig_a}
\end{subfigure}

\begin{subfigure}[t]{.50\textwidth}
\centering
\includegraphics[width=\linewidth]{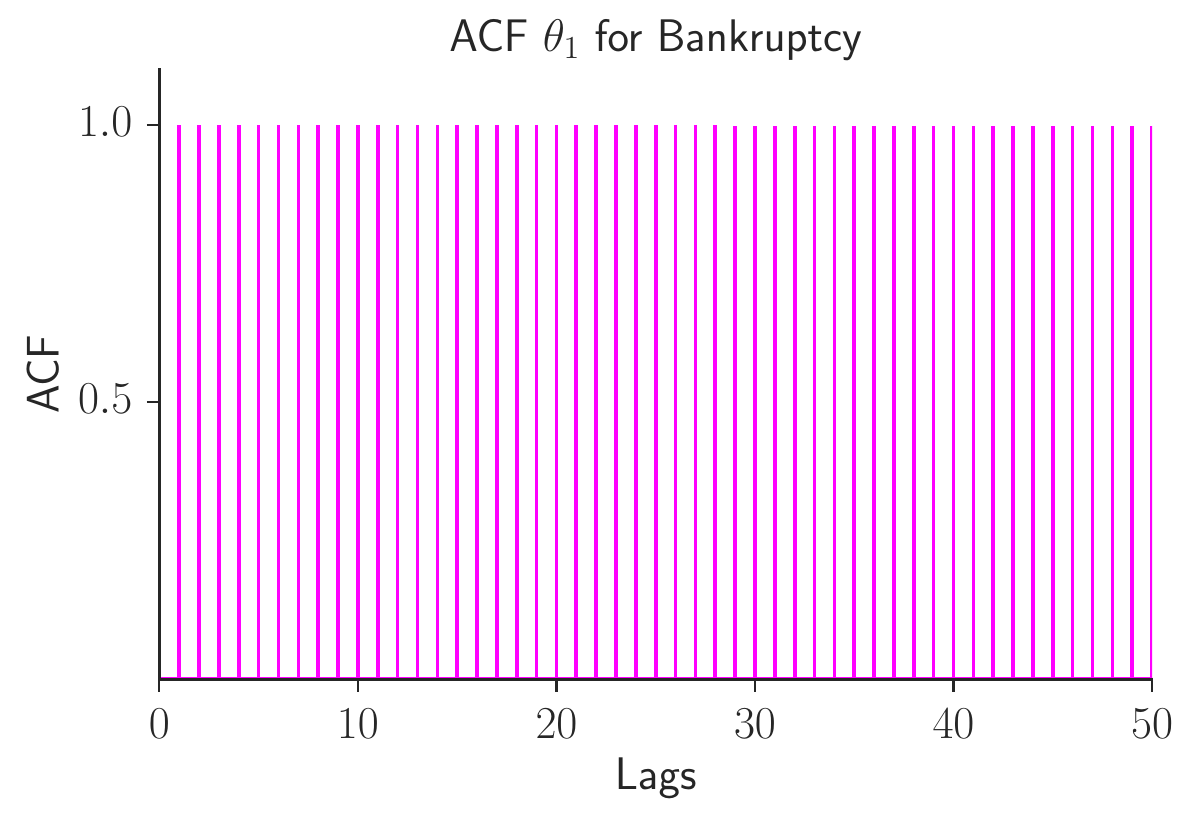}
        %\caption{$\Pr(\widehat{L}>0)$ as a function of $a$.}\label{fig:CT_fig_a}
\end{subfigure}

\begin{subfigure}[t]{.50\textwidth}
\centering
\includegraphics[width=\linewidth]{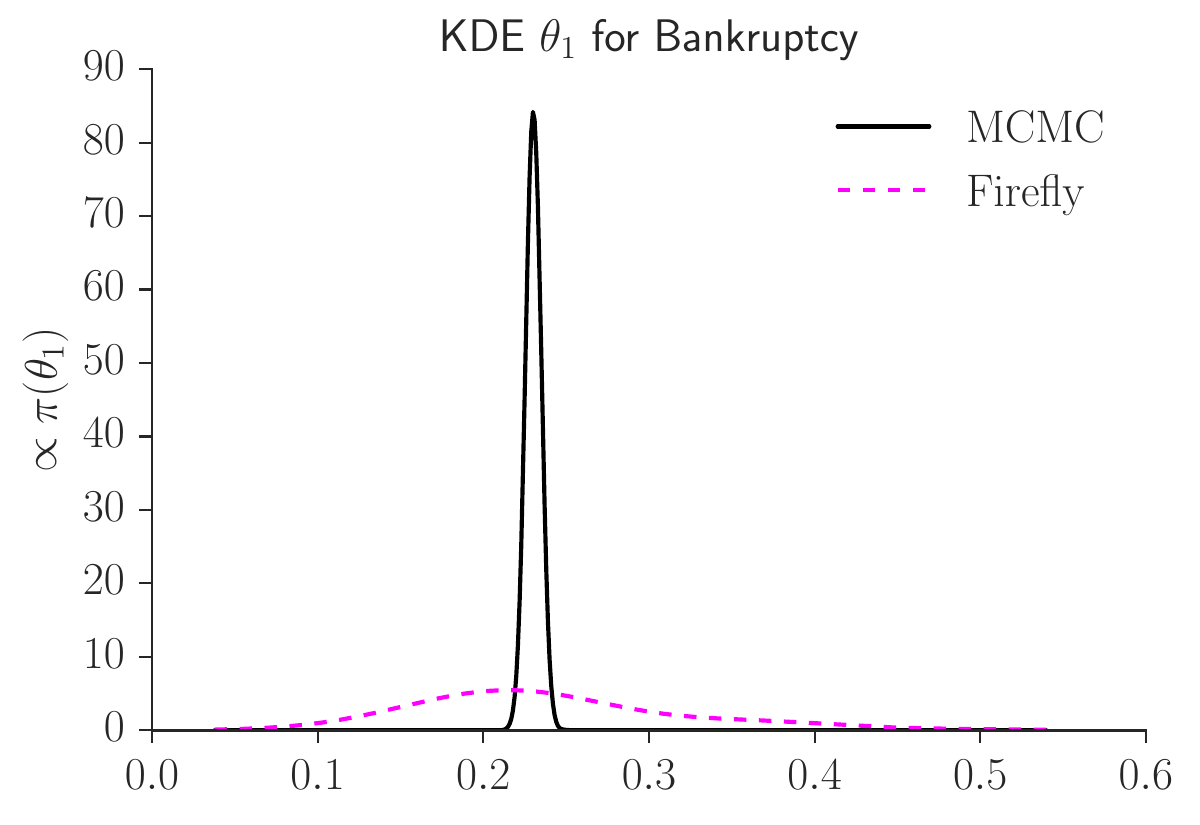}
\end{subfigure}

\caption{Results for Firefly Monte Carlo. Trace plots (upper panel), estimate of autocorrelation function (middle panel) and kernel density estimate (lower panel) of the posterior samples using FireFly Monte Carlo for the bankruptcy dataset. The vertical line in the upper panels marks the end of the burn-in period.}\label{fig:FireFly_Draws_KDE_ACF_bankruptcy}
\end{figure}

\begin{figure}
\centering

\begin{subfigure}[t]{.50\textwidth}
\centering
\includegraphics[width=\linewidth]{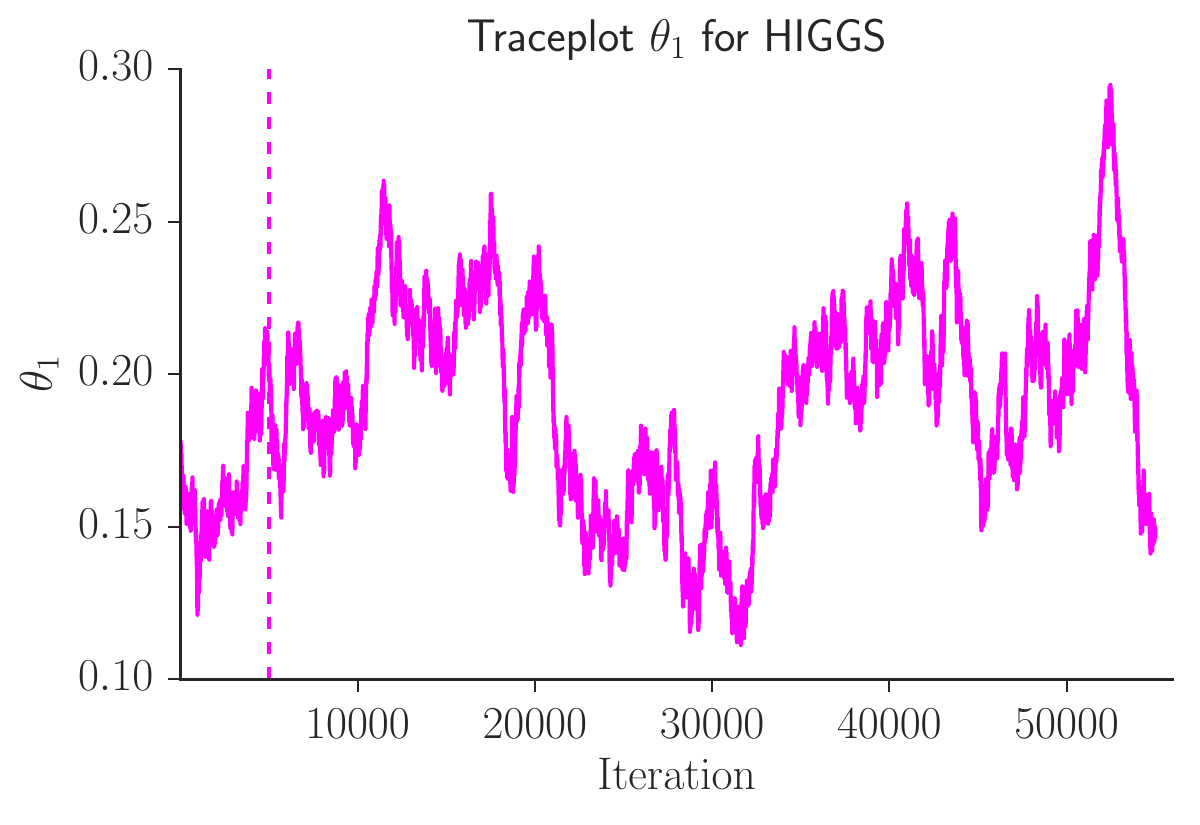}
        %\caption{$\Pr(\widehat{L}>0)$ as a function of $a$.}\label{fig:CT_fig_a}
\end{subfigure}

\begin{subfigure}[t]{.50\textwidth}
\centering
\includegraphics[width=\linewidth]{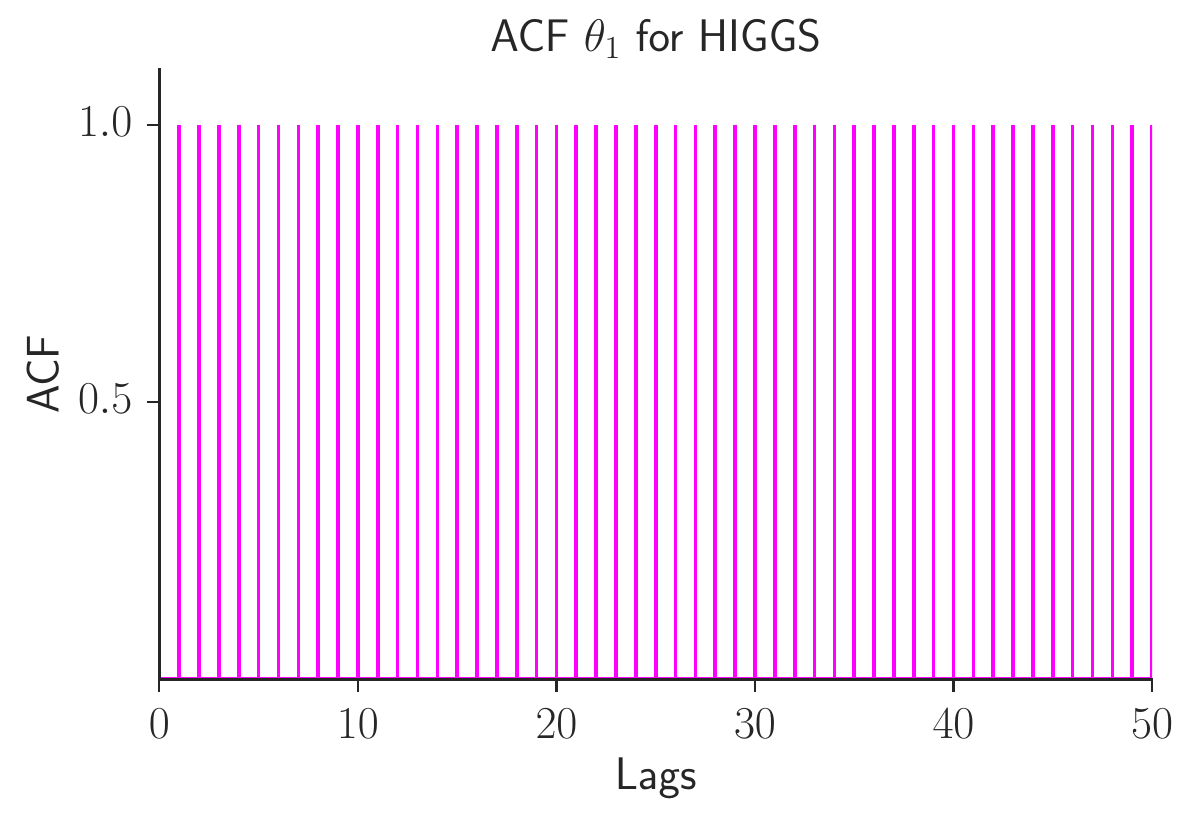}
        %\caption{$\Pr(\widehat{L}>0)$ as a function of $a$.}\label{fig:CT_fig_a}
\end{subfigure}

\begin{subfigure}[t]{.50\textwidth}
\centering
\includegraphics[width=\linewidth]{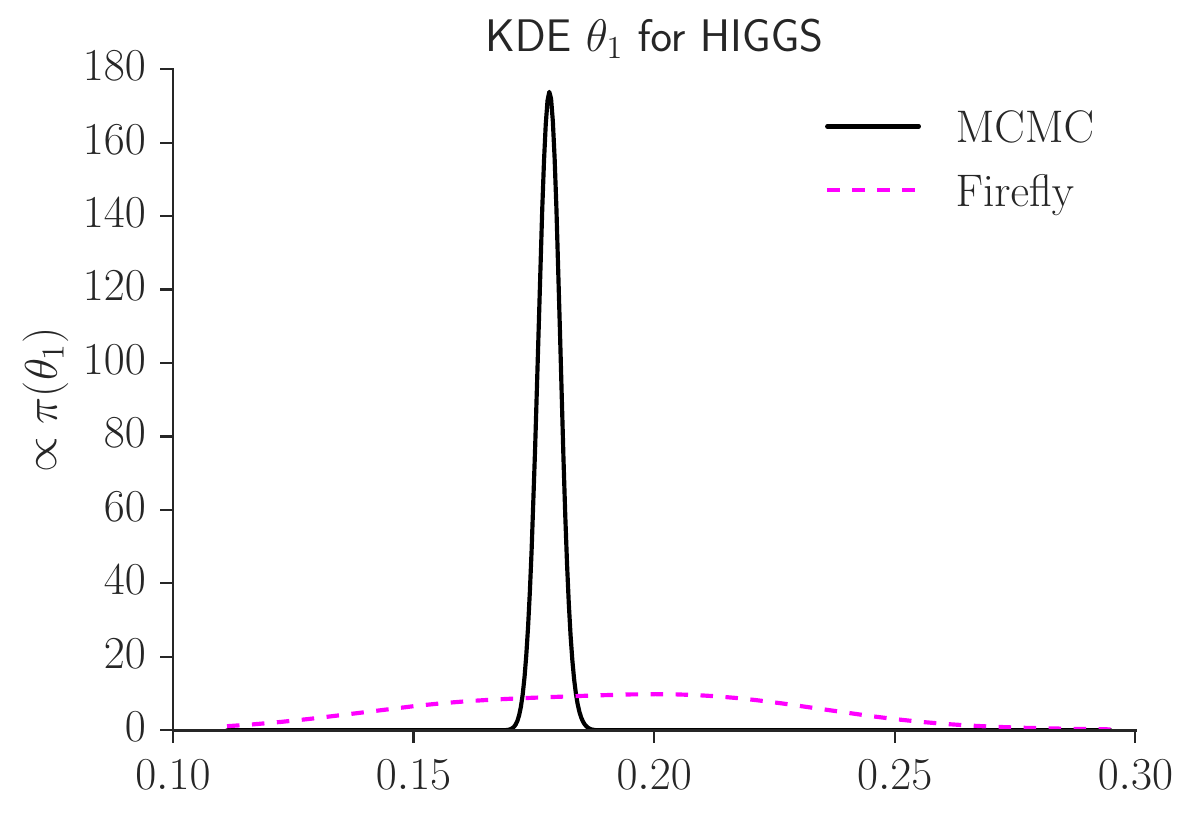}
\end{subfigure}

\caption{Results for Firefly Monte Carlo. Trace plots (upper panel), estimate of autocorrelation function (middle panel) and kernel density estimate (lower panel) of the posterior samples using FireFly Monte Carlo for the HIGGS dataset. The vertical line in the upper panels marks the burn-in $5{,}000$.}\label{fig:FireFly_Draws_KDE_ACF_HIGGS}
\end{figure}

\section{Additional material for the experiments in Section ~\ref{sec:Experiments} \label{app:FireflySupplement}}

Section \ref{subsec:tuningProductPoisson} outlines a tuning strategy that requires determining the intrinsic variability of the control variates given by $\gamma(\theta)= n^2\sigma^2_{d_{u_i}}$. Moreover, the tuning strategy requires $\overline{d}$ to set the lower bound (see Step 4.), which is obtained by estimating $d(\theta)$. Figures \ref{fig:gammahat_and_dhat} -- \ref{fig:gammahat_and_dhat_HIGGS}
show how these quantities vary over the posterior draws for the three datasets used in Section \ref{sec:Experiments}.

Figures \ref{fig:FireFly_Draws_KDE_ACF} --  \ref{fig:FireFly_Draws_KDE_ACF_HIGGS} illustrate the slow convergence of Firefly Monte Carlo for the $\theta_1$ draws (the performance is similar for all parameters). A similar performance of Firefly Monte Carlo is reported by \cite{bardenet2015markov,quiroz2016speeding,li2020improving}.

%\pagebreak

\section{Proofs\label{app:ProofsCollected}}
\subsection{Proof of Lemma \ref{lem:PoissonEstimator} \label{sec:Appendix_proof_Poisson_estimator}}
The estimator is given by \eqref{eq:UnbiasedLikelihoodEstimator}; since the $\widehat{d}_{m}^{\,\,(h, l)}$ are independent for all $h$ and $l$, it follows that $\xi_1, \dots, \xi_\lambda$ are independent. The following lemma,  whose proof is straightforward, is useful for the proof.
\begin{lemma}
\label{lem:Poisson_exponent_moment}
Suppose that $X\sim \mathrm{Pois}(1)$ and that $A  < \infty$. Then
\begin{enumerate}[label={\emph{(\roman*)}}]
\item $\mathrm{E}_X[A^X] = \exp(A-1).$
\item $\mathrm{V}_X[A^X] = \exp(-1)\left(\exp(A^2)-\exp(2A - 1)\right)$.
\end{enumerate}
\end{lemma}

\begin{proof}[Proof of Lemma \ref{lem:PoissonEstimator}.]
Proof of Part~(i). By the law of iterated expectations,
\begin{eqnarray*}
\mathrm{E}[\xi] = \mathrm{E}_{\mathcal{X}}[\mathrm{E}_{\widehat{d}|\mathcal{X}}[\xi | \mathcal{X}]] & = & \exp(a/\lambda + 1)\mathrm{E}_{\mathcal{X}}\left[\left(\frac{d-a}{\lambda}\right)^{\mathcal{X}} \right] \\
~ & = & \exp(d/\lambda),
\end{eqnarray*}
where the last equality follows from part (i) of Lemma \ref{lem:Poisson_exponent_moment}. Since the $\xi_l$ are independent for $l=1,\dots, \lambda$,
\begin{equation*}
\mathrm{E}[\widehat{L}_B] = \exp(q)\prod_{l = 1}^{\lambda} \mathrm{E}[\xi_l] = \exp(q+d)= L.
\end{equation*}

Proof of Part (ii)
By \eqref{eq:UnbiasedLikelihoodEstimator},$\{ \widehat{d}^{(h,l)}_m \geq a, \, \forall h, l \} \subset \{ \widehat{L}_B \geq 0 \}$. Hence
\begin{equation*}
\Pr(\widehat{L}_B \geq 0) \geq \Pr(\widehat{d}^{\,\,(h,l)}_m \geq a, \, \text{for all } h, l ) = 1.
\end{equation*}
Proof of  Part (iii) Since $\mathrm{E}[|\widehat{L}_B|] \geq \left|\mathrm{E}[\widehat{L}_B]\right|$ we obtain \begin{eqnarray*}
\mathrm{V}\left[|\widehat{L}_B|\right] & = & \mathrm{E}\left[|\widehat{L}_B|^2\right] - \mathrm{E}\left[|\widehat{L}_B|\right]^2 \\
~ & \leq & \mathrm{E}\left[\widehat{L}_B^2\right] - \mathrm{E}\left[\widehat{L}_B\right]^2 \\
~ & = & \mathrm{V}[\widehat{L}_B].
\end{eqnarray*}
We now derive $V[\widehat{L}_B]$ and show that it is finite which proves the result. By the law of total variance
\begin{equation}
\mathrm{V}[\xi] = \mathrm{E}_{\mathcal{X}}[\mathrm{V}_{\widehat{d}|\mathcal{X}}[\xi | \mathcal{X}]] + \mathrm{V}_{\mathcal{X}}[\mathrm{E}_{\widehat{d}|\mathcal{X}}[\xi | \mathcal{X}]]. \label{eq:LawTotalVariance}
\end{equation}
To compute $\mathrm{V}_{\widehat{d}|\mathcal{X}}[\xi | \mathcal{X}]$, note that for a collection of independent random variables $X_1, \dots X_J$,
\begin{equation}
\label{eq:variance_of_ind_product}
\mathrm{V}\left[\prod_{j = 1}^J X_j\right] = \prod_{j = 1}^J\left(\mathrm{V}[X_j] + \mathrm{E}[X_j]^2 \right) - \prod_{j = 1}^J \mathrm{E}[X_j]^2.
\end{equation}
Hence,
\begin{eqnarray*}
\mathrm{V}_{\widehat{d}|\mathcal{X}}[\xi | \mathcal{X}]& = & \exp\left(2\frac{a+\lambda}{\lambda}\right) \left(  \left(\frac{\sigma_{\widehat{d}}^2}{\lambda^2} + \frac{(d - a)^2}{\lambda^2}\right)^\mathcal{X} -  \left(\frac{(d - a)^2}{\lambda^2} \right)^\mathcal{X} \right)
\end{eqnarray*}
and taking the outer expectations and applying Lemma \ref{lem:Poisson_exponent_moment},
$$\mathrm{E}_{\mathcal{X}}[\mathrm{V}_{\widehat{d}|\mathcal{X}}[\xi | \mathcal{X}]]  = \exp\left(\frac{2a}{\lambda} + \frac{(d-a)^2}{\lambda^2} + 1\right)\left(\exp \left(\frac{\sigma^2_{\widehat{d}}}{\lambda^2}\right) - 1 \right).$$
Next,
\begin{eqnarray*}
\mathrm{V}_{\mathcal{X}}[\mathrm{E}_{\widehat{d}|\mathcal{X}}[\xi | \mathcal{X}]] & = &
\exp\left(2\frac{a+\lambda}{\lambda}\right)\mathrm{V}_{\mathcal{X}}\left[\left(\frac{d-a}{\lambda}\right)^{\mathcal{X}} \right] \\
~ & = & \exp\left(\frac{2a}{\lambda} + 1 \right)\left(\exp\left(\frac{\left(d-a\right)^2}{\lambda^2}\right)-\exp\left(2\frac{d-a}{\lambda}  - 1\right)\right),
\end{eqnarray*}
by part (ii) of Lemma \ref{lem:Poisson_exponent_moment} which, by \eqref{eq:LawTotalVariance} and simplification, yields
$$\mathrm{V}[\xi] = \exp\left(\frac{1}{\lambda^2}\left(\sigma_{\widehat{d}}^2 + (d-a)^2\right) + \frac{2a}{\lambda} + 1 \right) - \exp\left(\frac{2d}{\lambda}\right).$$
To compute $\mathrm{V}[\widehat{L}_B] = \exp(2q)\mathrm{V}\left[\prod_{l=1}^{\lambda} \xi_l \right]$, we use \eqref{eq:variance_of_ind_product} with  $\mathrm{E}[\mathrm{\xi}]^2 = \exp\left(2d/\lambda\right)$ to obtain
\begin{eqnarray*}
\mathrm{V}\left[\prod_{l=1}^{\lambda} \xi_l \right] & = & \exp\left(\sum_{l=1}^\lambda \frac{1}{\lambda^2}\left(\sigma_{\widehat{d}}^2 + (d-a)^2\right) + \frac{2a}{\lambda} + 1 \right) - \exp\left(\sum_{l=1}^\lambda \frac{2d}{\lambda}\right)\\
~ & = & \exp\left(\frac{1}{\lambda}\left(\sigma_{\widehat{d}}^2 + (d-a)^2\right) + 2a + \lambda \right) - \exp\left(2d\right).
\end{eqnarray*}
This expression exists as long as $\sigma^2_{\widehat{d}} < \infty$.

Proof of  Part~(iv).
To minimize $\mathrm{V}[\widehat{L}_B]$ above for a fixed $\lambda > 0$, it is sufficient to minimize the exponent $$f(a) = \frac{1}{\lambda}\left(\sigma_{\widehat{d}}^2 + (d - a)^2 + 2a + \lambda\right).$$
Since $f^{\prime}(a) = -2(d-a)/\lambda + 2$ and $f^{\prime\prime}(a)=2/\lambda > 0$, $a = d-\lambda$ is the minimum.

Proof of  Part~(v). This
follows from the expression of the variance $\mathrm{V}[\widehat{L}_B]$ derived in the proof of part (iii) of Lemma \ref{lem:PoissonEstimator} and that of \eqref{eq:UnbiasedLikelihoodEstimator} derived in \cite{papaspiliopoulos2009methodological}.
\end{proof}

\begin{proof} [Proof of Lemma \ref{lem:VarAbs}]
The proof follows from the more general Lemma \ref{lem:VarAbsL_finite_mixture}.
\end{proof}

\begin{proof}[Proof of Lemma \ref{lem:PrNonnegative}]
Note that
$$\Pr\left(\widehat{L}_B \geq 0\right) = \Pr\left(\prod_{l = 1}^\lambda \xi_l \geq 0\right)$$
and $\{\widehat{L}_B < 0\}$ can only occur if there is an odd number of negative terms $\xi_l$, $l=1,\dots.\lambda$. \cite[p. 277]{feller2008introduction} gives an expression for the probability of an odd number of negatives terms among a total of $\lambda$ terms that depends on the probability of a single $\xi_l$ being negative, i.e. $\Pr(\xi_l < 0)$. From this we obtain
$$\Pr(\widehat{L}_B \geq 0) = \frac{1}{2}\left(1 + \left(1 - 2 \Pr(\xi_l < 0) \right)^\lambda\right).$$
Notice that
\begin{equation*}
\Pr(\xi_l < 0 ) = \sum_{j = 1}^\infty \Pr\left(\prod_{l = 1}^j A_m^{(l)} < 0\right)\Pr(\mathcal{X}_l = j), \quad \mathcal{X}_l \sim \mathrm{Pois}(1), %\label{eq:Pr_xi_l_neg}
\end{equation*}
and we can again apply the result in \cite{feller2008introduction} to obtain
$$\Pr\left(\prod_{l = 1}^j A_m^{(l)} < 0\right) = \frac{1}{2}\left(1 - \left(1 - 2 \Pr(\xi_l < 0) \right)^j\right),$$
concluding the proof.
\end{proof}

\begin{proof}[Proof of Lemma~\ref{lemma: reversible u}]
We need to show that  $
p_U(du) q_U(du'|u)  = p_U(du') q_U(du|u')$.
Now, \begin{align*}
p_U(du) q_U(du'|u) & = \prod_{k=1}^G p_{U_k}(du_k) \frac1G\sum_{i=1}^G p_{U_i}(u_i)\prod_{j\neq i} \delta_{u_j}(du_j^\prime)\\
& = \prod_{k=1}^G p_{U_k}(du^\prime_k) \frac1G\sum_{i=1}^G p_{U_i}(u_i^\prime)\prod_{j\neq i} \delta_{u^\prime_j}(du_j)
\end{align*}
because $g(du')\delta_{u'}(du) = g(du)\delta_{u}(du')$ for any measure $g(\cdot)$.
\end{proof}

\begin{proof}[Proof of Theorem~\ref{lemma: signed pmmh sampling scheme}]
Part~(i). Reversibility follows because we are dealing with a MH sampler.

Part~(ii). The proof is essentially that of Theorem~1 of \cite{andrieu2009pseudo}, but under slightly different conditions. Consider first the case $G = 1$.
Let $\mathcal{B}(\Theta)$ be the Borel sets of $\Theta$ and $\mathcal{B}( \mathcal{U})$ the Borel sets of $\mathcal{U}$. We will first show that
if Algorithm~\ref{alg: theta based sampling scheme} can reach the set $A \in \mathcal{B}(\Theta)$ from $\theta \in \Theta$ in one step, i.e. $P_\Theta(\theta; A)> 0 $,
then $\ov P_{\Theta,U} (\theta, u; A \times B) > 0 $ for any $u \in \mathcal{ U}$ and $B \in \mathcal{B}(\mathcal{U})$ with $P_U(B) > 0 $, where $\overline{P}_{\Theta,U}$ is the transition kernel of Algorithm \ref{alg: signed block pmmh} and is given by
$$\overline{P}_{\Theta,U}(\theta, u; d\theta', du') = \overline{K}_{\Theta, U}(\theta, u; d\theta', du') + \delta_{\theta, u}(d\theta', du') \left(1-\int  \overline{K}_{\Theta, U}(\theta, u; d\theta', du')\right),$$
with $\ov K_{\Theta, U}(\theta, u; d\theta', du') = \alpha_{\Theta, U} (\theta,u, \theta',u')q_{\Theta}(\theta; d\theta')q_U(u; du')$.

Let $\alpha_Z(z,z') = 1 \wedge \exp(z'-z)$, where $z$ and $z'$ are defined in Section \ref{app:IF}.
We first note that $r_{\Theta,U}(\theta,u;\theta',u')=e^{z'-z}r_\Theta(\theta;\theta')$,
and that $1 \wedge (xy)\geq (1 \wedge x)(1 \wedge y)$.
Hence,  $\alpha_{\Theta, U} (\theta,u, \theta',u') \geq \alpha_\Theta(\theta, \theta') \alpha_Z(z,z')$, so that
$\overline K_{\Theta,U}(\theta,u; d\theta',du') \geq K_\Theta(\theta; d\theta') \alpha_Z(z,z') q_U(u; du') $ and that
$$\overline P_{\Theta,U} (\theta,u; d\theta',du') \geq \overline K_{\Theta,U}(\theta,u; d\theta',du').$$ Thus, if $\ov P_{\Theta,U} (\theta, u; A \times B) = 0 $,
then $K_\Theta(\theta; d\theta') \alpha_Z(z,z') q_U(u; du') = 0 $ almost everywhere $\theta' \in A$ as $\alpha_Z(z,z') q_U(u; du') > 0 $ by
Part~(i) of Assumption~\ref{ass: ergodic assumpy signed pmmh}. But this contradicts
Part~(ii) of Assumption~\ref{ass: ergodic assumpy signed pmmh}. This proves the one step result. Now we can similarly
show by induction that if
$P^{i}_\Theta(\theta; A)> 0 $ for $i=1, \dots, k$ implies that $\ov P_{\Theta,U} (\theta, u; A \times B) > 0 $
for any $u \in \mathcal{ U}$ and $B \in \mathcal{B}(\mathcal{U})$ with $P_U(B) > 0 $, then the same holds true for $i=k+1$. This completes the proof for
$G = 1$.

We now consider the $G=2$ case.
We will show that for $k \geq 2$, if $P_\Theta^{k}(\theta; A)> 0 $ then $\ov P^{k}_{\Theta,U} (\theta, u_{1:2}; A \times (B_1 \times B_2)) > 0 $
for $u_{1:2} \in \ov{\mathcal{U}}^2$ with $p_U(B_1)> 0 $ and $p_U(B_2) > 0 $. Let $\wt B_{-2} = B_1 \times \{u_2\} $ and $\wt B_{-1} = \{u_1\} \times B_2$.
Then, we can show similarly to the $G = 1$ case that if $P_\Theta(\theta; A)> 0 $, then $\ov P_{\Theta,U} (\theta, u_{1:2}; A \times (\wt B_{-1} \cup \wt B_{-2}))> 0 $.
The result for $ k \geq 2 $ follows as in the $G = 1$ case.

We can similarly obtain the result for a general $G$.
If  $k \geq G$ and  $P_{\Theta}^k(\theta; A)> 0 $ then $\ov P^{k}_{\Theta,U} (\theta, u_{1:G}; A \times (B_1 \times B_2 \times \cdots \times  B_G)) > 0 $,
where $B_i \in \mathcal{B}(\mathcal{U})$ with $p_U(B_i) > 0 $ for all $i=1, \dots, G$. Suppose that the result is true for $G=1, \dots, g$. Now define
$\wt B_{-i} = B_1 \times B_2 \times B_{i-1} \times \{u_i\} \times B_{i+1} \times \cdots \times B_{g+1}$. Then, by the induction hypothesis,
if $P^{g}_\Theta(\theta; A) > 0 $, then $\ov P^g_{\Theta,U} (\theta, u_{1:g+1}; A \times ( B_{-1} \cup \cdots \cup B_{-g-1}) > 0 $ assuming that $p_U(B_i) > 0 $ for $i=1, \dots, g+1$. The required result now follows as in the $G=1$ case.

The irreducibility and aperiodicity of the sampling scheme now follows from that of Algorithm~\ref{alg: theta based sampling scheme}.

Part (iii) follows from the strong law of large numbers for Markov chain, see e.g. \citet[Theorem 17.0.1]{meyn2012markov}.

To prove Part~(iv), we first consider the numerator of \eqref{eq:ISestimator}. By Theorem 27 in \cite{roberts2004general},\begin{align*}
\sqrt N \Big( N^{-1} \sum_{i=1}^N S_i \psi_i - \E _{\ov \pi} (S\psi) \Big) \rightarrow N\Big(0, \Var_{\ov \pi} (\psi S)\mathrm{IF}_{\ov \pi, \psi S} \Big).
\end{align*}
Next, by the strong law of large numbers for Markov chains $N^{-1} \sum_{i=1}^N S_i \rightarrow \E_{\ov \pi} (S)$ $\overline{\pi}$-almost surely, i.e.
the denominator  of \eqref{eq:ISestimator} converges to $\E_{\ov \pi} (S) \neq 0$.
The result now follows from Slutsky's theorem. An alternative proof is given in \cite{lyne2015russian} based on a bivariate CLT for the numerator and denominator of \eqref{eq:ISestimator}, followed by the delta method to prove the result. The delta method is valid under the assumption $\E_{\ov \pi} (S) \neq 0$.
\end{proof}

\begin{proof}[Proof of Lemma~\ref{lemma: u_i are indep pi bar}]
Using the notation in \eqref{eq:Z_defined}, it is straightforward to show that
\begin{align*}
\ov \pi(d u_{1:G}|\theta) & = \prod_{i=1}^G \exp\Big(\wt z_i(u_i, \theta)\Big)p_{U_i} (u_i) .
\end{align*}
\end{proof}
%\newpage
\subsection{Proofs for Section \ref{app:IF}}\label{subsec:ProofIF}

\begin{proof}[Proof of Lemma \ref{lem:change_of_measure}]
By Assumption \ref{ass:change_of_measure} and $\overline{\pi}(d\theta, du)=\bar \pi(\theta)\overline{\pi}(du|\theta)$,
\begin{align*}
\ov \pi(\theta, dv, d z, dw) &= \ov \pi(d\theta, du) |J(\theta,u)|
= \exp(z) C(\theta) p(d\theta, d v, d z, dw).
\end{align*}
Hence, $\ov \pi(d\theta, dv, d z )  = \exp(z) C(\theta) p(d\theta, d v, d z)$ and
\begin{align*}
\ov \pi(d\theta, ds, d z ) & = \exp(z) C(\theta) \int _{s =\textrm{sign} (v)} p(d\theta, d v, d z)\\
& = \exp(z) \bar \pi(\theta) p(ds|\theta) p(d z |\theta),
\end{align*}
with $p(ds|\theta) \coloneqq \int_{s = \mathrm{sign} (v)}p(dv|\theta)$.
\end{proof}

\begin{proof}[Proof of Lemma \ref{lemma: idealized ineff}]
Part (i) follows because
\begin{align*}
\frac{\ov \pi (d \theta',d s', d z')}{\ov \pi (d \theta,d s, d z)} \frac{\bar \pi (d\theta) p(ds|\theta) q(z';dz| \sigma^2, \rho)} {\bar \pi (d\theta') p(ds'|\theta') q(z;dz'| \sigma^2, \rho)},
\end{align*}
where the perfect proposals for $\theta$ and $s$ cancel the corresponding terms in $\ov \pi$.
Moreover, we can show that $$p(dz)q(z;dz'| \sigma^2, \rho) = p(dz')q(z';dz| \sigma^2, \rho).$$
Part (ii) follows from Part (i) and the fact that $q(z;dz'|\sigma^2, \rho)$ does not depend on $\theta$ and $s$.
Part (iii) then follows from Lemma 4 in \cite{pitt2012some}, but with the probability of accepting a proposal conditional on $z$ which arises from the correlated proposal.
\end{proof}

\begin{proof}[Proof of Lemma \ref{lem:bivariate_dist_z}]
$\mathrm{E}[Z^\prime] = (G-2)\sigma^2/2G$ since, under Assumption \ref{ass:more_restrictive_assumptions},
$$\sum_{i\not=j,i=1}^G Z_i(\theta',U_i) \sim \mathcal{N}\left(\frac{(G - 1)\sigma^2}{2G}, \frac{(G - 1)\sigma^2}{G} \right)
\quad \text{ and }\quad  Z_j(\theta',U_j') \sim \mathcal{N}\left( -\frac{\sigma^2}{2G}, \frac{\sigma^2}{G}
\right).$$
It also follows that $\mathrm{V}[Z'] = \sigma^2$. Moreover, $Z \sim \mathcal{N}(\sigma^2/2, \sigma^2)$, concluding the proof.
\end{proof}

%\subsection{Proofs for Section \ref{app:IFBlockPoisson}}\label{app:ProofsAppendixB}
To prove Lemma \ref{lem:VarAbsL_finite_mixture},
 we first need the preliminary Lemmas~\ref{lem:Mixture} to
\ref{lem:momentsLogNonCentral}.
\begin{lemma}
\label{lem:Mixture}(Non-central $\chi^{2}$ is a Poisson mixture
of central $\chi^{2}$). If $J\sim\mathrm{Pois}(\mu/2)$ and $W\vert J\sim\chi^{2}(k+2J)$,
then marginally
$
W\sim\chi^{2}\left(k,\mu\right).
$
\end{lemma}
\begin{proof}
See \citet{walck1996hand}.
\end{proof}

\begin{lemma}
\label{lem:momentsLogCentral}(Moments of log central $\chi^{2}$.\citep{pav2015moments} ).
If $W\sim\chi^{2}\left(k\right)$ and $Y=\log W$, then
$
\mathrm{E}Y  =\log2+\psi(k/2) $ and $\mathrm{V}Y  =\psi^{(1)}(k/2)$.
\end{lemma}

\begin{lemma}
\label{lem:momentsLogNonCentral}(Moments of log non-central $\chi^{2}$).
If $W\sim\chi^{2}\left(k,\mu\right)$ and $Y=\log W$, then
\begin{align*}
\mathrm{E}[Y] & =\log2+\mathrm{E}_{J}\left(\psi^{(0)}(k/2+J)\right)\\
\mathrm{V}[Y] & =\mathrm{E}_{J}\left[\psi^{(1)}(k/2+J)\right]+\mathrm{V}_{J}\left[\psi^{(0)}(k/2+J)\right],
\end{align*}
where $J\sim\mathrm{Pois}(\mu/2)$ and $\psi^{(q)}$ is the polygamma
function of order $q$.
\end{lemma}
\begin{proof}
Follows from \citet{pav2015moments}. From the mixture representation
in Lemma \ref{lem:Mixture} we know that we can represent $W\sim\chi^{2}\left(k,\mu\right)$
as $J\sim\mathrm{Pois}(\mu/2)$ and $W\vert J\sim\chi^{2}(k+2J)$.
By the law of iterated expectations and Lemma \ref{lem:momentsLogCentral}
\[
\mathrm{E}[Y]=\mathrm{E}_{J}[\mathrm{E}_{W\vert J}\left[Y\right]]=\log2+\mathrm{E}_{J}\left[\psi^{(0)}(k/2+J)\right].
\]
Also, from the law of total variance and Lemma \ref{lem:momentsLogCentral}
\begin{align*}
\mathrm{V}[Y] & =\mathrm{E}_{J}[\mathrm{V}_{W\vert J}[Y]]+\mathrm{V}_{J}[\mathrm{E}_{W\vert J}[Y]]
 =\mathrm{E}_{J}\left[\psi^{(1)}(k/2+J)\right]+\mathrm{V}_{J}\left[\psi^{(0)}(k/2+J]\right).
\end{align*}
\end{proof}
%\vspace{-1.5in}
\begin{proof}[Proof of Lemma \ref{lem:VarAbsL_finite_mixture}]
When $a=d-\lambda$ we have
\begin{equation*}
\log\left|\widehat{L}_B\right|=q+d+\sum_{l=1}^{\lambda}\sum_{h=1}^{\mathcal{X}_{l}}\log\left(\left|\frac{\widehat{d}_{m}^{\,\,(h,l)}-d}{\lambda}+1\right|\right)=q+d+\sum_{l=1}^{\lambda}\sum_{h=1}^{\mathcal{X}_{l}}\log\left(\left|\frac{\sqrt{\frac{\gamma}{m}}\bar{d}_{m}^{(h,l)}}{\lambda}+1\right|\right),
\end{equation*}
where $\mathcal{X}_{l}\sim\mathrm{Pois}(\text{1})$, $l=1,...,\lambda$.
Let $I^{(h,l)}\in\{1,...,C\}$ be indicators such that $I^{(h,l)}=c$
means that observation $\bar{d}_{m}^{(h,l)}$ comes from the $c$th
mixture component $\mathcal{N}(\mu_{c},\sigma_{c}^{2})$ with $\mathrm{Pr}(I^{(h,l)}=c)=\omega_{c}$.
Now, since the $\bar{d}_{m}^{(h,l)}$ are $iid$ and the total
number of $\bar{d}_{m}^{(h,l)}$ is $\sum_{l=1}^{\lambda}\mathcal{X}_{l}$
we have
\[
\mathrm{V}\left(\log\left|\widehat{L}_B\right|\vert\mathcal{X}_{1:\lambda}\right)=\left(\sum_{l=1}^{\lambda}\mathcal{X}_{l}\right)\mathrm{V}\log\left(\left|\frac{\sqrt{\frac{\gamma}{m}}\bar{d}_{m}^{(h,l)}}{\lambda}+1\right|\right).
\]
Define
\[
X^{(h,l)}=\frac{\sqrt{\frac{\gamma}{m}}\left(\bar{d}_{m}^{(h,l)}-\mu_{I^{(h,l)}}\right)}{\lambda},
\]
and note that
\[
X^{(h,l)}\vert\left(I^{(h,l)}=c\right)\sim N\left(0,\tilde{\sigma}_{c}^{2}\right),
\]
where $\tilde{\sigma}_{c}^{2}=\frac{\sigma_{c}^{2}}{\lambda^{2}}\frac{\gamma}{m}$.
Now, conditional on $I^{(h,l)}=c$, we have
\begin{align*}
\log\left(\left|\frac{\sqrt{\frac{\gamma}{m}}\bar{d}_{m}^{(h,l)}}{\lambda}+1\right|\right) & =\log\left(\left|X^{(h,l)}+\frac{\sqrt{\frac{\gamma}{m}}\mu_c+\lambda}{\lambda}\right|\right)\\
 & \overset{d}{=}\log\left(\left|\tilde{\sigma}_{c}Z+\frac{\sqrt{\frac{\gamma}{m}}\mu_c+\lambda}{\lambda}\right|\right),\text{ where }Z\sim \mathcal{N}(0,1)\\
 & =\log\tilde{\sigma}_c+\log\left(\left|Z+\frac{\sqrt{\frac{\gamma}{m}}\mu_c+\lambda}{\lambda\tilde{\sigma}_c}\right|\right)\\
 & =\log\left(\frac{\sigma_{c}}{\lambda}\sqrt{\frac{\gamma}{m}}\right)+\log\left(\left|Z+\frac{\mu_c+\sqrt{\frac{m}{\gamma}}\lambda}{\sigma_{c}}\right|\right)\\
 & =\log\left(\frac{\sigma_{c}}{\lambda}\sqrt{\frac{\gamma}{m}}\right)+\frac{1}{2}\log\left(\left(Z+\frac{\mu_c+\sqrt{\frac{m}{\gamma}}\lambda}{\sigma_{c}}\right)^{2}\right)\\
 & \overset{d}{=}\log\left(\frac{\sigma_{c}}{\lambda}\sqrt{\frac{\gamma}{m}}\right)+\frac{1}{2}\log\left(W^{(h,l)}\right),
 \\
 & \text{ where }W^{(h,l)}\sim\chi^{2}\left(1,\frac{(\mu_c+\sqrt{\frac{m}{\gamma}}\lambda)^{2}}{\sigma_{c}^{2}}\right).
\end{align*}
where $\chi^{2}\left(k,\lambda\right)$ denotes the non-central $\chi^{2}$
distribution with $k$ degrees of freedom and non-centrality parameter
$\lambda$. So $\log\left(\left|\frac{\sqrt{\frac{\gamma}{m}}\bar{d}_{m}^{(h,l)}}{\lambda}+1\right|\right)$
is a mixture of log of non-central $\chi^{2}$ variables with component
means and variances given by Lemma \ref{lem:momentsLogNonCentral}
\[
\eta_{c}\coloneqq\mathrm{E}\log\left(\left|\frac{\sqrt{\frac{\gamma}{m}}\bar{d}_{m}^{(h,l)}}{\lambda}+1\right|\vert I^{(k,l)}=c\right)=\log\left(\frac{\sigma_{c}}{\lambda}\sqrt{\frac{\gamma}{m}}\right)+\frac{1}{2}\left(\log2+\mathrm{E}_{J_{c}}\left(\psi^{(0)}(1/2+J_{c})\right)\right)
\]
and
\[
\nu_{c}^{2}\coloneqq\mathrm{V}\log\left(\left|\frac{\sqrt{\frac{\gamma}{m}}\bar{d}_{m}^{(h,l)}}{\lambda}+1\right|\vert I^{(k,l)}=c\right)=\frac{1}{4}\left[\mathrm{E}_{J_{c}}\left(\psi^{(1)}(1/2+J_{c})\right)+\mathrm{V}_{J_{c}}\left(\psi^{(0)}(1/2+J_{c})\right)\right]
\]
where the $J_{c}$ are independent and $J_{c}\sim\mathrm{Pois}\left(\frac{(\mu_c+\sqrt{\frac{m}{\gamma}}\lambda)^{2}}{2\sigma_{c}^{2}}\right)$.
By the mean and variance of a finite mixture we then have
\[
\eta\coloneqq\mathrm{E}\log\left(\left|\frac{\sqrt{\frac{\gamma}{m}}\bar{d}_{m}^{(h,l)}}{\lambda}+1\right|\right)=\sum_{c=1}^{C}\omega_{c}\eta_{c}
\]
\[
\mathrm{V}\log\left(\left|\frac{\sqrt{\frac{\gamma}{m}}\bar{d}_{m}^{(h,l)}}{\lambda}+1\right|\right)=\sum_{c=1}^{C}\omega_{c}(\nu_{c}^{2}+(\eta_{c}-\eta)^{2}).
\]
 Finally,
\begin{align*}
\mathrm{V}\left(\log\left|\widehat{L}_B\right|\right) & =\mathrm{E}_{\mathcal{X}_{1:\lambda}}\mathrm{V}\left(\log\left|\widehat{L}_B\right|\vert\mathcal{X}_{1:\lambda}\right)+\mathrm{V}_{\mathcal{X}_{1:\lambda}}\mathrm{E}\left(\log\left|\widehat{L}_B\right|\vert\mathcal{X}_{1:\lambda}\right)\\
 & =\mathrm{E}_{\mathcal{X}_{1:\lambda}}\left[\left(\sum_{l=1}^{\lambda}\mathcal{X}_{l}\right)\sum_{c=1}^{C}\omega_{c}(\nu_{c}^{2}+(\eta_{c}-\eta)^{2})\right]+\mathrm{V}_{\mathcal{X}_{1:\lambda}}\left(\left(\sum_{l=1}^{\lambda}\mathcal{X}_{l}\right)\eta\right)\\
 & =\lambda\sum_{c=1}^{C}\omega_{c}(\nu_{c}^{2}+(\eta_{c}-\eta)^{2})+\lambda\eta^{2}.
\end{align*}
$\mathrm{V}\left(\log\left|\widehat{L}_B\right|\right)$is finite since
$\psi^{(1)}(1/2+J)\leq\pi^{2}/2$ for all $J\geq0$ and for all $J\geq0$

\[
\left(\psi^{(0)}(1/2+J)\right)^{2}=\left(\psi^{(0)}(1)-2\log2+2\sum_{k=1}^{J}\frac{1}{2k-1}\right)^{2}<\left(\psi^{(0)}(1)-2\log2+2J\right)^{2}
\]
 and the Poisson has finite first and second moments.
\end{proof}

\subsection{Proofs for Section \ref{app: block PMMH L geq 0}\label{subsec:ProofsAppendixC}}
\begin{proof}[Proof of Lemma \ref{lem:Optimal_sigma2_pos_estimator}]
The result follows from numerically optimizing the expression
$$\mathrm{CT}(\sigma^2, \rho) \coloneqq \frac{\mathrm{IF}(\sigma^2,  \rho)}{\sigma^2},$$
with IF defined in Part (iii) of \eqref{lemma: idealized ineff}.
\end{proof}

\end{document}